\documentclass[twocolumns]{aa}

\usepackage[T1]{fontenc}
\usepackage[utf8]{inputenc}

\usepackage{graphicx}
\usepackage{natbib}
\usepackage{ulem}
\usepackage{textcomp}
\usepackage{gensymb}
\usepackage{longtable}
\usepackage{threeparttable}
\usepackage{multicol}
\usepackage{multirow}
\usepackage{textgreek}
\usepackage{float}
\usepackage{todonotes}
\usepackage[Symbol]{upgreek}
\usepackage{amsmath}
\usepackage{etoolbox}
\usepackage{txfonts}
\usepackage{url}
\usepackage{xcolor}

\usepackage[most]{tcolorbox}
\usepackage{hyperref}
\usepackage{xcolor}
\hypersetup{
  colorlinks   = true, 
  urlcolor     = blue, 
  linkcolor    = cyan,
  filecolor    = cyan,
  citecolor   = cyan 
}
\usepackage[all]{hypcap}

\newcommand\blfootnote[1]{%
  \begingroup
  \renewcommand\thefootnote{}\footnote{#1}%
  \addtocounter{footnote}{-1}%
  \endgroup
}
\usepackage{academicons}

\begin{document} 
	\title{Radio surface fluctuations in radio relics}
	\titlerunning{Radio surface fluctuations in radio relics}
	\authorrunning{Dom\'inguez-Fern\'andez et al.}

\author{P. Dom\'inguez-Fern\'andez\href{https://orcid.org/0000-0001-7058-8418}{\hskip2pt\includegraphics[width=9pt]{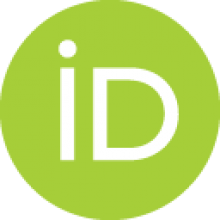}}\inst{1,2,3}, D.Ryu\href{https://orcid.org/0000-0002-5455-2957}{\hskip2pt\includegraphics[width=9pt]{orcid-ID.png}}\inst{4}, H. Kang\href{https://orcid.org/0000-0002-4674-5687}{\hskip2pt\includegraphics[width=9pt]{orcid-ID.png}}\inst{5}}

\institute{Harvard-Smithsonian Center for Astrophysics, 60 Garden Street, Cambridge, MA 02138, USA
\and
Dipartimento di Fisica e Astronomia, Universit\`a di Bologna, via P. Gobetti 93/2, 40129, Bologna, Italy
\and
INAF - Osservatorio di Astrofisica e Scienza dello Spazio di Bologna, via Gobetti 93/3, 40129 Bologna, Italy 
\and
Department of Physics, College of Natural Sciences UNIST, Ulsan 44919, Republic of Korea
\and
Department of Earth Sciences, Pusan National University, Pusan 46241, Republic of Korea
}

  
\abstract
{
Recent observations have revealed detailed structures of radio relics in a wide range of frequencies.
In this work,  
we perform three-dimensional magnetohydrodynamical simulations of merger shocks propagating through a turbulent magnetized intracluster medium, and employ on-the-fly Lagrangian particles to explore the
physical processes originating radio substructures and their appearances in high and low-frequency observations. We employ two cosmic-ray (CR) electron acceleration models: the fresh injection of electrons from the thermal pool and the re-acceleration of mildly relativistic electrons. We use the relative surface brightness fluctuations, $\delta S_{\nu}$, to define a ``degree of patchiness''.
We find that: 1) Patchiness is produced if the shock's surface has a distribution of Mach numbers, rather than a single Mach number;
2) Radio relics appear patchier if the Mach number distribution consists of a large percentage of low Mach numbers ($\mathcal{M}\lesssim2.5$);
3) As the frequency increases, the patchiness also becomes larger. Nevertheless, if radio relics are patchy at high frequencies (e.g., 18.6 GHz), they are necessarily also at low frequencies (e.g., 150 MHz);
4) To produce noticeable differences in the patchiness at low and high frequencies, the shock front should have a Mach number spread of $\sigma_{\mathcal{M}}\gtrsim0.3$--0.4;
5) The amount of the patchiness depends on the Mach number distribution as well as the CR acceleration model.
We propose $\delta S_{\nu}$ as a potential tool for extracting merger shock properties and information about particle acceleration processes at shocks in radio observations.
}


   \maketitle
%

\section{Introduction} \label{sec:intro}

\blfootnote{Corresponding author: Paola Dom\'inguez-Fern\'andez \\
\email{paola.dominguezfernandez@cfa.harvard.edu}}

Galaxy clusters reveal spectacular Mpc-scale diffuse radio emission. It seems clear that the observed emission is a consequence of merger events 
accelerating the synchrotron-emitting cosmic ray (CR) electrons (see \citealt{2011JApA...32..505V} or \citealt{2019SSRv..215...27B} for a review). 
The primary focus of this work is on \textit{radio relics}, which are typically found at the cluster periphery and exhibit elongated morphology with a high polarization fraction (see \citealt{br11} and \citealt{2019SSRv..215...16V} for reviews).

The elongated shape of radio relics is believed to be a result of
mild merger shocks ($\mathcal{M}_s \sim$ 1.7 -- 4) propagating in the intracluster medium (ICM) \citep[e.g.][]{ce06,vw10,2012A&A...546A.124V}. Nevertheless, recent high-resolution radio observations are revealing a plethora of additional complex substructures in radio relics \citep[e.g.][]{2018ApJ...852...65R,2020A&A...636A..30R,2014ApJ...794...24O,2017ApJ...835..197V,2018ApJ...865...24D,2021arXiv211106940D}. The origin of these radio substructures is challenging to understand due to the lack of our full understanding of the particle acceleration mechanisms causing the diffuse radio emission.

The preferred mechanism that accelerates relativistic particles at a collisionless shock is the first-order Fermi (Fermi-I) acceleration process of diffusive shock acceleration (DSA) \citep[e.g.][]{be87,1983RPPh...46..973D}. 
In this case, particles need to reach high enough momenta to cross
the shock transition zone, which
has a thickness of the order of the gyroradius of post-shock thermal
ions, and then they can fully participate in the DSA process. It has been widely discussed that there are two types of electron populations in the ICM that can participate in the DSA: thermal electrons and pre-existing mildly relativistic ($10 \lesssim \gamma \lesssim 10^4$) electrons \citep[e.g.][]{ka12,2013MNRAS.435.1061P}.

Following the idea that thermal electrons are directly injected into the DSA, or what is more commonly referred to as \textit{fresh injection}, presents certain challenges in our current understanding to explain observed radio relics such as: (i) an unrealistic shock acceleration efficiency in order to explain the observed radio power of some relics (see \citealt{2019SSRv..215...16V,2020A&A...634A..64B}); (ii) a discrepancy between the Mach numbers inferred from X-ray observations assuming density and/or temperature jump conditions and those inferred from radio observations assuming DSA \citep[e.g.][]{2020A&A...634A..64B}; (iii) a mismatch between our expectations from plasma physics and radio/X-ray observations of very weak shocks. 
In fact, recent Particle In Cell (PIC) simulations have shown that thermal electrons in the ICM can only be energized to the full DSA regime only in supercritical shocks with a sonic Mach number greater than a critical Mach number, $\mathcal{M}_{cr} \sim 2.3$ \citep[see e.g.][]{2019ApJ...876...79K,2021ApJ...915...18H}, which is at odds with several $\mathcal{M}_{radio} \sim$ 1.5 -- 2.3 shocks detected in the ICM \citep[e.g.][]{2019SSRv..215...16V}. Additional factors, such as the presence of pre-existing fossil CR electrons could potentially explain (i) and (iii) \citep[however, see discussion in][]{2021arXiv211014236H}, while pre-existing turbulence could explain (ii) \citep[e.g.][]{dominguezfernandez2020morphology,Wittor_2021}. Yet, our understanding on the involved processes still falls short and the details should be further investigated.

Regarding the substructure within relics, there are 
some 
observed features that cannot be easily understood. For example,
filamentary structures in the form of threads, twisted ribbons and/or bristles have been consistently observed in various radio relics. Some examples where these filamentary structures of few kpc scales have been found in the
Toothbrush \citep{2018ApJ...852...65R,2020A&A...636A..30R}, Sausage
\citep{2018ApJ...865...24D}, MACS J7017+35 \citep{2017ApJ...835..197V}, Abell 2256 \citep{Kamlesh_Abell2256}, and Abell 3667 \citep{2021arXiv211106940D} relics, despite of the very diverse merger scenarios of the host galaxy clusters in which each of these relics have formed. The majority of these filamentary structures point towards sites of shock acceleration \citep[see e.g.][]{2021arXiv211106940D}. While repeated DSA events at work in the downstream or in projection could potentially explain their origin, a multiple weak shock scenario in the ICM needs yet to be understood.

On the other hand, 
shock surfaces and their substructure in the ICM are challenging to be numerically reproduced. A limited knowledge on the merger history of each galaxy cluster, on the ICM conditions at clusters outskirts and/or on the CR budget of clusters add to the challenge.
Yet, some radio relics born in binary cluster merger events and observed close
to edge-on (i.e. small projection on the plane of the sky) serve as the simplest examples
for testing our current
theories. 
For example, the so-called Sausage relic in the CIZA J2242.8+5301 galaxy cluster showing an almost perfect $\sim 2$ Mpc arc-like structure and seen close
to edge-on, is one of the most studied radio relics and it has been observed in a very wide range of radio frequencies.
The Sausage relic has been primarily studied at low and medium radio frequencies, e.g. from $\sim$150 MHz to $\sim$6.5 GHz
\citep[see e.g.][]{vw10,Stroe2013,2016MNRAS.455.2402S,Loi_2017,Hoang2017,2017A&A...600A..18K,2018ApJ...865...24D}.
On the other end, at high frequencies, the Sausage relic has been observed with single-dish telescopes such as the Effelsberg telescope observing at 14.25 GHz and the Sardinia Radio Telescope (SRT) observing at 18.6 GHz \citep[][]{Loi_2020}, and
with the Arcminute Microkelvin Imager (AMI) and Combined Array for Research in Millimeter-wave Astronomy (CARMA) interferometers observing at 16 GHz and 30 GHz \citep[][]{2014MNRAS.441L..41S,2016MNRAS.455.2402S}. We refer the reader to Fig. 20 in Section 4.1.3 of \citealt{Stroe2013}, for a very illustrative example of the variations along the shock front in the Sausage relic at various frequencies.

This wealth of information inspires us to
study the surface brightness distribution of a shock front at high and low radio frequencies.  
Our primary objective in this study is to gain insights into the radio substructures generated by physical conditions. Specifically, radio substructures that arise as a consequence of our current particle shock acceleration theories across a wide range of radio frequencies. Our central emphasis lies in the analysis of radio surface brightness variations or fluctuations.

In \citealt{dominguezfernandez2020morphology,dominguez2021} it has been shown that a model of thermal electrons injected into DSA in the presence of a turbulent pre-shock magnetized ICM can reproduce some of the morphological small-scale characteristics of radio relics and also their most important observed features in continuum and polarized radio emission. In this work, we follow the numerical approach presented in \citealt{dominguezfernandez2020morphology} where an ensemble of tracer (Lagrangian) particles, represent a whole distribution of electrons embedded in a magneto-hydrodynamical (MHD) fluid. 
This type of hybrid modelling has been previously applied in post-processing of Eulerian cosmological MHD simulations (e.g. \citealt{sk13,2015ApJ...812...49H,2017MNRAS.470..240N,wittor2019,2019ApJ...883..138R}) and more recently, implemented on-the-fly in Lagrangian cosmological MHD simulations \citep[see e.g.][]{2022arXiv220705087B}. In this work, we model the cooling processes of each tracer particle at run-time using the MHD code PLUTO \citep[][]{pluto1,2018ApJ...865..144V}.

We consider a simplified 
set-up where  
a shock tube
is present in a turbulent medium that is representative of a small region of the ICM. We then assume that CR electrons are injected instantly at the shock discontinuity and acquire spectral properties according to the DSA theory. We focus on the two plausible models for ``injection'', namely thermal electrons injected into DSA (fresh injection) and fossil CR electrons injected into DSA (re-acceleration). In general, the re-acceleration model relies on a source providing the population of fossil electrons such as a radio jet from an AGN or a previous episode of shock/turbulence acceleration \citep{1999ApJ...520..529S,ka12}. Such electron population is therefore not expected to be distributed uniformly in the ICM, but concentrated in e.g. fossil clouds or bubbles. 

The paper is structured as follows. In Section \ref{section:num_set-up}, we describe our numerical set-up and initial conditions. In Section \ref{section:methods}, we include a description of the particles' initial spectral distribution and evolution. In this Section, we also explain how we obtain the emission maps. Section \ref{sec:results} shows our results and we summarize in Section \ref{sec:conclusions}.

\section{Numerical set-up}\label{section:num_set-up}

\subsection{Initial conditions: modelling the turbulent ICM with FLASH}
\label{sec:turb}

\begin{figure}
    \centering
    \includegraphics[width=0.8\columnwidth]{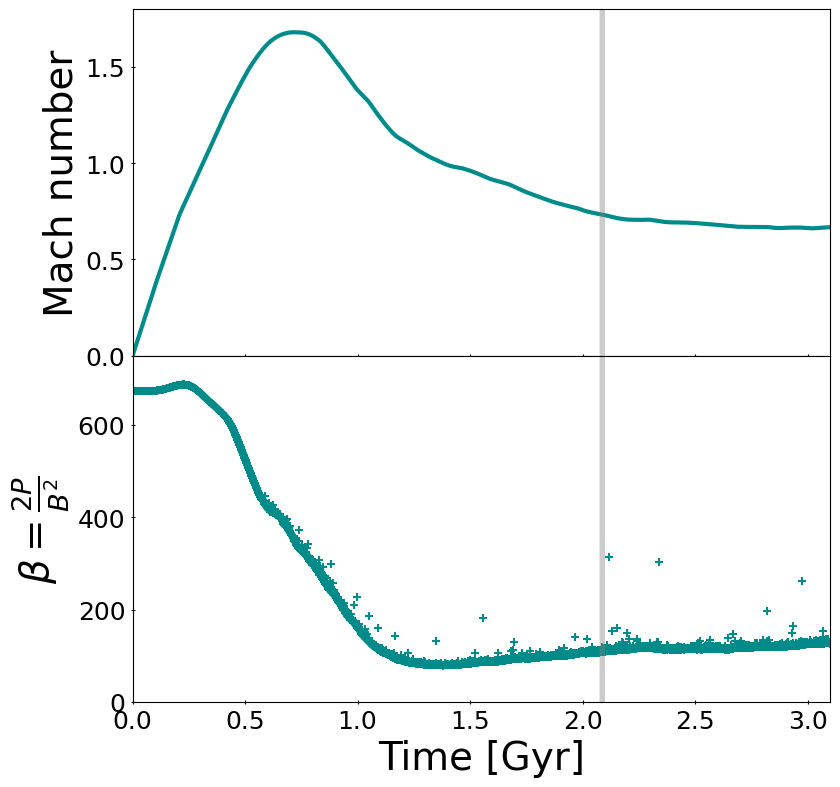}
    \includegraphics[width=0.8\columnwidth]{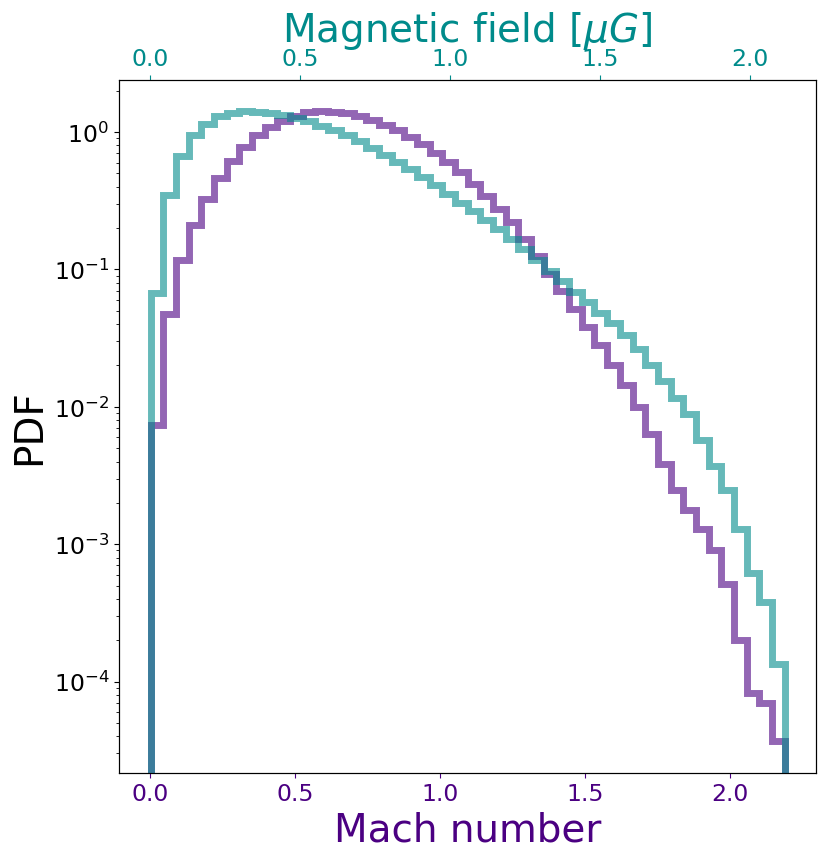}
    \caption{Upper: Evolution of rms Mach number (\textit{top panel}) and plasma beta (\textit{bottom panel}). The vertical gray line indicates the selected time in the FLASH simulation to be our initial turbulent ICM condition. The selected snapshot has an rms Mach number of $\mathcal{M}_s \sim$0.7 and a plasma beta of $\beta_{p} \sim$110. Lower: Magnetic field strength (\textit{darkcyan}) and Mach number (\textit{purple}) PDFs at the selected time in the FLASH simulation.}
    \label{fig:PDFs_B1}
\end{figure}

The turbulent ICM initial conditions were produced using the MHD FLASH code version 4.6.1 \citep{2000ApJS..131..273F, 2002ApJS..143..201C}. We use the same set-up used in \citealt{dominguezfernandez2020morphology,dominguez2021}: the unsplit staggered mesh (USM) MHD solver, which employs a constrained transport (CT) method at cell interfaces to maintain the divergence-free magnetic field property on a staggered grid \citep[e.g.][]{2009ASPC..406..243L}. The simulation domain is chosen to be a cubic box of size $L=L_x=L_y=L_z$, uniformly spaced over a $256^3$ cells grid, with periodic boundary conditions. We assumed an ideal gas equation of state with $\gamma_{0}=5/3$.

We employ a spectral forcing method based on the stochastic Ornstein-Uhlenbeck (OU) process with a finite autocorrelation time to generate turbulence \citep[e.g.][]{1988PhFl...31..506E,2006A&A...450..265S,2010A&A...512A..81F}. We use a solenoidal subsonic turbulence forcing with an injection scale of $2L/3$. The forcing amplitude, $f_0$, was set to be a paraboloid in Fourier space only containing power on the largest scales: $1 \leq kL/2\pi \leq 2$. The autocorrelation timescale was set equal to the dynamical timescale on the scale of energy injection, $t_{2L/3}=2L/3\sigma_v$, where $\sigma_v=125$ km/s is the rms velocity amplitude of the fluctuations.

We set a magnetic field seed of 0.05 $\mu$G on 
the density 
$10^{-4} m_p\,\mathrm{cm}^{-3}$. The evolution of the rms Mach number and the volume weighted mean plasma beta, $\beta_p$, is shown on the upper panel of Fig.~\ref{fig:PDFs_B1}. 
We selected a snapshot corresponding to $2t_{2L/3}$
to act as our initial condition (see Sec.~\ref{sec:PLUTO} and main set-up in Fig.~\ref{fig:init}). 
On the lower panel of Fig.~\ref{fig:PDFs_B1}, we show the probability distribution function (PDF) of the magnetic field strength and Mach number at that time.

We note that the physical evolution of these turbulence-in-a-box simulations can be
fully described by the dimensionless sonic and Alfv\'en Mach numbers, $\mathcal{M}$ and $\mathcal{M}_{A}$. Therefore, as long as 
$\mathcal{M}$ and $\mathcal{M}_{A}$ remain unchanged, all dimensional variables, such as $L$, $\rho$, $v$, $B$, etc., can
be scaled arbitrarily.
In the following, we work with $L=200$ kpc. We would like to emphasize that our analysis of a final shock with an extent of 3 Mpc is specifically discussed in Sec.~\ref{sec:beam}. In this particular case, we employed the stacking of our simulation boxes, made feasible due to our periodic boundary conditions. For further details, we direct the reader to that section.


\subsection{Main PLUTO simulations}
\label{sec:PLUTO}

\begin{figure}
    \centering
    \includegraphics[width=\columnwidth]{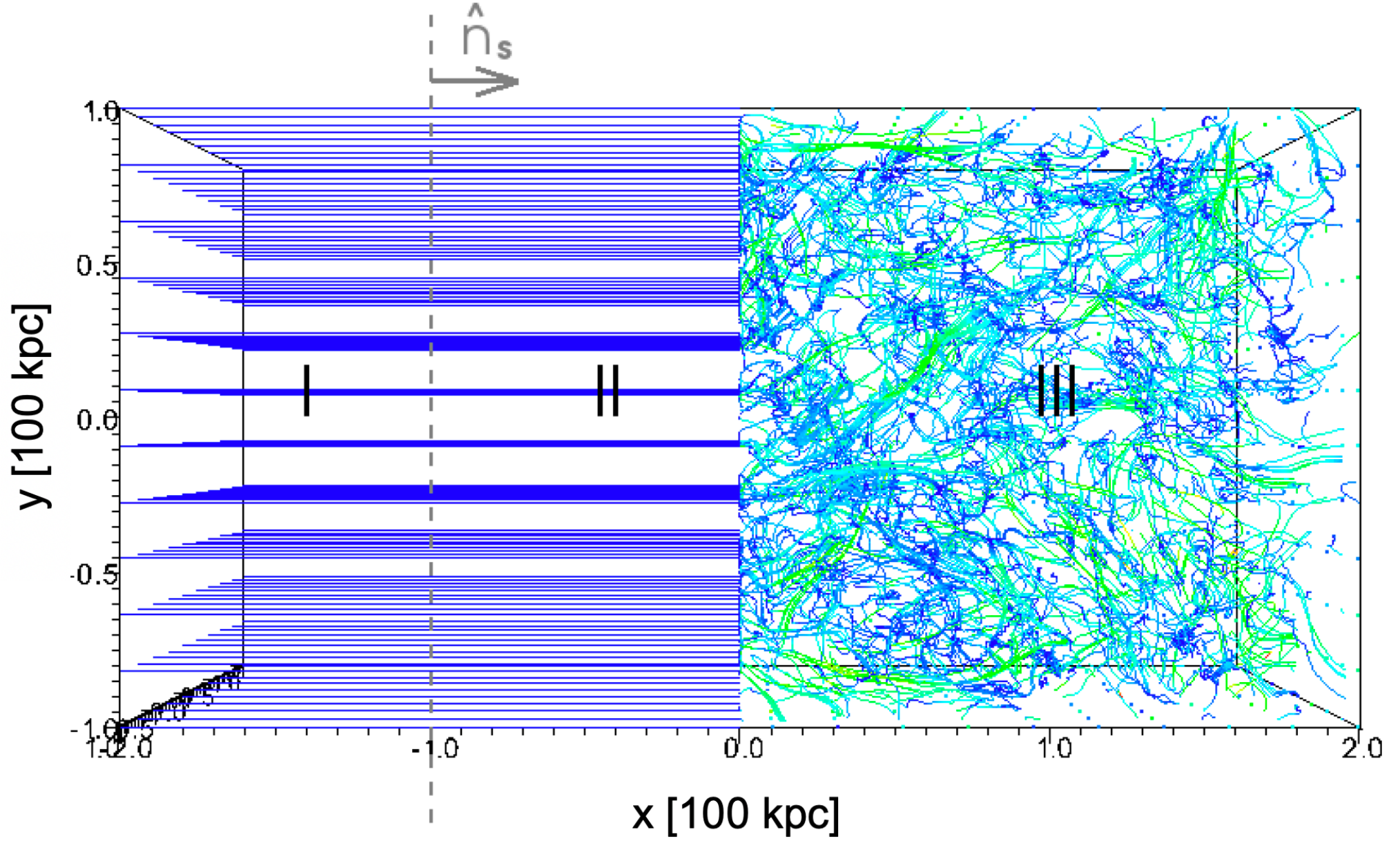}
    \caption{Initial magnetic field configuration in the PLUTO code taken from \citealt{dominguezfernandez2020morphology}. The streamlines are coloured according to the magnitude of the magnetic field: green, light colours denote higher values, while dark blue colour indicates lower values. I denotes the post-shock region, II the uniform pre-shock region, and III the turbulent pre-shock region  (see Sec.~\ref{sec:turb}). The left side is a uniform medium with a $B_x$ component matching the mean value of the $B_x$ of the turbulent medium. We have one Lagrangian particle per cell placed in the whole regions II and III.}
    \label{fig:init}
\end{figure}

In order to study the synchrotron emission in an MHD shock tube, we use the code PLUTO \citep{pluto1}, and follow the same set-up used in \citealt{dominguezfernandez2020morphology,dominguez2021}. The PLUTO code solves the following conservation laws for ideal MHD:
\begin{equation}
    \frac{\partial \rho}{\partial t} + \mathbf{\nabla} \cdot (\rho \mathbf{v}) = 0, 
\end{equation}
\begin{equation}
    \frac{\partial \mathbf{m}}{\partial t} + \mathbf{\nabla} \cdot \left[ \mathbf{m} \mathbf{v} - \frac{1}{4\pi}\mathbf{B}\mathbf{B} + \mathbf{I} \left( p + \frac{B^2}{8\pi} \right) \right]^T = 0,
\end{equation}
\begin{small}
\begin{equation}
    \frac{\partial (E_t)}{\partial t} + \mathbf{\nabla} \cdot \left[ \left( \frac{\rho v^2}{2} + \rho e + p  \right)\mathbf{v} - \frac{1}{4\pi}(\mathbf{v} \times \mathbf{B}) \times \mathbf{B} \right] = 0, 
\end{equation}
\end{small}
\begin{equation}
    \frac{\partial \mathbf{B}}{\partial t} - \mathbf{\nabla} \times (\mathbf{v} \times \mathbf{B}) = 0,
\end{equation}
\begin{equation}
    \mathbf{\nabla} \cdot \mathbf{B} = 0,
\end{equation}
\begin{equation}
    \rho e =  \frac{p}{\gamma_0 - 1},
\end{equation}
where $\rho$ is the gas mass density, $\mathbf{m}=\rho \mathbf{v}$ is the momentum density, $p$ is the thermal pressure, $\mathbf{B}$ is the magnetic field, 
$e$ the specific internal energy, here the adiabatic index is $\gamma_{0} = 5/3$ as well 
and $E_t$ is the total energy density defined as
\begin{equation}
    E_t = \rho e + \frac{\mathbf{m}^2}{2\rho} + \frac{\mathbf{B}^2}{8\pi},
\end{equation}

For more detailed information on the PLUTO code, we recommend referring to \citealt{pluto1}.

Our computational domain is a rectangular box (400 kpc $\times$ 200 kpc $\times$ 200 kpc with $256 \times 128 \times 128$ cells, respectively), where $x\in$[-200,200] kpc, $y \in$[-100,100] kpc, and $z\in$[-100,100] kpc (see Fig.~\ref{fig:init}). The right-hand half of the domain is filled with a turbulent medium
(see Sec.~\ref{sec:turb}), representing
a realistic ICM, while the left-hand half contains a uniform medium in which the shock is launched. We define a shock discontinuity at $x=-100$ kpc (see Fig.~\ref{fig:init} for the initial configuration of the magnetic field). This defines three regions in our simulation box: a post-shock uniform region (I), a pre-shock uniform region (II) and a pre-shock turbulent region (III).

The turbulent medium is produced externally, with the procedure outlined in Sec.~\ref{sec:turb}. The turbulent fields are then read into PLUTO and interpolated onto the numerical mesh used to evolve shocks with a  bi- or tri-linear interpolation at the desired coordinate location at the initial time.
The initial boundary conditions of the computational domain are \textit{outflow} in $x$ (zero gradient across the boundary) and \textit{periodic} in $y$ and $z$. We used a piecewise parabolic method (PPM) for the spatial integration, whereas a $2^{nd}$ order TVD Runge-Kutta method for the time stepping with a Courant-Friedrichs-Lewy (CFL) condition of $0.3$. The Riemann solver for the flux computation that we used is the Harten-Lax-van Leer-Discontinuities (HLLD) solver \citep[see][]{2005JCoPh.208..315M}. We control the $\nabla \cdot \mathbf{B} = 0$ condition with the hyperbolic divergence cleaning technique where the induction equation is coupled to a generalized Lagrange multiplier (GLM) \citep[e.g.][]{ded02}. We refer the reader to Appendix A of \citealt{dominguezfernandez2020morphology} for more details on the effects of the
initial interpolation of the external input and handling of the magnetic field divergence under the GLM and constrained transport techniques. \\
The initial conditions for the density, pressure and velocity in region II (\textit{pre-shock} uniform region at [-100,0] kpc) 
 are set to the mean value of the corresponding turbulent fields. In the case of the magnetic field in region II, we set it to be the mean value of the $B_x$ component of the turbulent medium. The initial conditions for region I (\textit{post-shock} region) are selected according to the Rankine-Hugoniot conditions \citep[e.g.][]{Landau1987Fluid}. 
 We study shocks propagating 
  into the turbulent medium
 with different initial sonic Mach numbers. Shocks can be classified as quasi-parallel and quasi-perpendicular if $\theta_{bn} \leq 45^{\circ}$ or $\theta_{bn} > 45^{\circ}$, respectively, where $\theta_{bn}$ is the angle between the upstream magnetic field and normal of the shock. In this work, we consider only the quasi-parallel limit, i.e. $\theta_{bn} = 0^{\circ}$ \footnote{Note that here we define the direction of the upstream magnetic field as the direction of the mean magnetic field of the turbulent medium}. 
 Note that this only defines the initial configuration. Once the shock front reaches region III, $\theta_{bn}$ can vary. Nevertheless, we will not consider the effects of $\theta_{bn}$ changing in our CR modelling (see Sec.~\ref{section:methods}).

Finally, we fill the computational domain from the shock discontinuity up to the right side of the box with one \textit{Lagrangian} particle per cell. This gives us a total number of 3,145,728  Lagrangian particles for each run. Each particle has its independent energy evolution as will be described in detail in the following section. 

\begin{table*}
\centering
\begin{tabular}{cccccc}
    &&& \\ \hline
    && \textbf{Fresh injection model}
    & \\ \hline
     Run ID & $\mathcal{M}_i$ & $\mathcal{M}_{cr}$ &$\alpha_{\eta}$ & 
       $\eta_e$ \\ \hline
      Inj\_3M\_Mcr1  & 3.0  & 1.0 & 1 & $10^{-3}$ \\
      Inj\_2p5M\_Mcr1  & 2.5  & 1.0 & 1 & $10^{-3}$ \\ 
      Inj\_2M\_Mcr1 & 2.0  & 1.0 & 1 & $10^{-3}$ \\ 
      Inj\_3M\_Mcr1\_pow2  & 3.0  & 1.0 & 2 & $10^{-3}$ \\
      \hline
      Inj\_3M\_Mcr2p3  & 3.0  & 2.3 & 1 & $10^{-3}$ \\
      Inj\_2p5M\_Mcr2p3  & 2.5  & 2.3 & 1 & $10^{-3}$ \\  
      Inj\_2M\_Mcr2p3  & 2.0  & 2.3 & 1 & $10^{-3}$ \\
      \hline
      && \textbf{Re-acceleration model}
& \\ \hline
     Run ID & $\mathcal{M}_i$ & $f_{pre}$ & 
       $\gamma_{min}$ & $s$ \\ \hline
      Re-DD\_3M\_1e2  & 3.0  & Dirac Delta & $10^{2}$ & - \\
      Re-DD\_2p5M\_1e2  & 2.5  & Dirac Delta  & $10^{2}$ & - \\  
      Re-DD\_2M\_1e2 & 2.0  & Dirac Delta & $10^{2}$ & -\\ 
      
      \hline
      
      Re-DD\_3M\_1e3  & 3.0  & Dirac Delta & $10^{3}$ & - \\
      Re-DD\_2p5M\_1e3  & 2.5  & Dirac Delta  & $10^{3}$ & - \\ 
      Re-DD\_2M\_1e3 & 2.0  & Dirac Delta  & $10^{3}$ & -\\
      
      \hline
      Re-PL\_3M\_s2p25 & 3.0  & Power-law  & $10$ & 2.25  \\
      Re-PL\_2p5M\_s2p25 & 2.5  &  Power-law  & $10$ &2.25 \\ 
      Re-PL\_2M\_s2p25 & 2.0  &  Power-law  & $10$ &2.25 \\

    \hline
      Re-PL\_3M\_s3 & 3.0  & Power-law  & $10$ & 3  \\
      Re-PL\_2p5M\_s3 & 2.5  &  Power-law  & $10$ & 3 \\ 
      Re-PL\_2M\_s3 & 2.0  &  Power-law  & $10$ & 3 \\
      
      \hline
\end{tabular}
\caption{Parameters used in all our simulation runs for the fresh injection and re-acceleration models. Note that run Inj\_3M\_Mcr1\_pow2 is discussed only in Appendix \ref{app:extra_models}}.
\label{table:init}
\end{table*}

\section{Non-thermal radio emission from shocks}\label{section:methods}

\subsection{Particle energy spectrum}
\label{sec:spectrum}

Each Lagrangian particle represents an ensemble of CR electrons and is characterized by an energy distribution function,
\begin{equation}
    F(E,\tau) = \frac{N(E,\tau)}{n_0},
\end{equation}
which gives the number of electrons per fluid number density. These particles evolve according to the CR transport equation
defined in Eq.~8 in \citealt{2018ApJ...865..144V},
\begin{equation}\label{cr_eq}
    \frac{d F}{d\tau} + \frac{\partial}{\partial E} \left[ \left(   - \frac{E}{3}\nabla_{\mu}u^{\mu} - c_rE^2 \right) F \right]=0.
\end{equation}
Here, the first term in square brackets accounts for energy losses due to adiabatic expansion, while the second one accounts for the synchrotron and inverse-Compton losses for electrons with isotropically distributed velocity vectors,
\begin{equation}\label{eq:cr}
    c_r = \frac{4}{3} \frac{\sigma_Tc\beta^2}{m_{e}^{2}c^{4}} \left[ \frac{B^2}{8\pi} + a_{\rm rad}T_{\mathrm{CMB}}^4(1+z)^4
    \right], 
\end{equation}
where $\beta=v_e/c$ is the velocity of the electrons, $m_e$ their mass and $a_{\rm rad}$ the radiation constant. For the present work, we assume a constant redshift of $z=0$.
The reader may refer to \citet{2018ApJ...865..144V} for a complete description of the numerical implementation and the semi-analytical scheme used for solving Eq. (\ref{cr_eq}).

We study a simplified scenario where the non-thermal spectra of the particles are activated instantly at the shock discontinuity. While the particles follow the fluid flow since $t=0$, the particle's energy distribution starts to evolve only when the particles are at the shock discontinuity. We use a shock finder based on converging flows and pressure jumps in order to tag shock cells and identify the propagating shock.  
The reader may refer to \citealt{dominguezfernandez2020morphology} for more details on the shock finder implementation.

Once the Lagrangian particles are \textit{activated} at the shock discontinuity, they get assigned an initial energy distribution. In this work, we considered two different models. Below, we outline the characteristics of the two initial energy distributions, namely, $F_{inj}(E)$ and $F_{re}(E)$, associated with each of these models:
\begin{itemize}
    \item[1)] \underline{\textit{Fresh injection model}}:
\begin{equation}\label{eq:energy_dist}
F_{inj}(E) = \frac{N(E)}{n_0} = \frac{f_{0,inj}}{n_0}\, E^{-p},     
\end{equation}
where $p=q-2$ is what it is usually called the \textit{injection spectral index}, $f_{0,inj}$ is the normalization constant and $n_0$ is the fluid number density. The power-law index, $q$, for non-relativistic shocks used
in our model is that obtained from the steady state theory
of DSA \citep{1983RPPh...46..973D}:
\begin{equation}\label{spectral_index}
    q = \frac{3(\gamma_0 + 1)\mathcal{M}^2}{2(\mathcal{M}^2-1)} = \frac{4\mathcal{M}^2}{\mathcal{M}^2-1},
\end{equation}
or
\begin{equation}\label{eq:p}
    p = 2 \left( \frac{\mathcal{M}^2 + 1}{\mathcal{M}^2 - 1} \right),
\end{equation}
where we considered the test particle acceleration at a shock of Mach number $\mathcal{M}$, and for the second equality in Eq.~(\ref{spectral_index}), we  
use $\gamma_0=5/3$. It is relevant to mention that for relativistic shocks, the parameter $q$ will exhibit different behavior with respect to the shock compression ratio \citep[see e.g.][]{PhysRevLett.94.111102}. However, as emphasized in the introduction, this study is exclusively concerned with non-relativistic shocks, which are commonly observed in the ICM (see also Table~\ref{table:init}).
The normalization factor, $f_{0,inj}$, is assigned according to the energy contained in the shock. That is, we considered that the total energy per fluid number density is
\begin{equation}\label{eq:CR_energy}
    \int F_{inj}(E)\, E \, dE = \frac{E_{{CR}}}{n_0},
\end{equation}
where the CR energy will be considered to be a function of the Mach number, $E_{\rm CR}(\mathcal{M})$. 
With $E_{\rm CR}(\mathcal{M})$, one can obtain the normalization factor as
\begin{equation}\label{eq:norm}
    f_{0,inj} = \left\{
	     \begin{array}{ll}
		 \frac{E_{\rm CR}(\mathcal{M})\,(4-q)}{\left[{E_{\rm max}}^{4-q}\, - \, {E_{\rm min}}^{4-q} \right]}  & \mathrm{if\ } q \ne 4 \\
		 & \\
		 E_{\rm CR}(\mathcal{M}) \log \left( \frac{E_{\rm max}}{E_{\rm min}}\right) & \mathrm{if\ } q=4 \\
		 
	       \end{array}
	     \right.
\end{equation}
The left-hand side of Eq.~(\ref{eq:CR_energy}) and the above Eq.~\ref{eq:norm} depend on the injection energy $E_{min}$ and the maximum allowed energy $E_{max}$. 
The energy flux of accelerated CRe should be a fraction of the energy dissipated by the shock:
\begin{equation}
    \phi_{CR} = \eta(\mathcal{M}) \phi_{sh} = \eta(\mathcal{M}) \frac{\rho_{pre}v_{sh}^3}{2},
\end{equation}
where $\phi_{CR}$ is given in units of [erg $cm^{-2}$ $s^{-1}$], $\rho_{pre}$ is the pre-shock (upstream) density, $v_{sh}$ is the velocity of the shock and $\eta(\mathcal{M})$ is the acceleration efficiency as a function that depends only on the shock's Mach number. The final expression for $E_{CR}$ is obtained by equating the cosmic ray energy flux crossing each particle volume element, and the total energy of cosmic rays advected with a post-shock (downstream) velocity,
\begin{equation}
    E_{CR}(\mathcal{M}) = \frac{1}{2} \eta(\mathcal{M}) \frac{\rho_{pre}}{v_{post}} v_{sh}^3,
\end{equation}
where $E_{CR}$ is in units of [erg $cm^{-3}$]. It is worth noting that the values of $\mathcal{M}$, $v_{sh}$, $\rho_{pre}$, and $v_{post}$ are determined through our shock-finding algorithm \citep[see Sec.~\ref{sec:spectrum} and][ for comprehensive information]{dominguezfernandez2020morphology}.
We will assume the following toy model for the $\eta({\mathcal{M}})$ function:

\begin{equation}
    \eta(\mathcal{M}) = \eta_e (\mathcal{M} - \mathcal{M}_{cr} )^{\alpha_{\eta}}
\end{equation}
where the critical Mach number $\mathcal{M}_{cr}$, power law index $\alpha_{\eta}$ and $\eta_e$ are parameters of our selection (see Table \ref{table:init}). 
In this work, we consider $E_{min}$ and $E_{max}$ corresponding to $\gamma_{min}=10$ and $\gamma_{max}=10^7$. The upper boundary is chosen to be arbitrarily high because for ICM shocks, $E_{max}/m_e c^2 \gg 1$, 
while the lower boundary is selected baring in mind that electrons with momenta $p \gtrsim 3p_{th,p}$\footnote{Here $p_{th,p}=(2m_pk_B T_{post})^{1/2}$ is the post-shock thermal proton momentum. For $T_{post}\sim 10^7$ K, we have $p_{th,p}/(m_e c)\sim 3-10$.}, could diffuse across the shock and participate in the DSA process
(see the discussion in \citealt{2020JKAS...53...59K}).
The value of $\eta_e$ is very uncertain for the weak ($\mathcal{M}_s \lesssim 5$)
shocks in the ICM \citep[see e.g.][]{2011ApJ...733...63R,2020ApJ...897L..41X}. The chosen value agrees with the expectations of DSA for strong shocks \citep[e.g.][]{ka12} and lies in the range of values required
to explain observations of radio relics \citep[see][]{2020A&A...634A..64B}. Nevertheless,
we note that, since $E_{min}$ and $E_{max}$ are initially fixed, $\eta_e$ can be treated as a free parameter to re-scale the particle energy distribution such that it matches the observed surface brightness. Although there has been limited advancement in this area, recent findings from PIC simulations indicate that shocks with Mach numbers weaker than $\mathcal{M}_s\simeq 2.3$ may not possess the capability to facilitate electron acceleration through DSA, as demonstrated by studies such as \citealt[][]{2019ApJ...876...79K} and \citealt{2021ApJ...915...18H}. This serves as the primary motivation behind our choice of the $\mathcal{M}_{cr}$ parameter. Furthermore, we explore the conservative scenario of $\mathcal{M}_{cr}=1$, which allows for the possibility of even the weakest shocks contributing to electron acceleration via DSA. Our selection of the $\alpha_{\eta}$ parameter is motivated by fitting functions derived and extensively studied in \citealt[][]{Kang_2007, kr13, 2019ApJ...883...60R}.

\vspace{5mm}

    \item[2)] \underline{\textit{Re-acceleration model}}:

\begin{equation}
    F_{re}(E) = \frac{1}{n_0} \left[ 
    (p+2)E^{-p} \int_{E_{min}}^{E} E'^{p-1} f_{pre}(E') dE'
    \right],
\end{equation}
with $p$ given in Eq.~\ref{eq:p}. Here, $f_{pre}$ is the spectrum of pre-existing fossil electrons. 
We will consider the following two cases:
    \begin{subequations}
    \begin{equation}\label{eq:pre_DD}
        f_{pre,1}(E) = f_{0,1} \, \delta(E - E_{cut}),
    \end{equation}
    \begin{equation}\label{eq:pre_PL}
        f_{pre,2}(E) = f_{0,2}\, \left( \frac{E}{E_{min}} \right)^{-s},
    \end{equation}
    \end{subequations}
where $f_{0,1}$ and $f_{0,2}$ are normalization constants and $s$ is an energy spectral index\footnote{Note that similarly to $p$, $s=s_p -2$, where $s_p$ is the spectral index of the distribution function in momentum space.}.
The Dirac delta function model represents an idealized scenario with a single-burst population of pre-existing electrons. However, it offers us better control over the characteristic energy of these fossil electrons, as discussed in previous studies \citep[see, e.g.,][]{ka12,2013MNRAS.435.1061P}. On the other hand, the power-law model is a more realistic representation of the seed population of electrons, as proposed in \citealt{2011ApJ...734...18K} (see Eqs.~7 and 8).
In the Dirac delta function model, we consider $E_{cut}$ corresponding to $\gamma_{min}=10^2$ and $10^3$, while in the power-law model, we set $E_{min}$ corresponding to $\gamma_{min}=10$ and $s=2.25$ and 3, 
with the spectrum covering the range of $\gamma_{min}=10$ and $\gamma_{max}=10^7$, 
as listed in Table~\ref{table:init}. The selected $s$ values are what is typically expected if pre-existing CRe were generated at previous shocks or as a result
of turbulent acceleration \citep[see e.g.][]{2005PhRvL..95z5004C}.

The energy distribution function of each particle for each corresponding case can be re-written as:

\begin{subequations}
    \begin{equation}\label{eq:re-acc_DD}
    F_{re}(E) = \frac{f_{0,1}}{n_0} \frac{(p+2)}{E_{cut}} \left( \frac{E}{E_{cut}}\right)^{-p}, 
    \end{equation}
        \begin{small}
        \begin{equation}
    F_{re}(E) = \left\{
	     \begin{array}{ll}
		 \frac{1}{n_0} \frac{(p+2)}{(p - s)} f_{pre,2}(E) \left[ 1 - \left( \frac{E}{E_{min}} \right)^{s-p} \right]       & \mathrm{if\ } s \ne p \\
		 & \\
		 \frac{1}{n_0} (p+2) f_{pre,2}(E) \log\left( \frac{E}{E_{min}} \right) & \mathrm{if\ } s=p. \\
		 
	       \end{array}
	     \right.
    \end{equation}
    \end{small}
\end{subequations}
If $s \ne p$ and for $E \gg E_{min}$, we can re-write the energy distribution function as \citep[see also Eq.8 of][]{2011ApJ...734...18K}
\begin{equation}\label{eq:re-acc_PL}
    F_{re}(E) = \frac{1}{n_0} \frac{p+2}{\lvert p-s \rvert} \, f_{0,2} \, \left( \frac{E}{E_{min}} \right)^{-r},
\end{equation}
where $r=\mathrm{min}(s,p)$.

Note that for both the Dirac delta function and power-law cases, the pre-shock and post-shock CR number densities, $N_{CR,pre} = \int f_{pre}(E) dE$ and $N_{CR,re} = n_0\int F_{re} (E) dE$, are mathematically related as
\begin{equation}
        N_{CR,re} = \frac{p + 2}{p - 1} \, N_{CR,pre}.
\end{equation}
This relation basically reflects the effect of compression, and it is only a function of $p$.
Analogously, the pre-shock and post-shock CR energy densities 
are related as
 \begin{equation}
        E_{CR,re} = \frac{p+2}{p-2} E_{CR,pre}.
\end{equation}
The normalization factors, $f_{0,1}$ and $f_{0,2}$, are usually set by the ratio of upstream CR electron pressure to
gas pressure, $R = P_{CR,pre}/P_{g,pre}$ \citep[see][]{ka12}, where $P_g = (\gamma_0 -1)\rho e$. Here, $R$ is a parameter that will be scaled to match the
observed fluxes of radio relics. 
Nevertheless, it will not be relevant for the research question that we pursue in this work because $R$ effectively only re-scales the particle energy spectrum.

Finally, the complete electron distribution at the shock location can be written as a sum of the re-accelerated and freshly injected electron populations,
\begin{equation}\label{eq:fresh_plus_re}
    F(E) = F_{re}(E) + F_{inj}(E).
\end{equation}
We point out that the relative significance of the freshly injected and re-accelerated electron populations is ruled by the normalization constants $f_0$ ($f_{0,1}$ and $f_{0,2}$ for the two cases considered in Eqs.~\ref{eq:pre_DD}-\ref{eq:pre_PL}) and $f_{0,inj}$ and the slopes $p$ and $s$ in our model.

\end{itemize}

\subsection{Synchrotron emission}
\label{sec:sync_emiss}

\begin{figure*}
    \centering
    \includegraphics[width=0.37\textwidth]{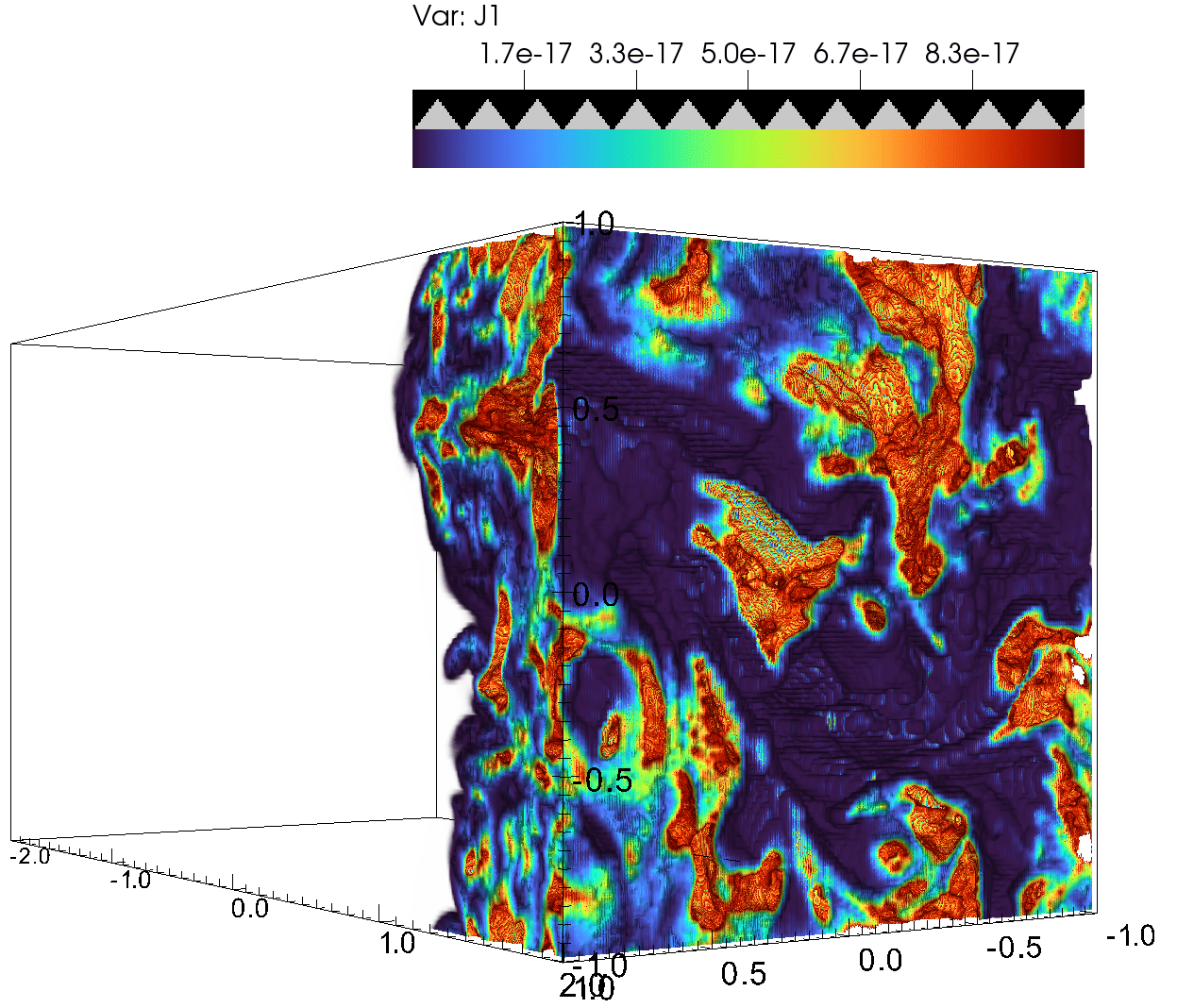}
    \includegraphics[width=0.37\textwidth]{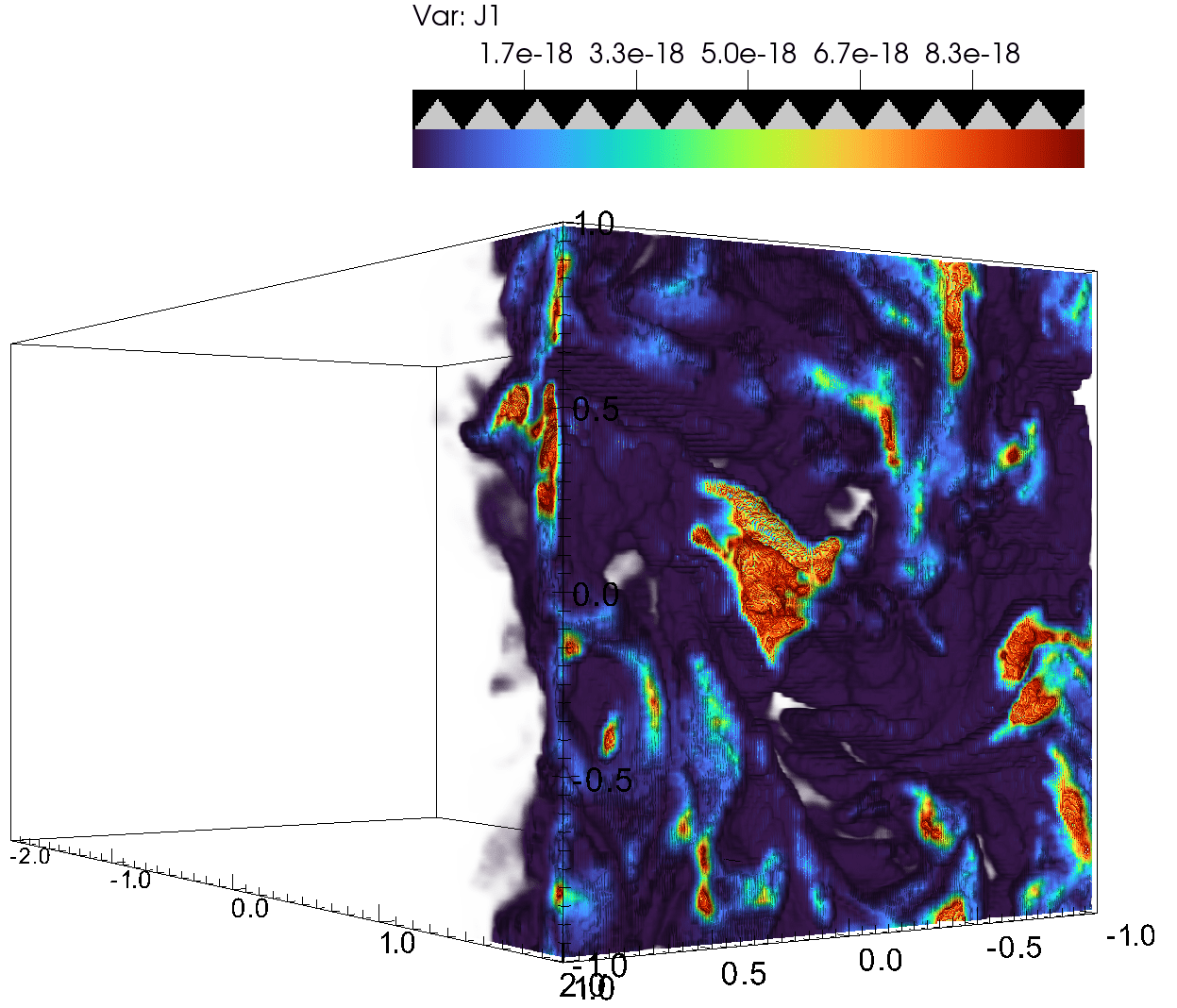}\\
    \includegraphics[width=0.37\textwidth]{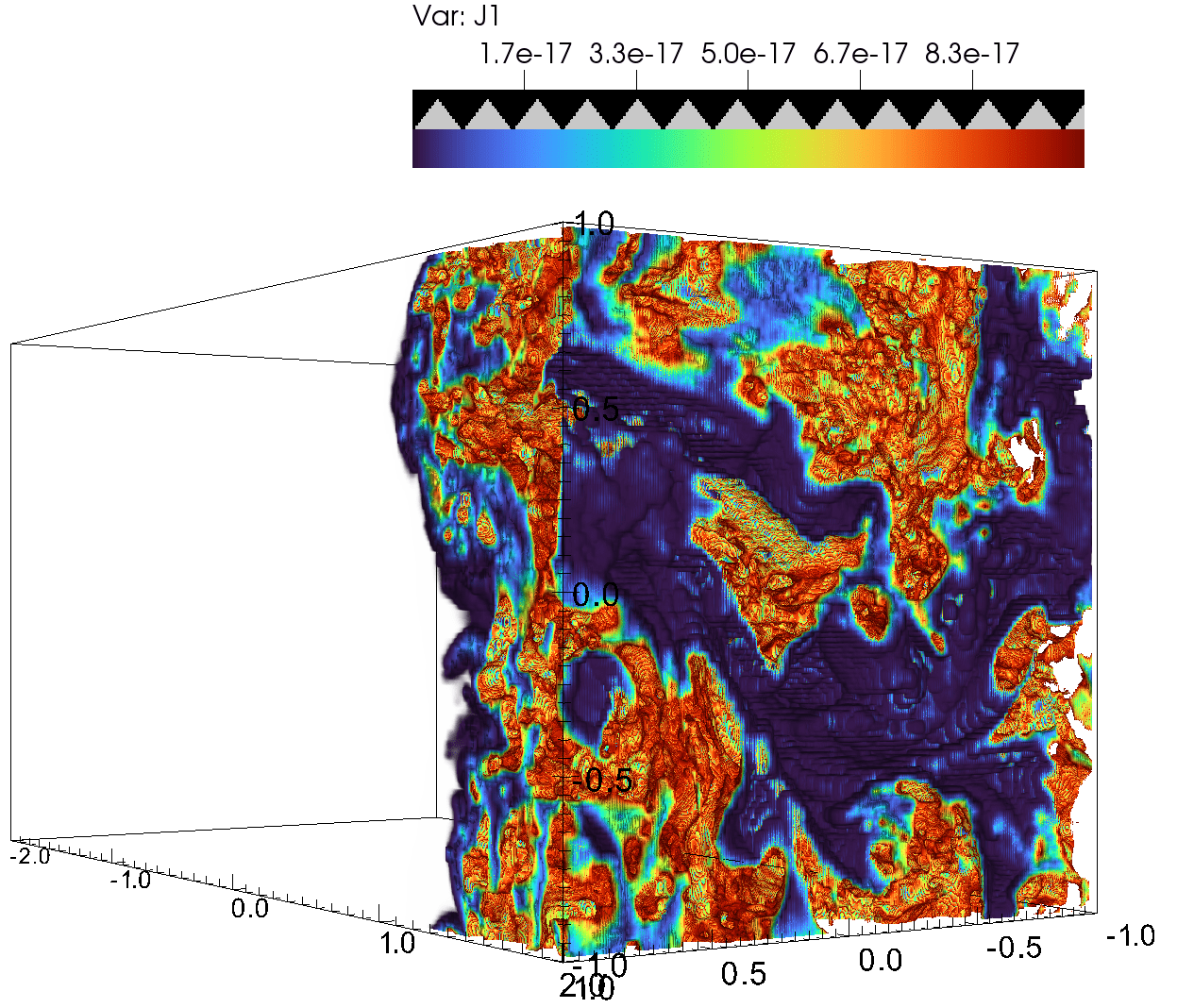}
    \includegraphics[width=0.37\textwidth]{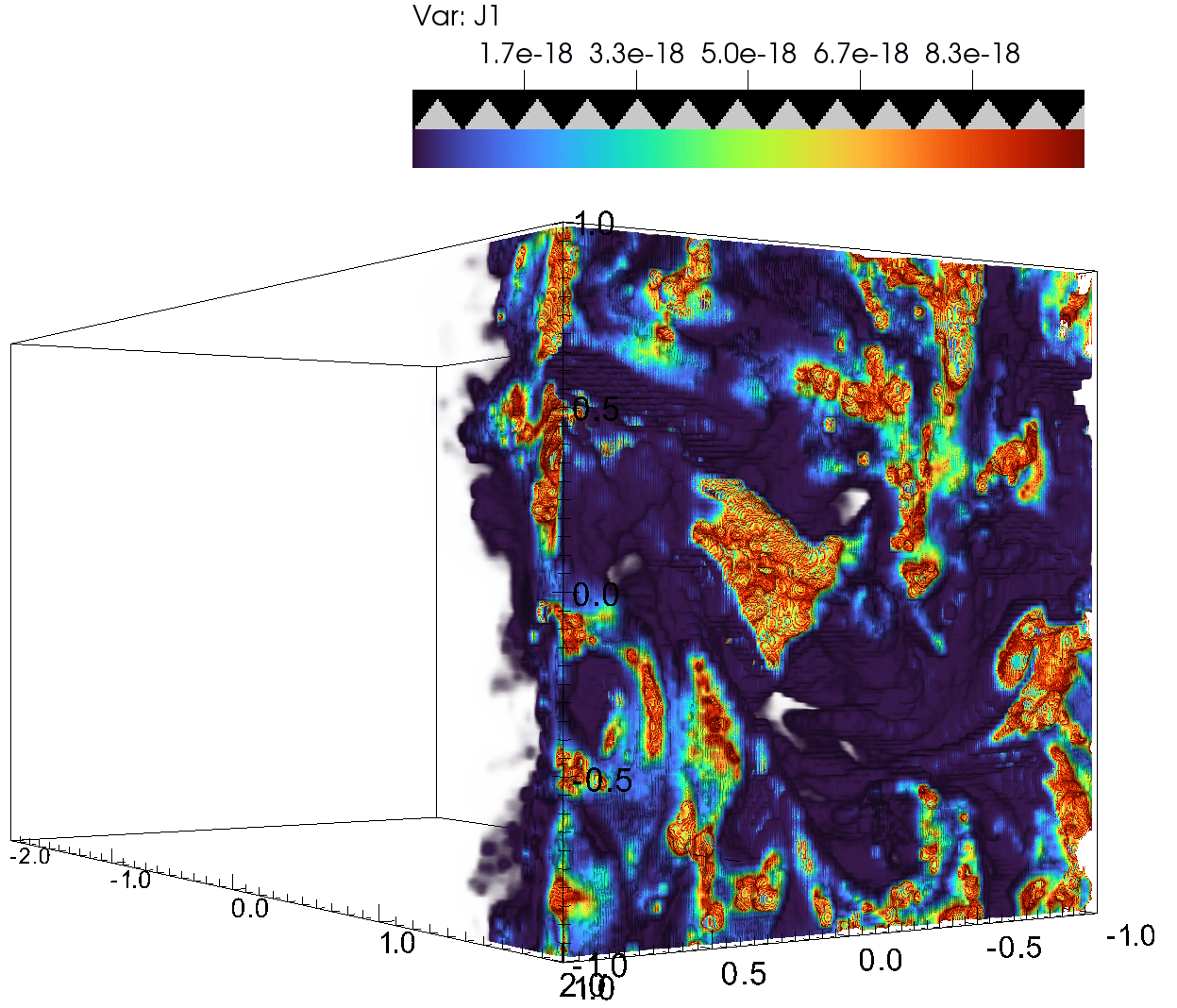}
    \caption{ Volume rendering of the synchrotron emissivity rendering in code units at 650 MHz (left) and 18.6 GHz (right) for the 
     fresh-injection model (\textit{upper row}; case $\mathcal{M}_{cr}=1$; see Sec.~\ref{sec:surf_brightness}) and the re-acceleration model (\textit{lower row}; case $s=2.25$). All renderings correspond to the $\mathcal{M}_i=3$ case at $t=$ 184.7 Myr.
    }
    \label{fig:3d}
\end{figure*}

\begin{figure*}
    \centering
    \includegraphics[width=0.31\textwidth]{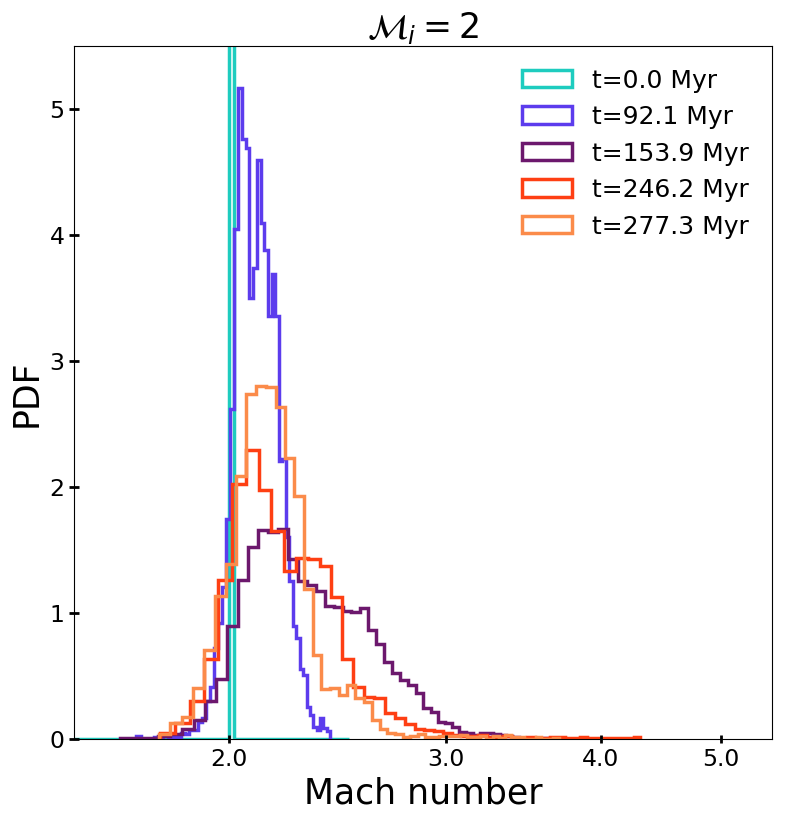}
    \includegraphics[width=0.32\textwidth]{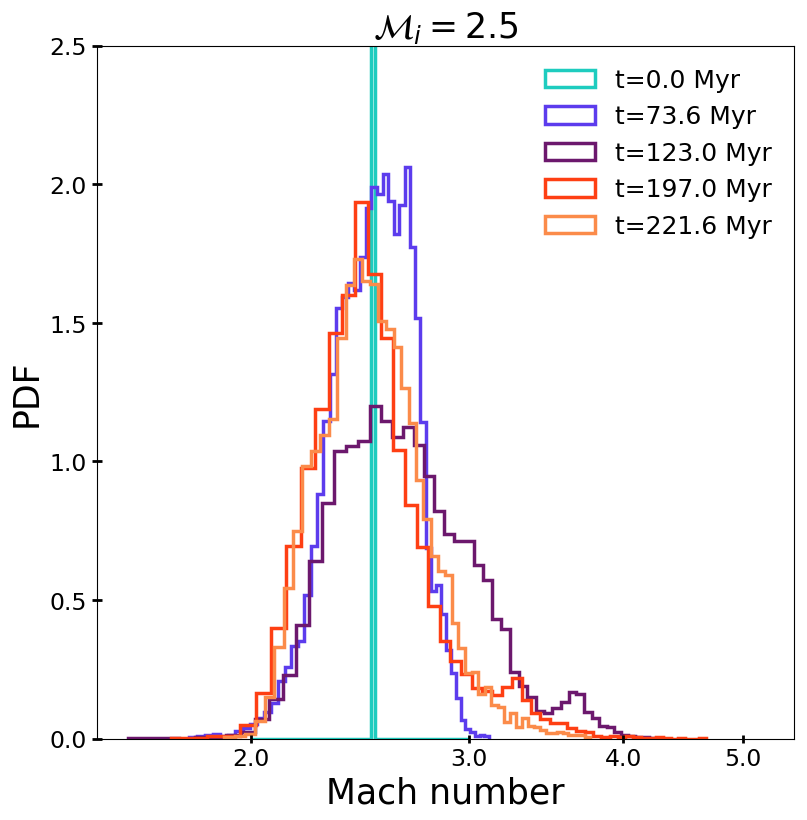}
    \includegraphics[width=0.32\textwidth]{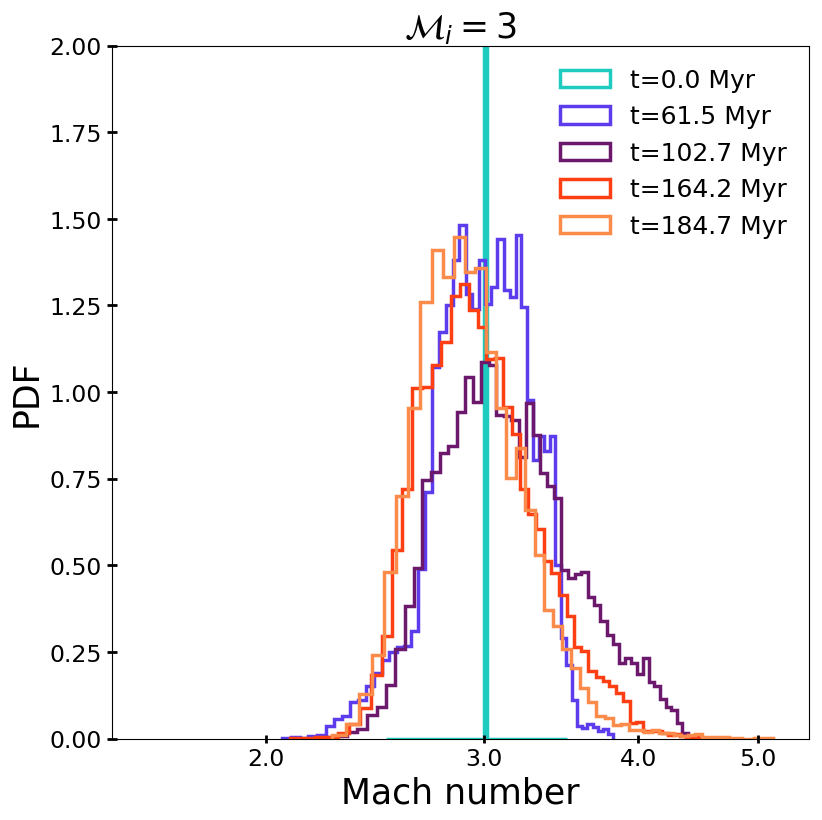}\\
    \includegraphics[width=0.32\textwidth]{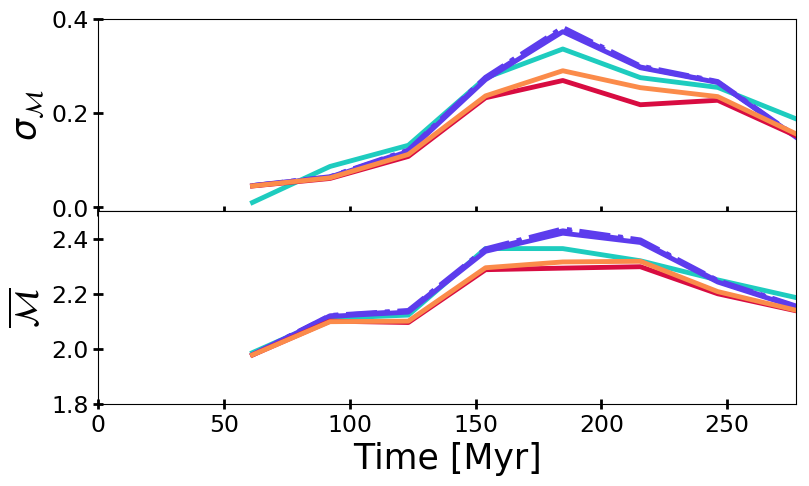}
    \includegraphics[width=0.32\textwidth]{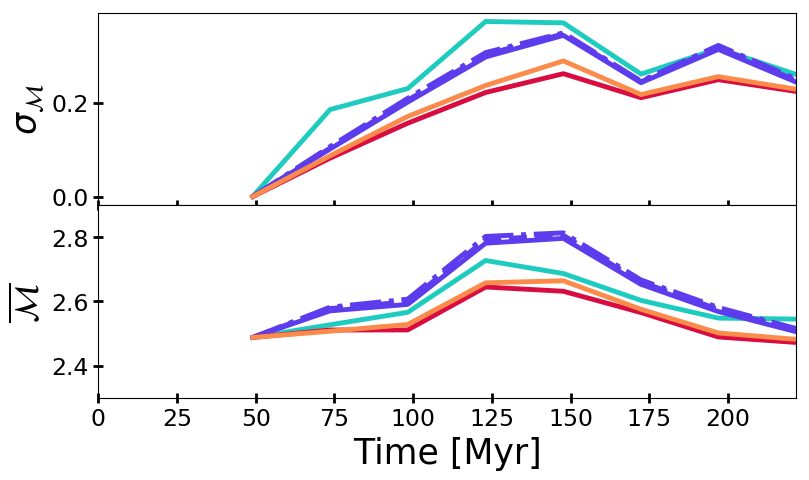}
    \includegraphics[width=0.32\textwidth]{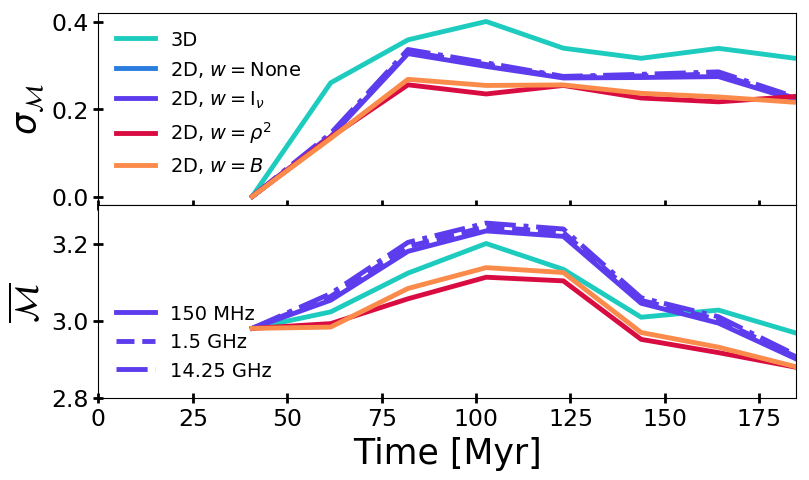}\\
    
    \caption{
    \textit{Upper row}: 3D Mach number distribution at the shock cells at selected simulation times. 
    \textit{Lower row}: Standard deviation of the Mach number PDF as function of time. Each column corresponds to the different initial Mach number of the shock. For the 2D Mach number distribution we used different weights for the projection (see legend). Note that here $I_{\nu}$ is taken from the fresh-injection model (case $\mathcal{M}_{cr}=1$; see Sec.~\ref{sec:surf_brightness}). See also Appendix~\ref{app:Eradio_Mach} for a comparison between weights corresponding to the emission coming from the fresh-injection and re-acceleration models.
    }
    \label{fig:mach_evol}
\end{figure*}

The synchrotron emissivity of a tracer particle in a local magnetic field $\mathbf{B}'$ in the direction $\mathbf{\hat n_{los}}'$\footnote{The emissivities in the observer frame are
obtained by applying the appropriate transformations: $\mathcal{J}_{\rm syn}=\mathcal{D}^2\mathcal{J}_{\rm syn}^{'}$, where $\mathcal{D}$ is the Doppler factor. Note that for the non-relativistic case, $\mathcal{D}\approx 1$. We refer the reader to Sec.~3 of \citealt{2018ApJ...865..144V} for more details. }, per unit solid angle, volume and frequency is given by
\begin{small}
\begin{equation}
    \mathcal{J}_{\rm syn}^{'}(\nu_{\rm obs}',\mathbf{\hat n_{los}'},\mathbf{B}')
    = \int N(E')
    \mathcal{P}(\nu_{\rm obs}',E',\phi') dE' d\Omega',
\end{equation}
\end{small}
where $\mathcal{P}(\nu_{\rm obs}',E',\phi')$ is the synchrotron power per unit frequency and unit solid angle emitted by a single particle that has energy $E'$ and $\phi'$ is the angle that the velocity vector of the particle makes with the direction $\mathbf{\hat n_{los}'}$. Following \citealt{1965ARA&A...3..297G}, the synchrotron emissivity in units of [erg cm$^{-3}$ s$^{-1}$ Hz$^{-1}$ str$^{-1}$] is
\begin{equation}\label{eq:17}
     \mathcal{J}_{\rm syn}^{'}(\nu_{\rm obs}',\mathbf{\hat n_{los}^{'}},\mathbf{B}')
     = \frac{\sqrt{3} e^3}{4\pi m_e c^2} 
     | \mathbf{B}' \times \mathbf{\hat n_{los}}' | \int N(E') F(\xi) \, dE',
\end{equation}
where all primed quantities are evaluated in the local comoving
frame, and
where $\mathbf{\hat n_{los}}'$ is the unit vector in the direction of the line of sight in the comoving frame and $F(\xi)$ is a Bessel function integral given by
\begin{equation}\label{eq:Bessel}
    F(\xi) = \xi \int_{\xi}^{\infty} K_{5/3}(z')\,dz',
\end{equation}
where
\begin{equation}
\label{J_pol}
    \xi = \frac{\nu_{\rm obs}'}{\nu_c^{'}} = \frac{4\pi m_{e}^3c^5 \nu_{\rm obs}'}{3eE'^2| \mathbf{B}' \times \mathbf{\hat n_{los}'}|},
\end{equation}
where $\nu_{c}'$ is defined as the critical frequency at which the emission peaks. Note that strictly speaking only those particles with a pitch angle coinciding with the angle between $\mathbf{B}'$ and $\mathbf{\hat n_{los}'}$ contribute to the emission along the line of sight in Eq. (\ref{eq:17}). Nevertheless, in this work, we assume an isotropic distribution of pitch angles.

\begin{figure*}
    \includegraphics[width=0.46\textwidth,height=7.5cm]{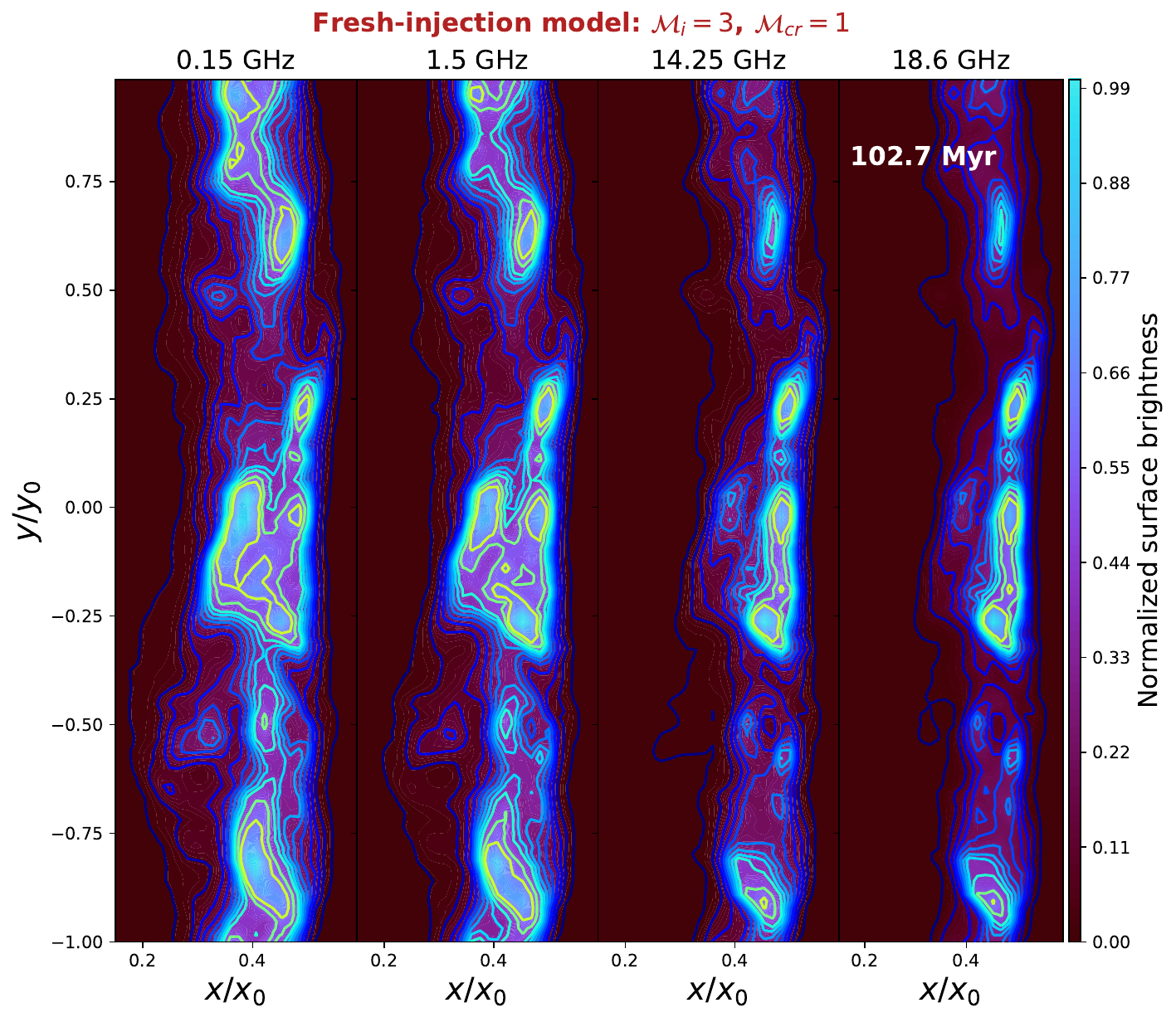}
    \includegraphics[width=0.46\textwidth,height=7.5cm]{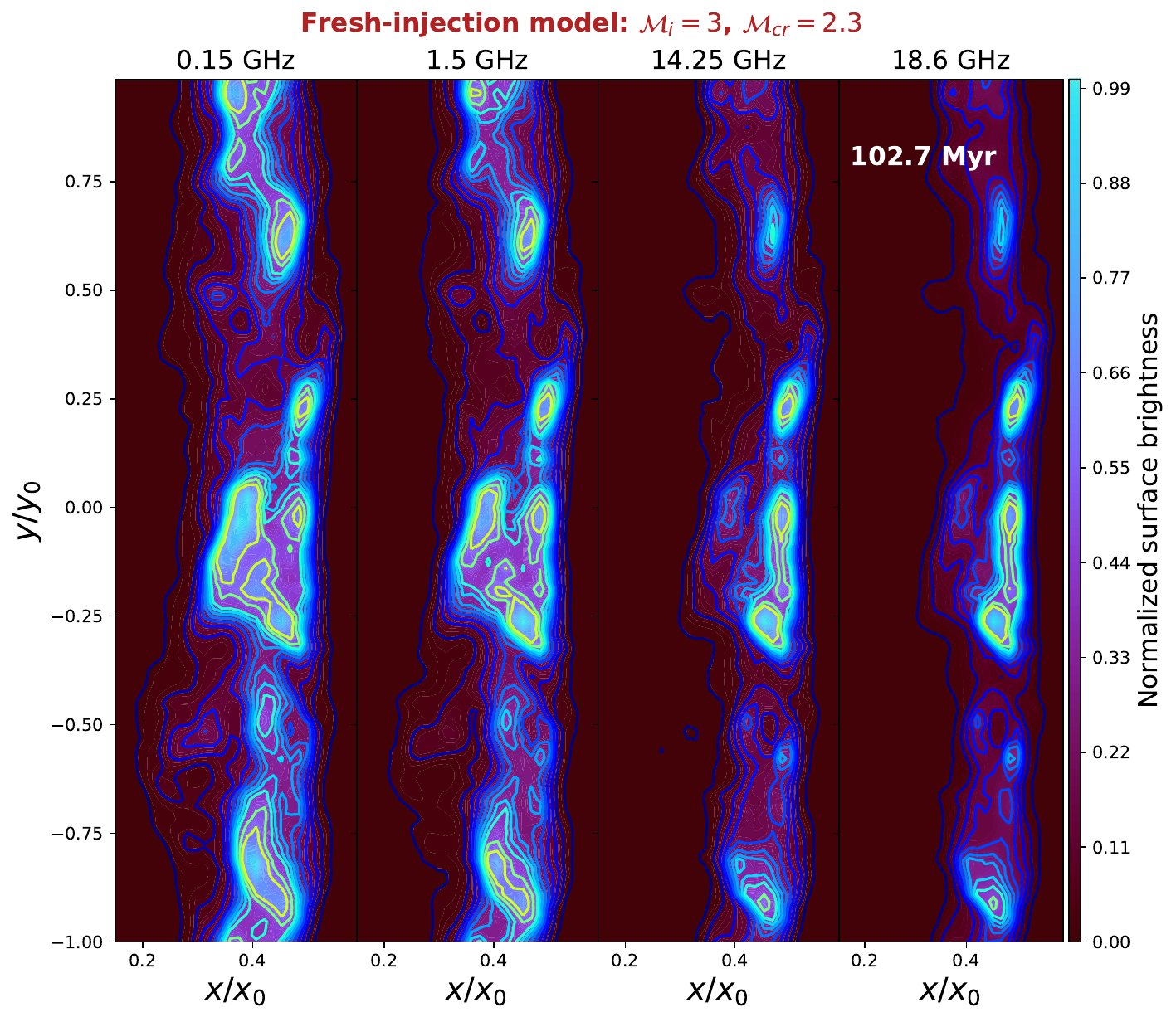}
    %
    \includegraphics[width=0.46\textwidth,height=7.5cm]{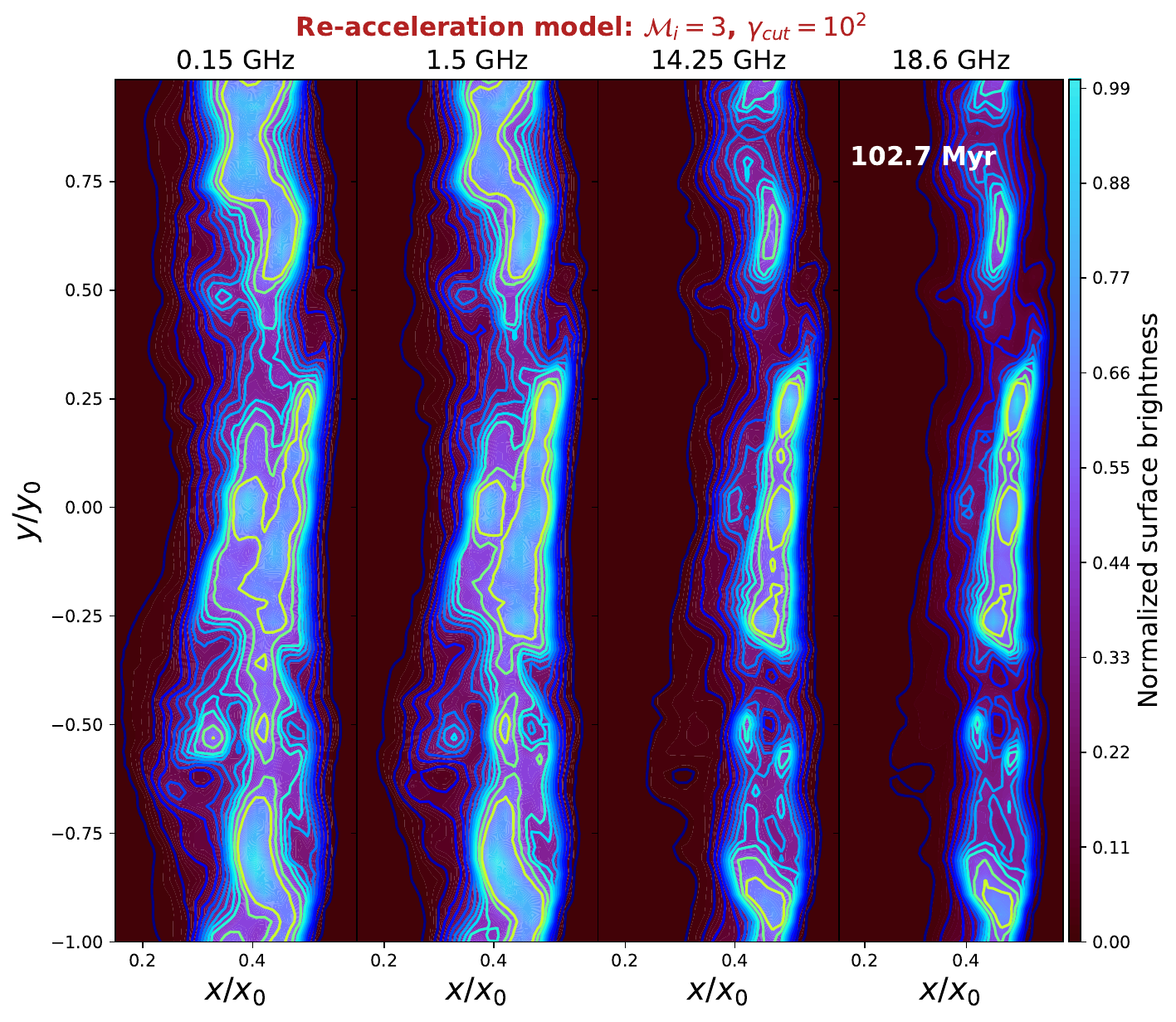}
    \includegraphics[width=0.46\textwidth,height=7.5cm]{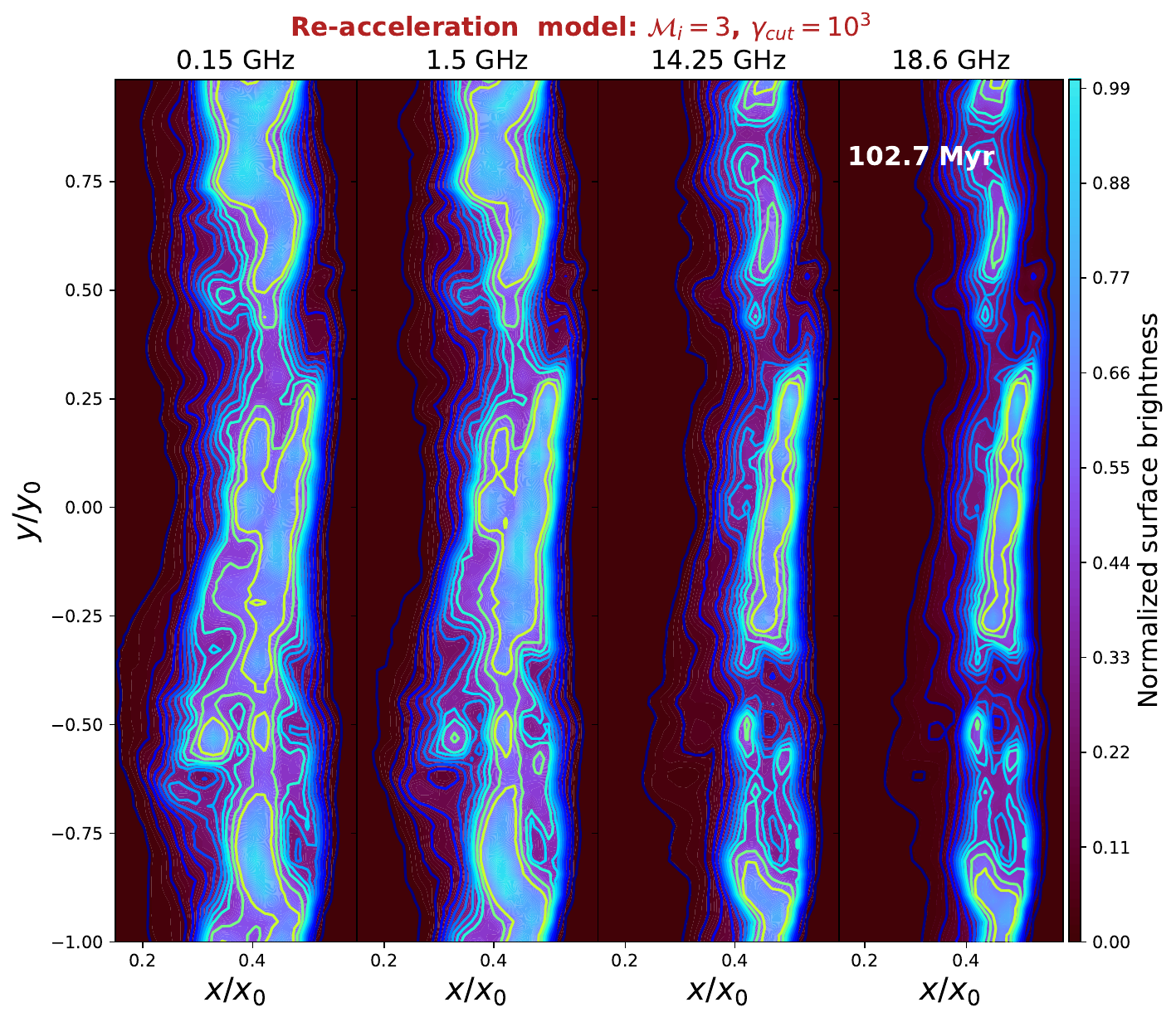}
    \includegraphics[width=0.46\textwidth,height=7.5cm]{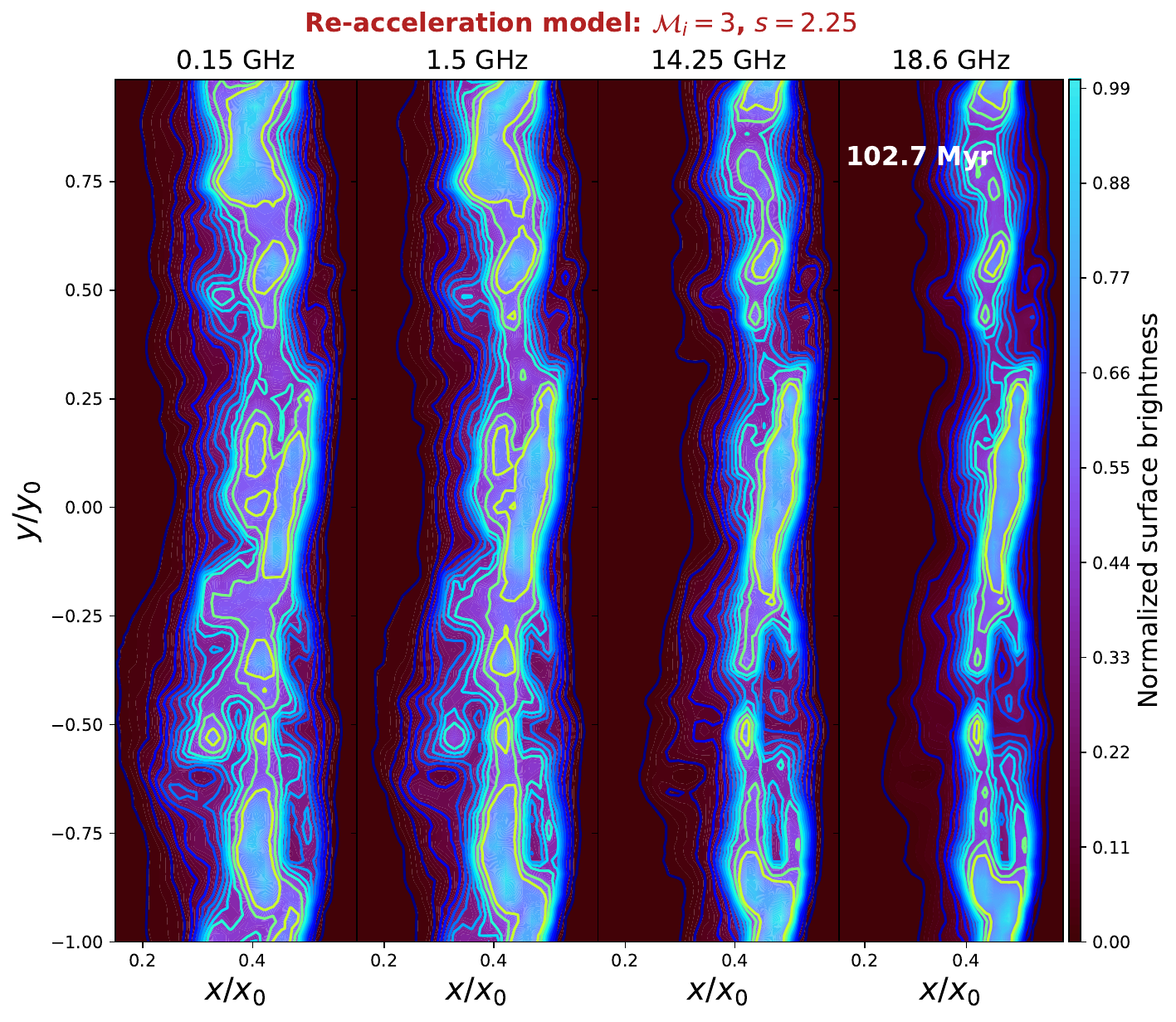} \hspace{11mm}
    \includegraphics[width=0.46\textwidth,height=7.5cm]{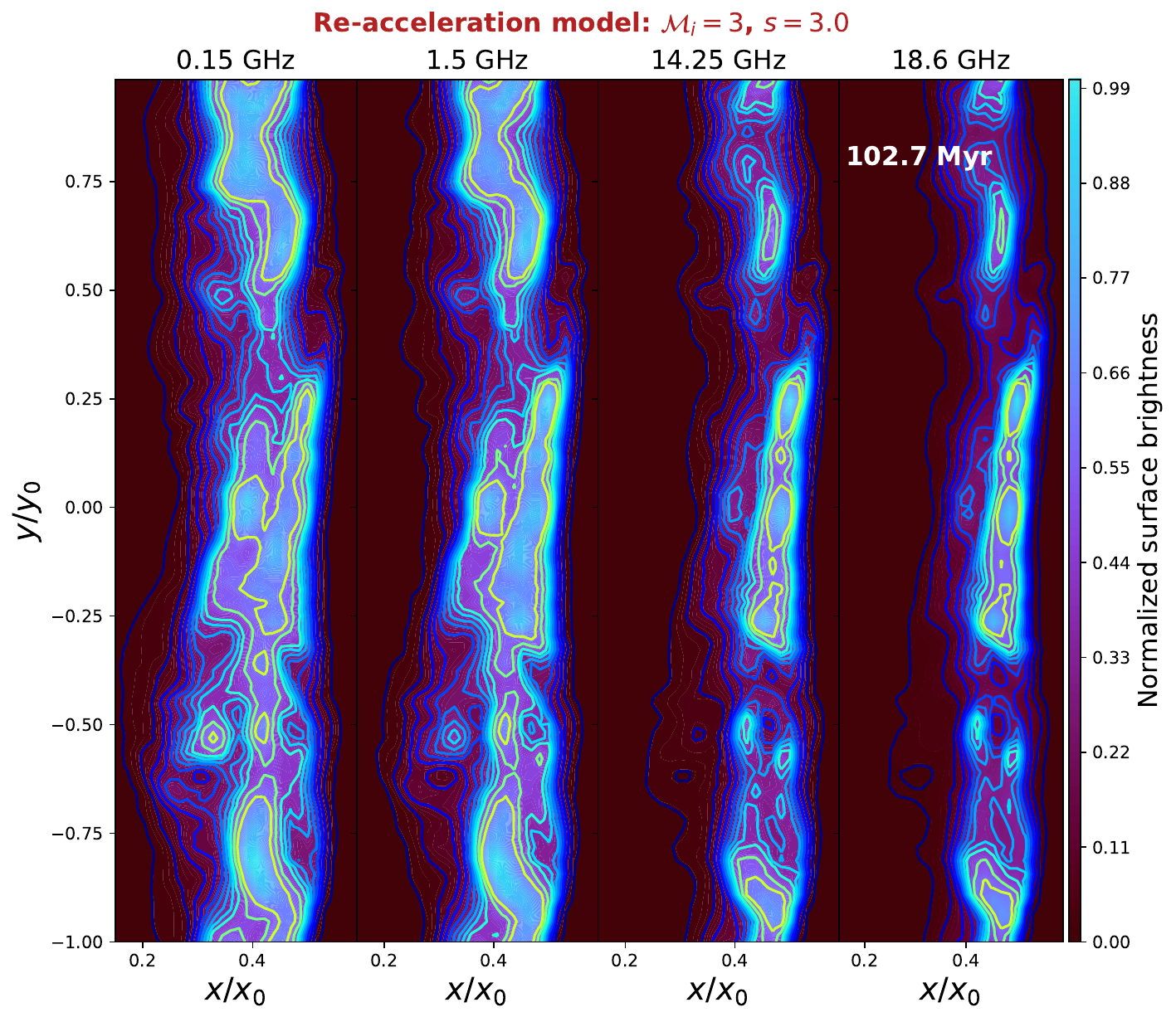}
\caption{Normalized surface brightness maps of the shock front for the $\mathcal{M}_i = 3$ case in the fresh-injection model (first row) and re-acceleration model with a Dirac Delta (second row) and a power-law (third row) fossil populations. Each sub-panel shows four different frequencies, namely 150 MHz, 1.5 GHz, 14.25 GHz and 18.6 GHz. The first (last) four columns correspond to $\mathcal{M}_{cr} = 1$ ($\mathcal{M}_{cr} = 2.3$), $\gamma_{cut} = 10^2$ ($\gamma_{cut} = 10^3$) and $s = 2.25$ ($s=3$), respectively. The time-step selected of the simulation is 120.7 Myr (see the top corner of each sub-panel). 
}
    \label{fig:maps_all_otherstyle}
\end{figure*}

Once the synchrotron emissivity is interpolated back onto the Eulerian grid, the specific intensity maps can then be obtained by integrating along a line of sight as
\begin{equation}\label{eq:intensity}
    I_{\nu} = \int \mathcal{J}_{\rm syn}(\nu_{\rm obs},x,y,z)dz,
\end{equation}
in units of [erg cm$^{-2}$ s$^{-1}$ Hz$^{-1}$ str$^{-1}$]. The observed flux is estimated assuming a Gaussian beam with $\theta$ width. The surface brightness maps in units of [mJy/beam] are obtained computing $S_{\nu} \approx I_{\nu} \theta^2$, where $\theta^2= \pi \theta_1 \theta_2/(4\ln(2))$ is the beam area of the telescope. Finally, the integrated spectra (or net flux) can be obtained by integrating the specific intensity $I_{\nu}$ over the area covered by the source in the plane of the sky, that is
\begin{equation}\label{eq:flux}
   F_{\nu}(\nu) = \int I_{\nu}(\nu, x, y)dxdy, 
\end{equation}
in units of [erg s$^{-1}$ Hz$^{-1}$ str$^{-1}$].

\section{Results}
\label{sec:results}

\subsection{The shock front}\label{sec:Mach_dist}

We start by showing three-dimensional renderings of the synchrotron emissivity produced by an initial $\mathcal{M}_i=3$ shock at $t=$ 184.7 Myr in the fresh-injection and re-acceleration models at low (650 MHz) and very high (18.6 GHz) frequencies in Fig.~\ref{fig:3d}. 
The turbulent pre-shock ICM (see Sec.~\ref{sec:turb}) naturally induces fluctuations at the shock front as it is propagating through the simulation box. 
These fluctuations determine the variation in the energy spectrum of Lagrangian particles at the shock. However, as illustrated in Fig.~\ref{fig:3d}, it is obvious that thermal and fossil electrons injected into DSA yield different fluctuations. This disparity will be thoroughly examined in the subsequent sections of this manuscript.

It would be instructive to first show how some fluid variables change at the shocked cells as function of time.
In the upper row of Fig.~\ref{fig:mach_evol}, we show the evolution of the shock's Mach number, $\mathcal{M}$, of the shock cells in the three-dimensional (3D) computational volume for the three different initial Mach numbers of the shock $\mathcal{M}_i$.  
This evolution is mainly determined by 1) the pre-shock rms Mach number and 2) $\mathcal{M}_i$. 
We find that higher $\mathcal{M}_i$'s produce more extended tails and higher standard deviations in the PDF. In particular, an increment of 0.5 in $\mathcal{M}_i$ results in an increment of $\sim 0.05$ in the standard deviation $\sigma_{\mathcal{M}}$ (see lower row of Fig.~\ref{fig:mach_evol}).

The Mach number we infer from observations is inherently influenced by the projection of phenomena onto the plane of the sky. Consequently, we present the statistical distribution of the projected Mach number in the lower row of Fig.~\ref{fig:mach_evol}. This perspective provides the reader with a supplementary understanding of how projected characteristics differ from the actual 3D data.
We define 
the Mach number on the two-dimensional (2D) $X-Y$ plane projected along the $Z$-axis as
\begin{equation}\label{eq:Mweighted}
   \mathcal{M}_w = \frac{ \int \mathcal{M} \, w \, dZ}{ \int w \, dZ}, 
\end{equation}
where $w$ is a selected weight. Note that here the projected Mach number weighted with the emission is not the same as the Mach number that can be obtained from radio observations.
We remind the reader that the Mach number, as determined from radio observations, relies on a two-step process. Initially, the spectral index of the radio spectra is computed across multiple frequencies. Subsequently, this spectral index is employed to infer the Mach number, utilizing DSA. (see Eq.~28 of \citealt{dominguezfernandez2020morphology}).

As can be seen from the lower row of Fig.~\ref{fig:mach_evol}, the projected Mach number weighted with the radio emission could give rise to a large discrepancy when compared to the projected Mach number weighted with other fluid quantities such as the density or the magnetic field. We see that the standard deviation, $\sigma_{\mathcal{M}}$, and mean, $\overline{\mathcal{M}}$, are larger, if the Mach number is weighted with the radio emission. 
On the other hand, we do not find significant differences in the projected Mach number when radio emissions at different frequencies are used as weights.
Nevertheless, as we will show in the following sections, there is a varying ``degree of patchiness'' in radio relics with frequency. 
In Appendix~\ref{app:Eradio_Mach} we additionally show that the $\sigma_{\mathcal{M}}$ obtained weighting with the emission from the fresh injection model is larger than the  $\sigma_{\mathcal{M}}$ where the radio emission from the re-acceleration model is used. This happens due to the different degrees of patchiness achieved in both the fresh injection and re-acceleration models. 
We expect that this effect as observed in surface brightness maps can introduce a bias between the real and radio inferred Mach numbers.  
However, we leave this type of analysis relating the current work to observations for future work.

\subsection{Surface brightness maps}\label{sec:surf_brightness}

We start by showing the normalized surface brightness maps for 
selected cases of the simulations described in Table~\ref{table:init} at 150 MHz, 1.5 GHz, 14.25 GHz and 18.6 GHz at two different time-steps. In this section, we only show the surface brightness maps corresponding to the $\mathcal{M}_i = 3$ case, while we remind the reader that our general results also include the $\mathcal{M}_i = 2$ and $\mathcal{M}_i = 2.5$ cases. 
A $\mathcal{M}\sim 3$ shock should be the most representative example because
several radio relics observations have reported similar Mach number values inferred either from radio or X-ray.  

In Figs.~\ref{fig:maps_all_otherstyle} and \ref{fig:maps_donwstream}, we show the corresponding maps for the fresh-injection model with the two critical Mach numbers studied, namely $\mathcal{M}_{cr} = 1$ (left-hand-side) and  $\mathcal{M}_{cr} = 2.3$ (right-hand-side) in the first row. In the second row, we show the corresponding surface brightness maps for the re-acceleration model assuming a fossil electron population represented by a Dirac delta $f_{pre,1}$ (see Eq.~\ref{eq:pre_DD}). On the left-hand-side, we show the maps where $\gamma_{cut}=10^2$ and on the right-hand-side, those where $\gamma_{cut}=10^3$. Finally, on the third row, we show the maps corresponding to the fossil electron population that is represented by a power law ($f_{pre,2}$ (see Eq.~\ref{eq:pre_PL}). On the left-hand-side, we show the case where the parameter $s=2.25$ is assumed, while on the right-hand-side, we show the $s=3$ case.

Fig.~\ref{fig:maps_all_otherstyle} illustrates the early propagation of the radio shock front at $t=102.7$ Myr. The shock front is located at $x/x_0\sim0.5$ (see also Fig.~\ref{fig:init} for a reference; note that here $x_0=100$ kpc) and the extent of the downstream region is $\sim 30$ kpc at 150 MHz. 
At this stage, we can already observe significant differences comparing the four frequencies, i.e. 150 MHz, 1.5 GHz, 14.25 GHz and 18.6 GHz, for each model considered. 
From this visualization, we argue that different observing frequencies may result in different pictures for 
the radio substructures generated at the shock front.
In particular, we observe that both models, i.e. the fresh-injection and re-acceleration models, can produce patchiness along the radio shock front. Yet, as it can be seen from Fig.~\ref{fig:maps_all_otherstyle}, the fresh-injection model produces more patchy substructures than the re-acceleration model.

It shall be noted that the shock front has a ``distribution of Mach numbers'', and it is the Mach numbers distribution
that defines the observed morphology in the radio emission;
the radio substructures are fully defined by the initial mean strength of the shock in our simulations, hence the mean of the Mach number distribution, i.e.  $\overline{\mathcal{M}}$, along with the spread of the Mach number distribution, $\sigma_{\mathcal{M}}$ (see Section~\ref{sec:Mach_dist}).  
As we will further discuss in Section~\ref{sec:patchiness},
the generation of patchy radio structures is more pronounced if the Mach number distribution at the shock front contains a large amount of $\mathcal{M}\lesssim 2.5$. This remains true for both the fresh-injection and re-acceleration models. 
However, the CRe energy spectrum has a different dependency on the Mach number values in the two models, which leads to slightly different radio substructures.

 We present identical surface brightness maps to those in Fig.~\ref{fig:maps_all_otherstyle}, albeit for a different snapshot, in Appendix~\ref{app:extra_features}. In Fig.~\ref{fig:maps_donwstream}, we show the radio shock at $t=184.7$ Myr. In this case, the shock front is located almost at the right end of region III at $x/x_0 \sim1.75$ and the extent of the downstream region is $\sim 80$ kpc at 150 MHz. Similar to Fig.~\ref{fig:maps_all_otherstyle}, the Mach number distribution of the shock surface determines the radio substructures generated at the shock front. Further downstream, these substructures are additionally defined by the mixing of emission with different spectral ages. %
 In the following, we use the surface brightness maps from Figs.~\ref{fig:maps_all_otherstyle} and \ref{fig:maps_donwstream} to examine two limiting cases of radio shocks in the ICM. These cases serve as comparative benchmarks, focusing on two distinct downstream widths, with the larger width being most representative of a typical radio relic scenario.
 We also refer the reader to Appendix \ref{app:extra_features} for additional information on the simulated radio emission such as spectral index maps and profiles.

\subsection{Degree of patchiness}
\label{sec:patchiness}
%

\begin{figure*}
    \centering
    \includegraphics[width=7.6cm]{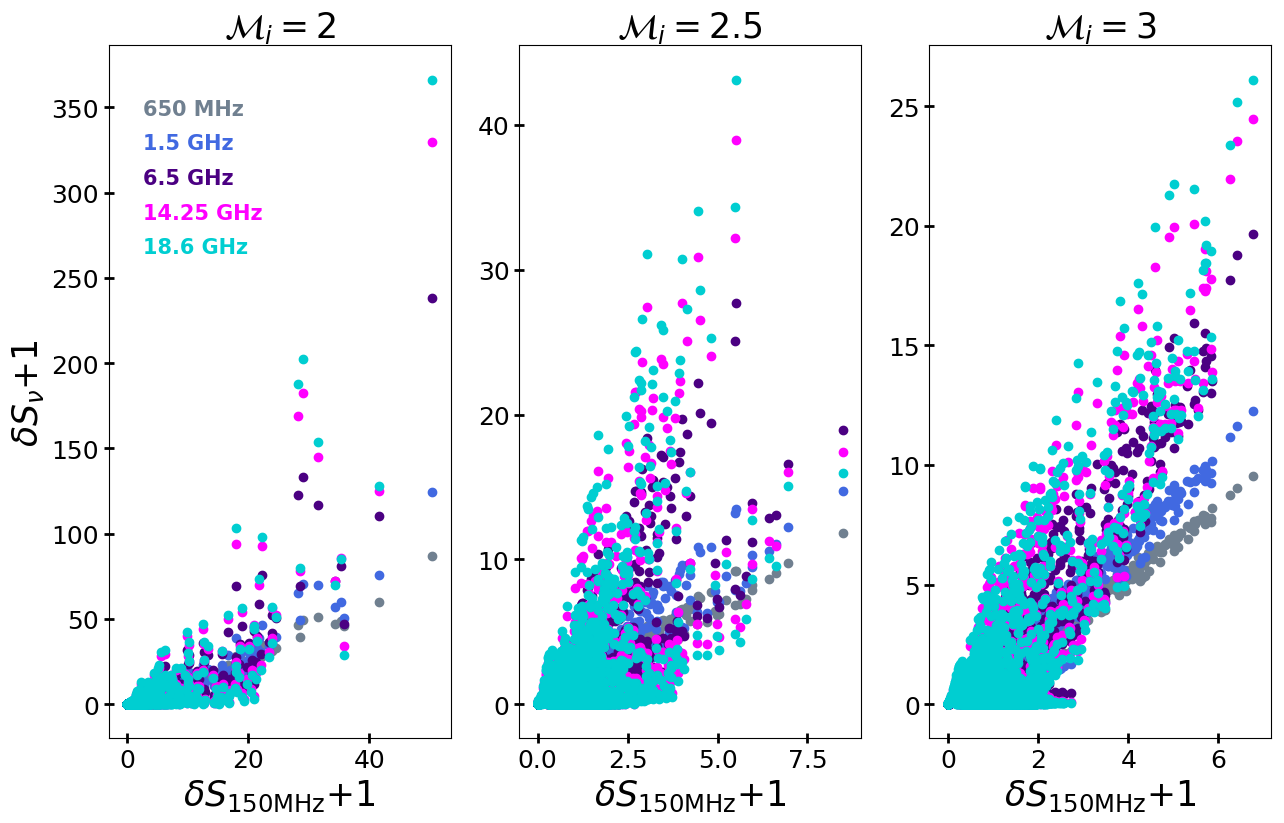}
    \includegraphics[width=7.6cm]{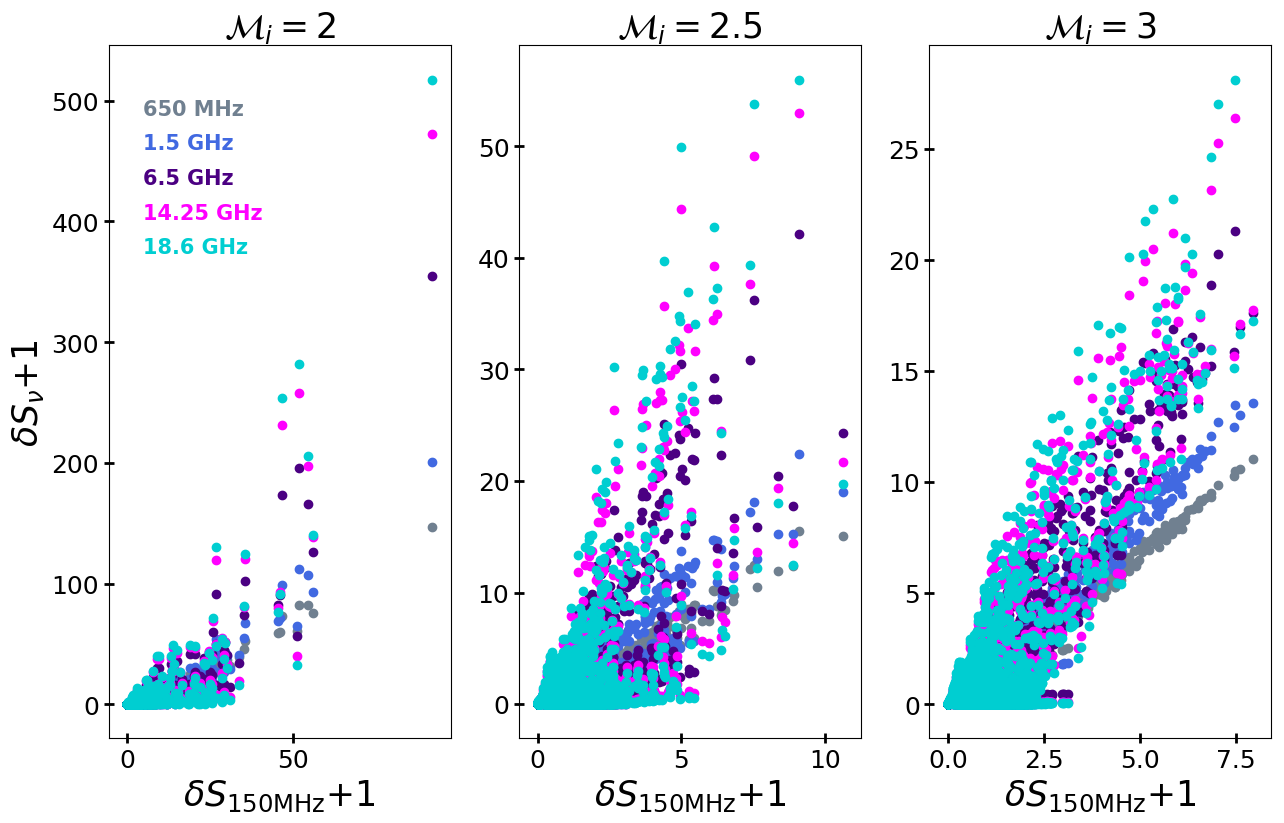} \\
    \includegraphics[width=7.6cm]{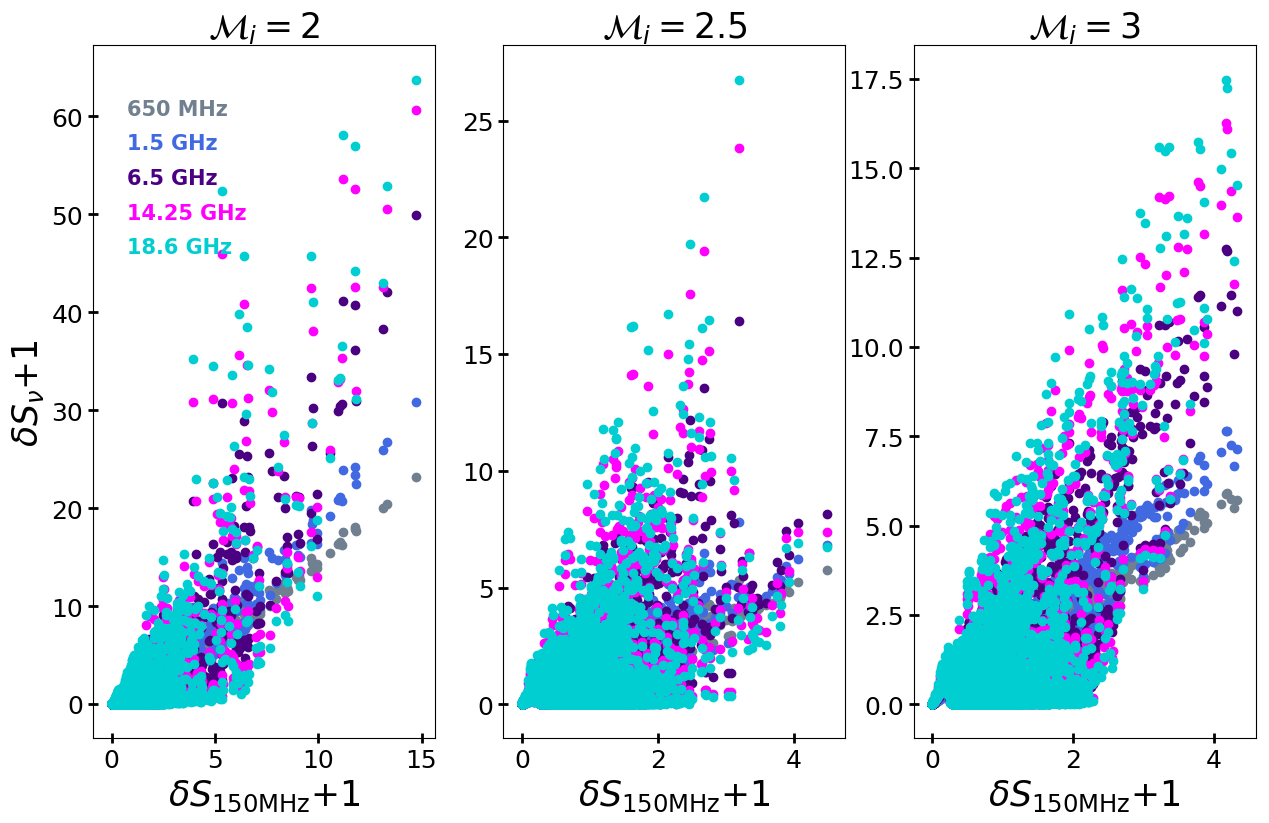}
    \includegraphics[width=7.6cm]{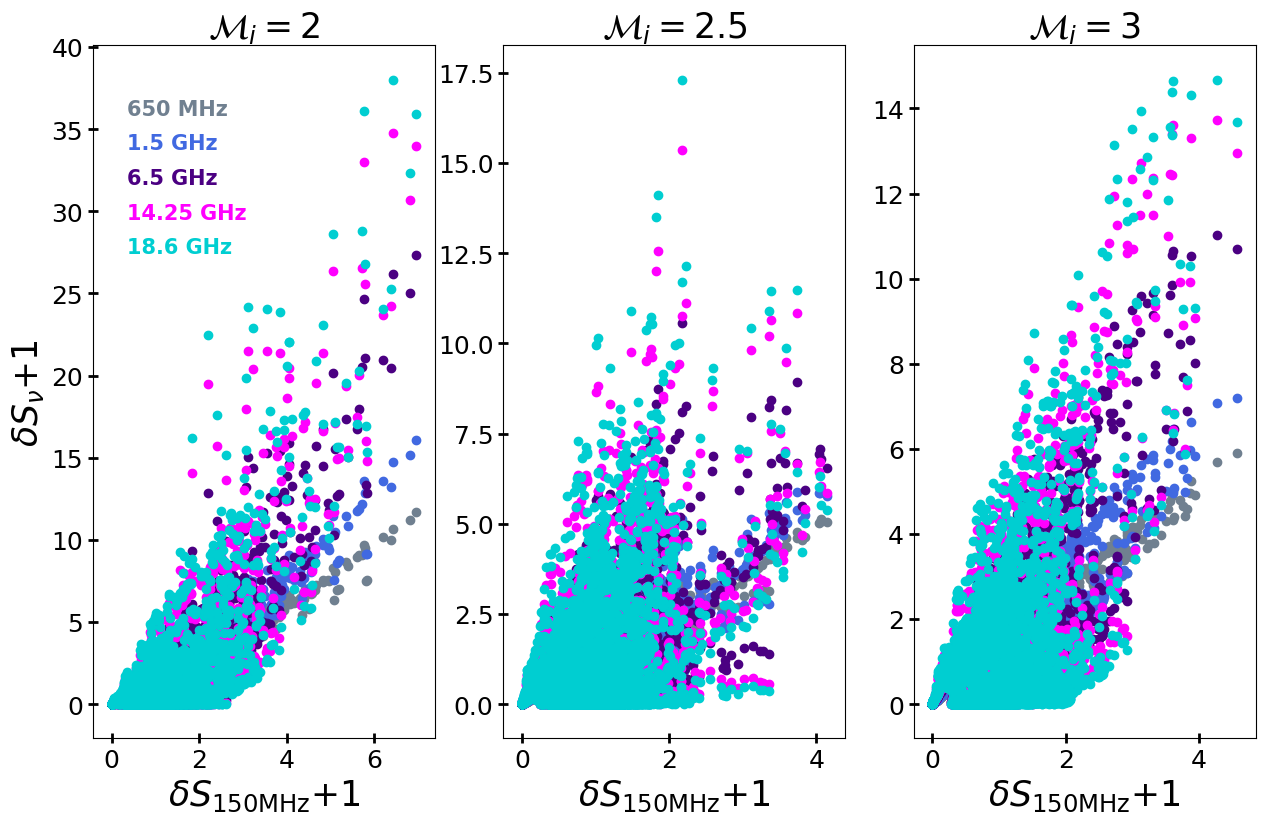}
    %
    \includegraphics[width=7.6cm]{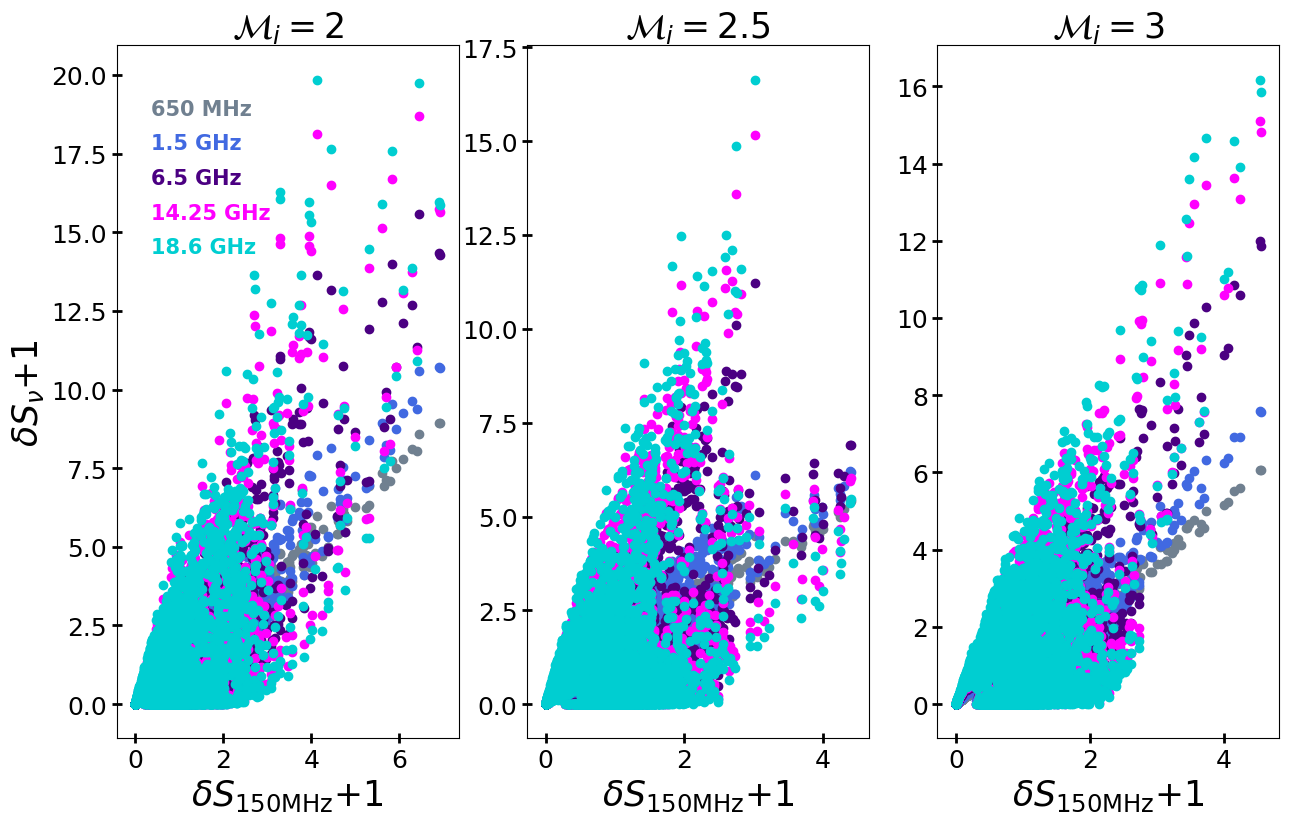}
    \includegraphics[width=7.6cm]{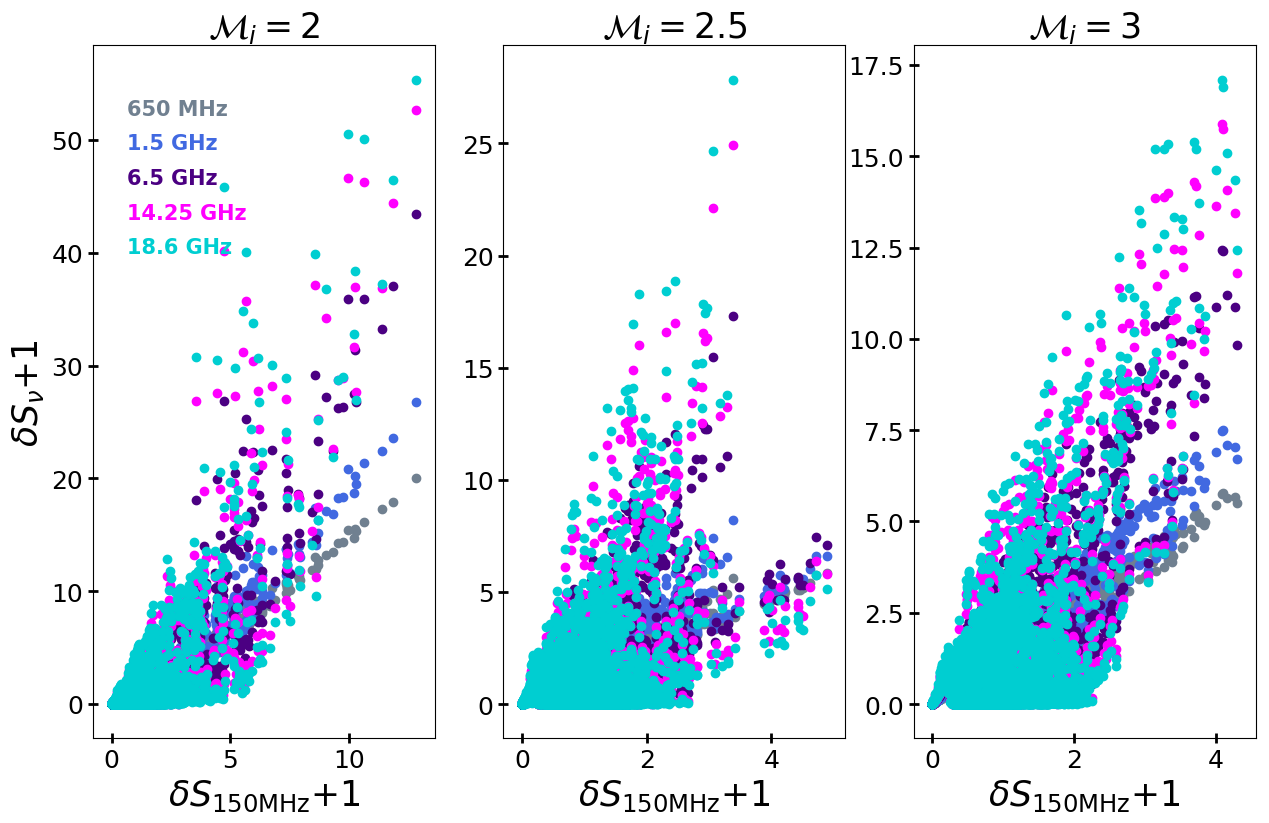}
    \caption{Phase plots of the $\delta S_{\nu}/\bar{S_{\nu}} +1$ distribution at $t=184.7$ Myr for all our models at all frequencies versus the same distribution but at 150 MHz. The top, middle and bottom rows correspond to the fresh injection model, the re-acceleration model assuming a Dirac Delta pre-existing distribution function and the re-acceleration model assuming a power-law
    pre-existing distribution function, respectively.
    The differences between the columns correspond to the different parameters studied first (second) column correspond to $\mathcal{M}_{cr} = 1$ ($\mathcal{M}_{cr} = 2.3$), ${\gamma}_{cut} = 10^2$ (${\gamma}_{cut} = 10^3$) and $s = 2.25$ ($s=3$) in the first, second and third rows, correspondingly. 
    }
    \label{fig:phase-plots}
\end{figure*}

\begin{figure*}
    \centering
    \includegraphics[width=16.5cm]{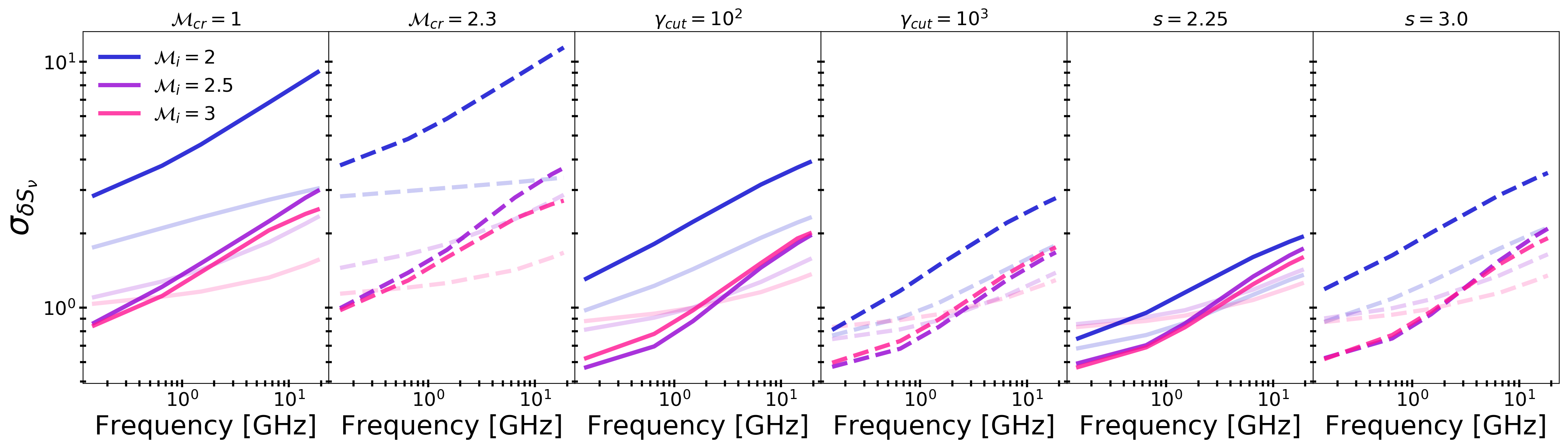}
    \caption{ Standard deviation of the $\delta S_{\nu}$ distribution for all our models at all frequencies. Two simulation time-steps are shown: $t=102.7$ Myr (higher transparency) and $t=184.7$ Myr (no transparency). The first two panels correspond to the fresh injection model ($\mathcal{M}_{cr} = 1$ and $\mathcal{M}_{cr} = 2.3$ parameters), the third and fourth panels to the re-acceleration model assuming a Dirac Delta pre-existing distribution function (${\gamma}_{cut} = 10^2$ and ${\gamma}_{cut} = 10^3$ parameters) and the fifth and sixth panels to the re-acceleration model assuming a power-law
    pre-existing distribution function ($s = 2.25$ and $s=3$ parameters).
    }
    \label{fig:sync_statistics_2}
\end{figure*}

A good first order approximation to quantify the amount of substructure in radio relics is by computing the relative surface brightness fluctuations or perturbations,
\begin{equation}
    \delta S_{\nu} = S_{\nu}/\bar{S_{\nu}} - 1,
\end{equation}
where $\bar{S_{\nu}}$ is the mean surface brightness of the selected area. 
In this work, we use $\delta S_{\nu}$ to define the ``degree of patchiness or substructure'' in radio relics.
Within this definition, radio relics will have a higher degree of patchiness or substructure if the spread of   
$\delta S_{\nu}$ is larger. We calculated $\delta S_{\nu}$ within a spatial domain of 35 kpc x 200 kpc x 200 kpc at $t=102.7$ Myr and extended it to 85 kpc x 200 kpc x 200 kpc at $t=184.7$ Myr. With this selection we ensure a through coverage of the downstream region in both snapshots.

In Fig.~\ref{fig:phase-plots} we show the  
$\delta S_{\nu}+1$ distributions at $t=184.7$ Myr
for all the 6 models and 6 frequencies: 150 MHz, 650 MHz, 1.5 GHz, 6.5 GHz, 14.25 GHz and 18.6 GHz. We compare each distribution corresponding to a different frequency with the distribution at the lowest frequency, i.e. 150 MHz. 
In all the models we find that the higher the frequency, the larger the spread of $\delta S_{\nu}$. In other words, the radio shock front has a higher degree of patchiness or substructure at higher frequencies.

As mentioned before, the substructure will be defined by the Mach distribution at the shock front which in turn depends on the pre-shock medium. 
Hence, a large standard deviation of $\delta S_{\nu}$, $\sigma_{\delta S_{\nu}}$, is a consequence of a large spread of the Mach number distribution, $\sigma_{\mathcal{M}}$.

\begin{figure*}
    \centering
    \includegraphics[width=14.8cm]{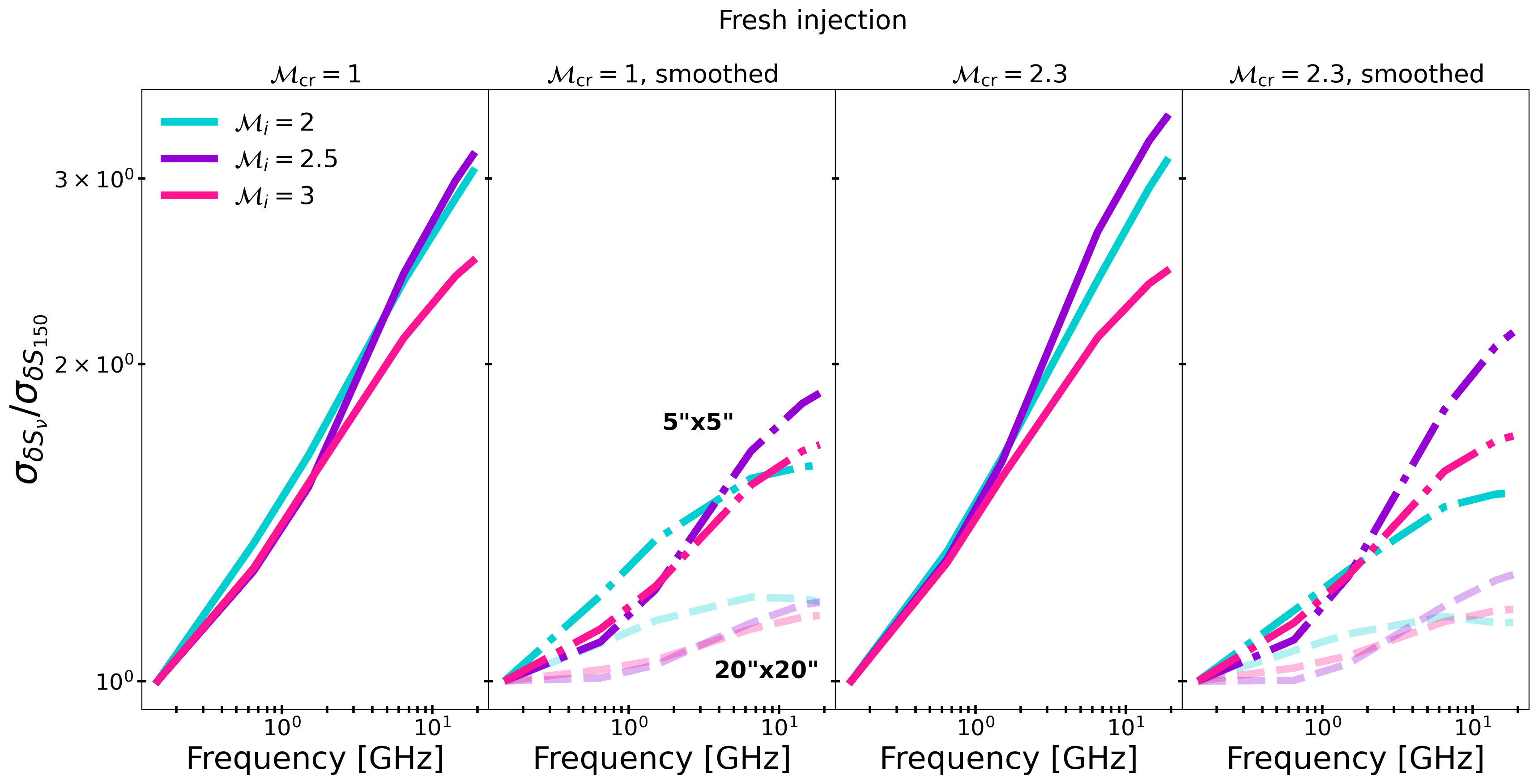}
    \includegraphics[width=14.8cm]{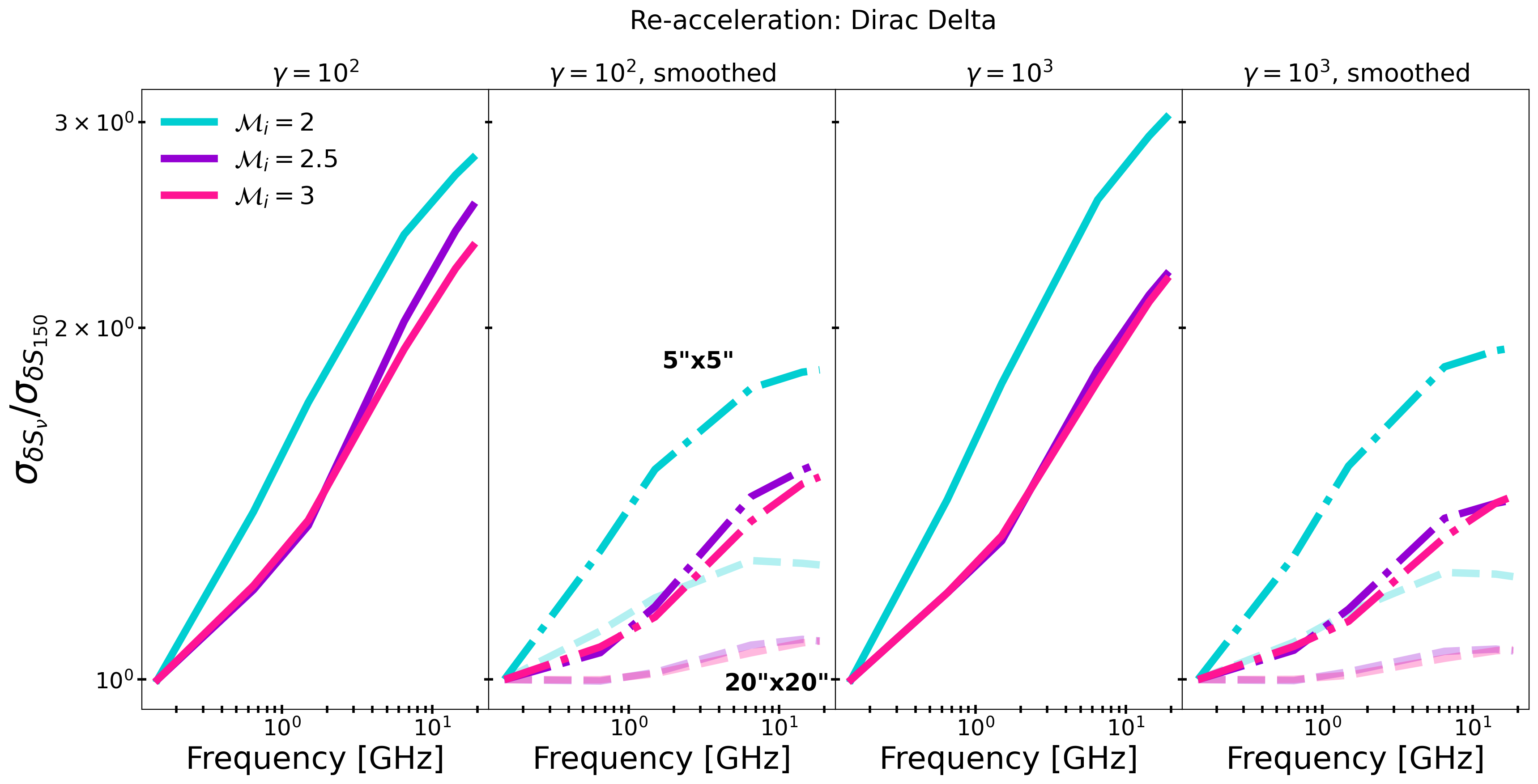}
    \includegraphics[width=14.8cm]{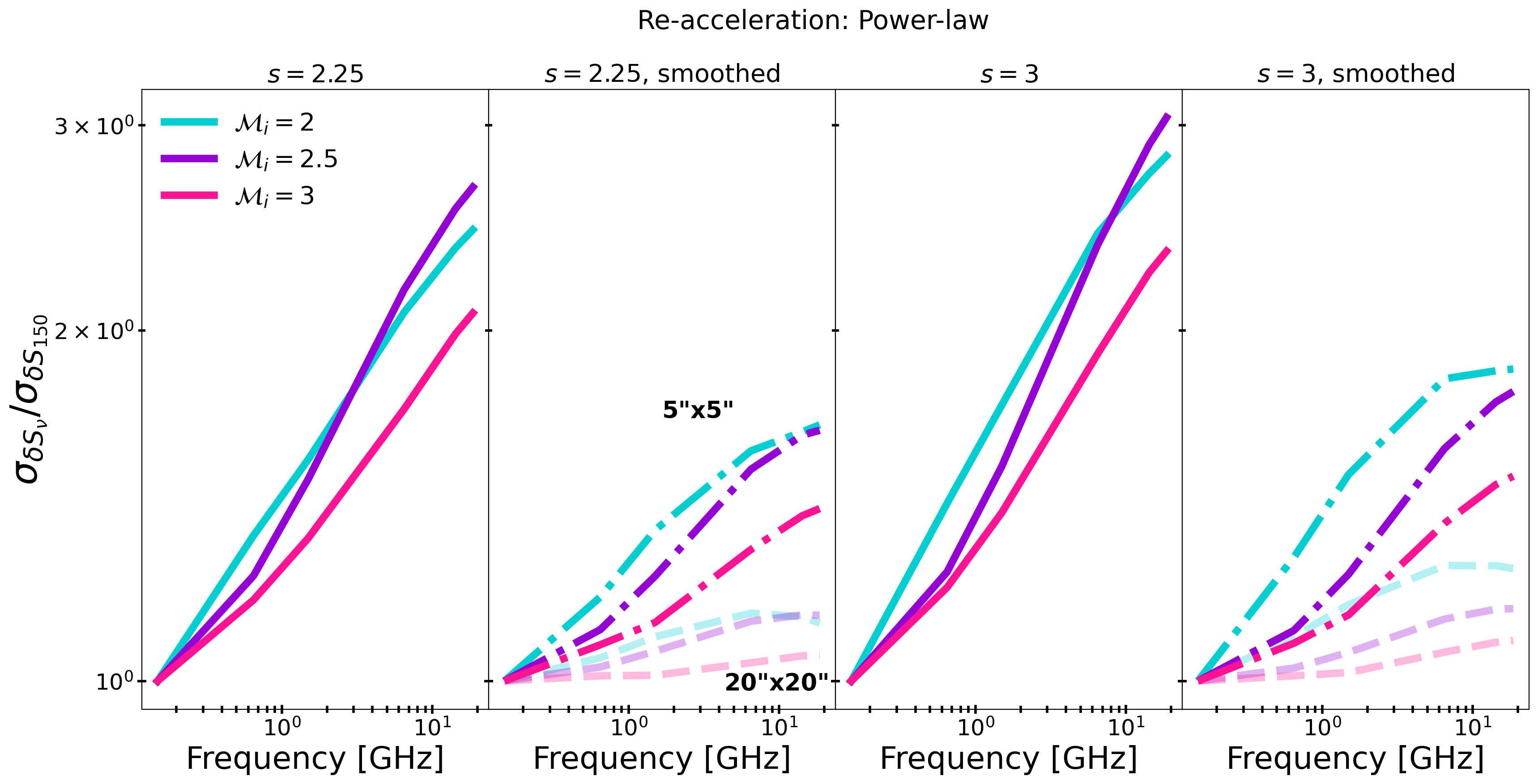}
    \caption{Standard deviation of the $\delta S_{\nu}$ distribution normalised by that at 150 MHz at the shock-front at 184.7 Myr for all our models at
all frequencies. The top, middle and bottom rows correspond to the fresh injection model, the re-acceleration model assuming
a Dirac Delta pre-existing distribution function and the re-acceleration model assuming a power-law pre-existing distribution function, respectively. The first and third columns show the results from the original maps and the second and fourth columns show the results of the map smoothed with 5"x5" (dot-dashed) and 20"x20" (dashed) beam sizes.}
    \label{fig:new_stacked}
\end{figure*}

\begin{figure*}
    \centering
    \includegraphics[width=16.5cm]{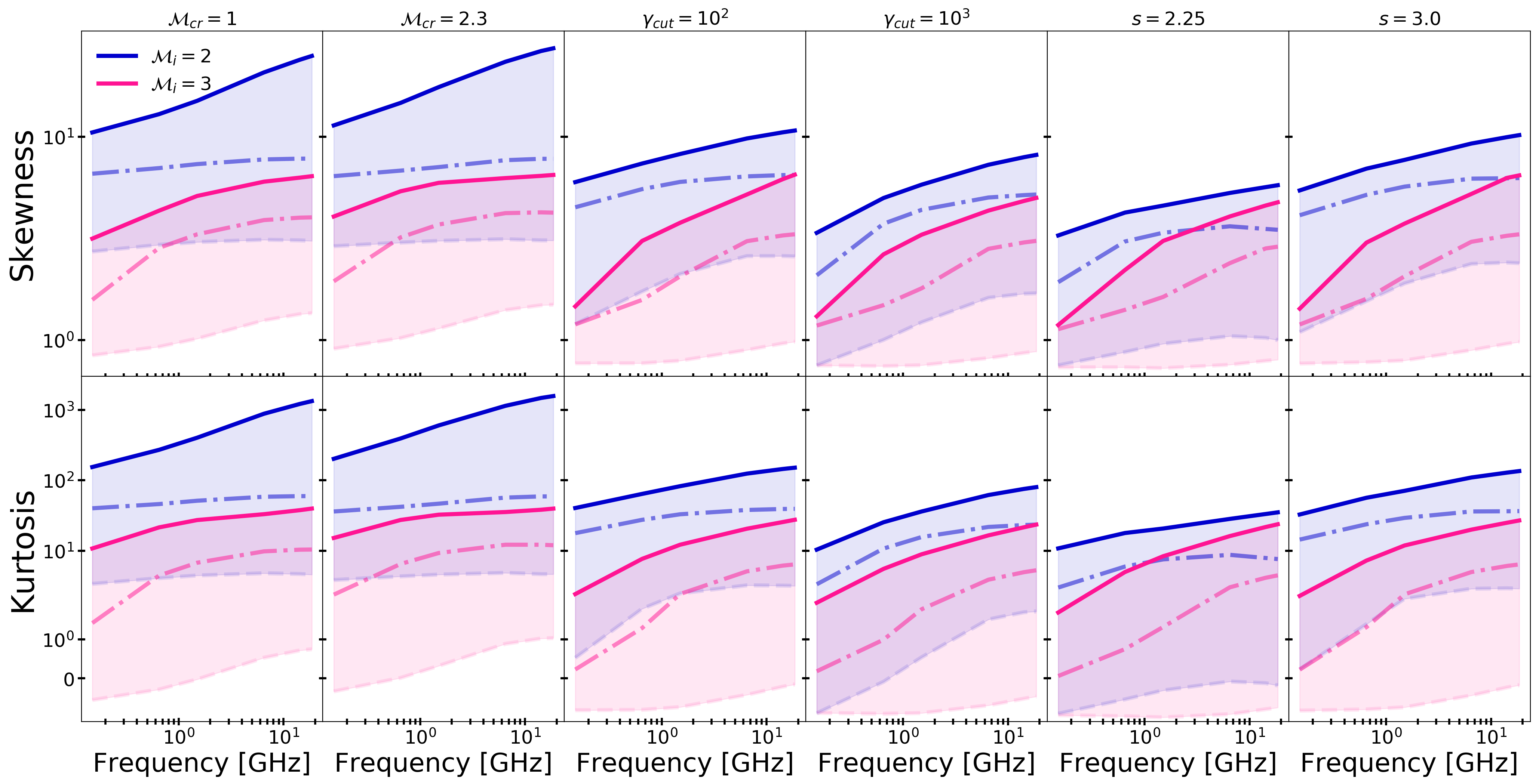}
    \caption{Skewness (\textit{upper row}) and kurtosis (\textit{lower row}) of the $\delta S_{\nu}/$ distribution for all our models at all frequencies. We only show the  $\mathcal{M}_{i}=2$ and $\mathcal{M}_{i}=3$ cases. The upper solid lines correspond to the unsmoothed data and the dot-dashed and dahsed lines correspond to the data smoothed with the 5"$\times$5" and 20"$\times$20" beam sizes, respectively. 
    }
    \label{fig:sync_statistics_skew_new}
\end{figure*}

\begin{figure}
    \centering
    \includegraphics[width=0.9\columnwidth]{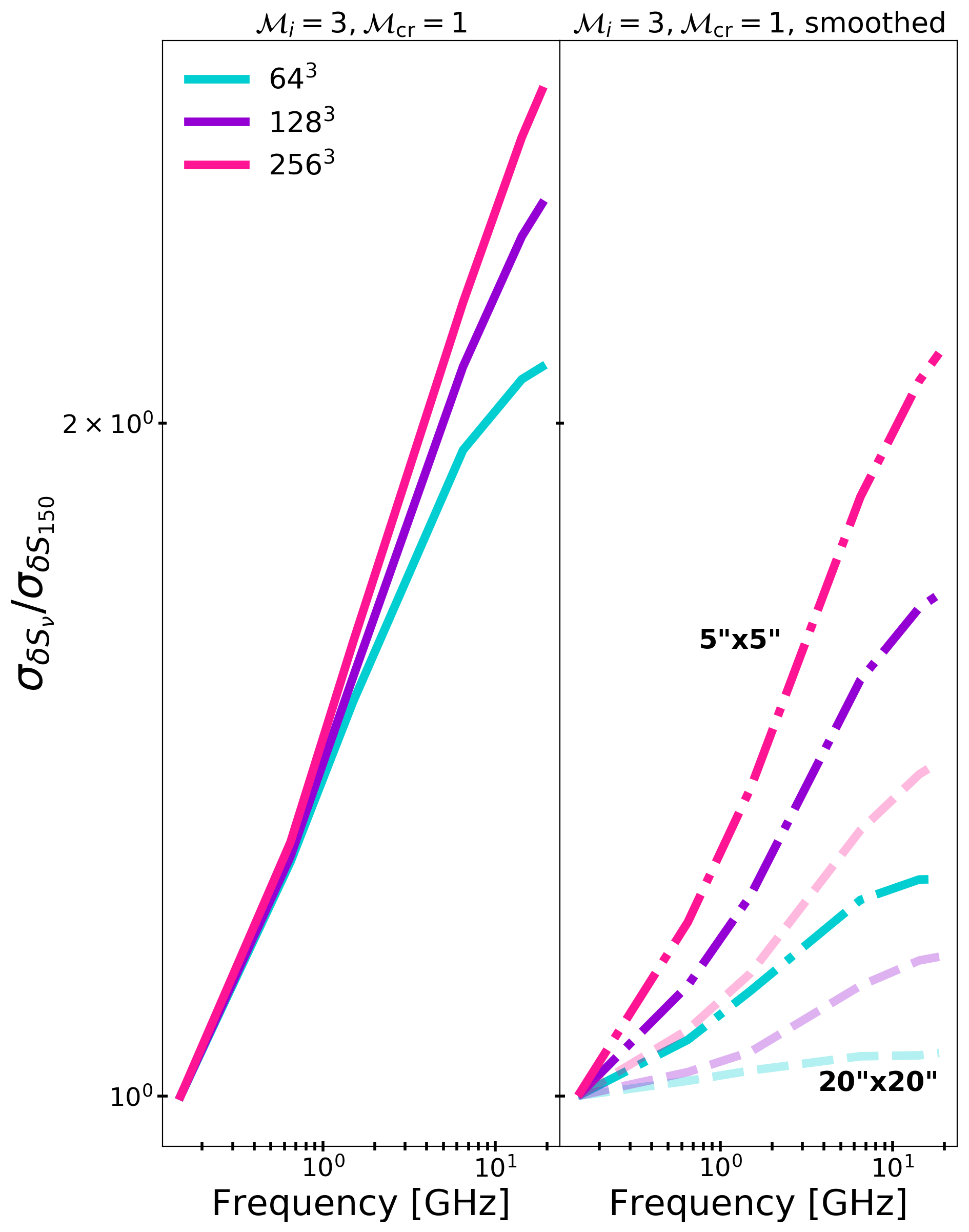}
    \caption{Standard deviation of the $\delta S_{\nu}$ distribution normalised by that at 150 MHz at the shock-front for the $\mathcal{M}_i=3$ case at all frequencies for the $\mathcal{M}_{cr}=1$ fresh injection model at $t=187.4$ Myr. We show the result for three different numerical resolutions. The first column shows the results from the original maps and the second column the results from the smoothed maps, similar to Fig.~\ref{fig:new_stacked}
    }
    \label{fig:std_num-res}
\end{figure}

In Fig.~\ref{fig:sync_statistics_2} we plot $\sigma_{\delta S_{\nu}}$ as function of frequency for all the studied models. We considered again two simulation time-steps, $t=102.7$ Myr and $t=184.7$ Myr. This figure summarizes all the information from all our models, and we discuss the main findings in the following:

\begin{itemize}
    \item[i)] \textit{Low/high frequencies}:  
    A visual inspection of Figs.~\ref{fig:maps_all_otherstyle}-\ref{fig:maps_donwstream} reveals a patchier emission at high frequencies compared to lower frequencies. We observe almost no distinction between the surface brightness maps at 14.25 GHz and 18.6 GHz in all our models. This is confirmed in the phase-plots shown in Fig.~\ref{fig:phase-plots} where we can see that $\delta S_{\nu}+1$ reaches the highest values at the highest frequencies for both the fresh-injection and re-acceleration models. The largest discrepancies between high (18.6 GHz) and low  (150 MHz) frequencies can be more clearly seen in Fig.~\ref{fig:sync_statistics_2} where the standard deviation of the corresponding distribution can be as large as $\sim 7$ times the standard deviation of the 150 MHz distribution in the fresh-injection model. The same trend follows in the re-acceleration models except that here, the standard deviation of the distribution can be as large as $\sim 2.5$ and $\sim 2$ times the standard deviation of the 150 MHz distributions, respectively for the Dirac delta and power-law fossil electron populations. \\

    \item[ii)] \textit{Initial Mach number}: The larger $\mathcal{M}_i$, the less patchy the radio emission. This is expected as a stronger shock can more easily compress and amplify the magnetic field along the shock. This is again reflected in Fig.~\ref{fig:phase-plots} and in Fig.~\ref{fig:sync_statistics_2}, where the $\mathcal{M}_i=3$ case leads to the lowest values of $\delta S_{\nu}+1$ and therefore, the lowest values of $\sigma_{\delta S_{\nu}}$. As mentioned in the point above, we observe increasing trends of the standard deviation with the frequency in Fig.~\ref{fig:sync_statistics_2} for the three $\mathcal{M}_i$ cases. However, the trend can especially steepen for the $\mathcal{M}_i = 2$ case in the fresh-injection model. This happens because in the fresh-injection model, the differences between the spectral normalization factors corresponding to $\mathcal{M}\sim 2$ and $\mathcal{M}\gtrsim 3$ regions can be very large (see Appendix \ref{app:Eradio_Mach}). Since in this model, the emission will ultimately be biased towards regions where there is higher compression (higher Mach number), then we expect larger variations/fluctuations in the surface brightness maps when the distribution of Mach numbers at the shock front contains a significant amount of low Mach numbers (see first row of Fig.~\ref{fig:mach_evol} in Section \ref{sec:Mach_dist}). Finally, we note that the re-acceleration models in the $\mathcal{M}_i = 2.5$ and $\mathcal{M}_i = 3$ cases can reproduce similar degrees of patchiness (see last four panels of Fig.~\ref{fig:sync_statistics_2}). \\

    \item[iii)] \textit{Fresh-injection model (Critical Mach number)}: At any $\mathcal{M}_i$ it will be hard (or almost impossible) to visually distinguish only from surface brightness variations different $\mathcal{M}_{cr}$ models. In the $\mathcal{M}_i = 2.5$ and $\mathcal{M}_i = 3$ cases, there is no significant distinction between a model forbidding an acceleration below $\mathcal{M}_{cr} = 1$ or $\mathcal{M}_{cr} = 2.3$ at all frequencies (see Fig.~\ref{fig:maps_all_otherstyle}). This is quantitatively better observed in the top row of Fig.~\ref{fig:phase-plots}, where the $\delta S_{\nu}+1$ values are comparable in both the $\mathcal{M}_{cr} = 1$ and $\mathcal{M}_{cr} = 2.3$ cases. This changes significantly in the $\mathcal{M}_i = 2$ case where we see the highest values of the standard deviation in comparison to the stronger shocks with $\mathcal{M}_i = 2.5$ and $\mathcal{M}_i = 3$ (see Fig.~\ref{fig:sync_statistics_2}). In fact, in this case we observe the largest discrepancies between the standard deviation of the $\mathcal{M}_{cr} = 1$ and $\mathcal{M}_{cr} = 2.3$ cases, with the latter yielding higher values. This last result happens due to the fact that only a low percentage of the Mach number distribution at the shock front (see left panel of Fig.~\ref{fig:mach_evol}) is contributing to the emission, i.e. the high-end tail of the distribution. Despite the different $\mathcal{M}_{cr}$, the discrepancies in the surface brightness are not very pronounced, primarily because the radio emissivity is known to be biased towards higher Mach numbers in the fresh-injection model \citep[see e.g.][]{2019ApJ...883..138R,dominguezfernandez2020morphology,Wittor_2021}. Therefore, we conclude that the $\mathcal{M}_{cr}$ parameter has a minor impact in our results. \\

    \item[iv)] \textit{Re-acceleration model (Energy of fossil electrons $\gamma_{cut}$)}: The smaller the $\gamma_{cut}$, the larger the fluctuations or the more patchy the substructure will look like. Indeed, the two columns of the middle row in Fig.~\ref{fig:sync_statistics_2} show that the $\delta S_{\nu}+1$ values are higher in the $\gamma_{cut}=10^2$ case by less than a factor 2 when compared to the $\gamma_{cut}=10^3$ case. These differences are naturally subtle. Therefore, we conclude that in a re-acceleration model with a fossil population characterized by a Dirac delta, the parameter $\gamma_{cut}$ does not impact our main two results, i.e. the higher the frequency and/or the lower the mean strength of the shock (or the parameter $\mathcal{M}_i$ in this study) the larger the surface brightness fluctuations.\\
    
    \item[v)] \textit{Re-acceleration model (Energy spectral index $s$)}: The steeper the power-law fossil spectrum is (larger value of $s$), the larger the fluctuations or the more patchy the substructure will look like. Similar to the other re-acceleration model where we assumed the fossil electron population is characterized by a Dirac delta function, the differences in the $\delta S_{\nu}+1$ values between the $s=2.25$ and $s=3$ cases are subtle with less than a factor 2 difference in their values. Yet here, the nature of these differences  
    lies in 1) the proportion of particles at the shock front which would acquire a re-acceleration spectrum with an $s$ or $p$ power-law index and 2) the normalization factor.  
    For example, for the $s=2.25$ case, spectra with index $r\sim s$ will mainly dominate the emission, whereas for the $s=3$ case, spectra with index $r\sim p$ will dominate. On the other hand, the normalization factors in the $s=2.25$ case are larger for the spectra with index $r\sim p$ while in the $s=3$ case, the spectra with index $r\sim s$ have larger normalization factors. \\
    
    \item[vi)] \textit{Evolution of the shock}: In general, the degree of surface brightness fluctuations at the shock front is not expected to be global and will be susceptible to the local environment within the galaxy cluster. For this reason, we analyzed two different time-steps of our simulations. The net result is drawn in Fig.~\ref{fig:sync_statistics_2} where 
    we show $\sigma_{\delta S_{\nu}}$ as function of frequency at $t=120.7$ Myr (high transparency) and $t=184.7$ Myr (no transparency).
    At $t=184.7$ Myr the shock 
    entails more surface brightness fluctuations and therefore, larger values of the $\sigma_{\delta S_{\nu}}$. This is observed for both the fresh-injection and re-acceleration models. Past studies of shocks in simulated galaxy clusters show that relic shocks should have a broad Mach number distribution \citep[see][]{ry03}. Therefore, we consider our results from this time-step
    to be a more realistic representation of how can surface brightness fluctuations vary with frequency in different models.
    \end{itemize}

We would like to emphasize that, as previously mentioned in the introduction, shocks with $\mathcal{M}_i=2$ are not theoretically expected to be sites of DSA, given that the critical Mach number is typically found to be around $\mathcal{M}_{cr}\sim 2.3$. However, observations often infer Mach numbers using simplified assumptions from both X-ray and radio data, suggest the involvement of such weak shocks, or even weaker, in the DSA process. Our results intriguingly suggest that if this is indeed the case, and the fresh injection model is at play, significant differences in radio morphologies between low and high frequencies should be observable. This is clearly illustrated in the first two panels of Fig.~\ref{fig:sync_statistics_2}, where we observe substantial disparities in $\sigma_{\delta S_{\nu}}$.

Finally, in the context of the fresh injection model, it is widely acknowledged in the literature that $\eta(\mathcal{M})$ is an increasing function with the Mach number \citep[][]{Kang_2007,kr13,Caprioli_2014,2019ApJ...883...60R}. In Appendix~\ref{app:extra_models}, we additionally explore various toy $\eta(\mathcal{M})$ models to investigate whether the already found trends can be altered. Physically, it is impossible for the efficiency to increase with weaker shocks, which means $\eta(\mathcal{M})$ should be a decreasing function with the Mach number. Therefore, in Appendix~\ref{app:extra_models}, our analysis is restricted to hypothetical models featuring efficiency functions that display more pronounced increases than the currently accepted theoretical models.

\setlength{\tabcolsep}{8pt}
\begin{table}
\centering
\begin{tabular}{cccc}
    & & &\\ \hline
      Beam & FWHM [kpc] & $\nu$ [GHz] & Telescope  \\ \hline
      5"$\times$5" &  16.6 & 0.15 &  LOFAR$^a$ \\ 
      5"$\times$5" &  16.6 & 0.65 & GMRT$^b$  \\ 
      5"$\times$5" &  16.6 & 1.5 & VLA$^c$   \\ 
      20"$\times$20" & 66.3 & 14.25 & Effelsberg$^d$  \\
      20"$\times$20"& 66.3 & 18.6  & SRT$^d$  \\
      \hline
\end{tabular}

\caption{Restoring beam sizes assumed for smoothing the emission maps at each observing frequency. We considered the redshift of the CIZA J2242.8+5301 cluster ($z=0.1921$) to quote the corresponding linear length of the full width half maximum (FWHM).
The last column mentions the telescope and reference that we considered for our mock radio maps. Note that the radio maps in \citealt{Loi_2020} quote a resolution of 72"$\times$72" (FWHM = 238.8 kpc) for the Effelsberg observation and 54"$\times$54" (FWHM = 179.1 kpc) for the SRT observation, which are too large compared to our simulation box. Therefore, we have selected a conservative beam of 20"$\times$20" for these two high frequencies. \\
$^a$ {\tiny \citealt{Hoang2017}} \\
$^b$ {\tiny \citealt{Stroe2013}} \\
$^c$ {\tiny \citealt{2018ApJ...865...24D}} \\
$^d$ {\tiny \citealt{Loi_2020}}
}
\label{table:beams}
\end{table}

We show that steeper models, which assign greater weight to larger Mach numbers, result in a higher degree of patchiness or substructure. Across most of our models, we observe that $\sigma_{\delta_{S_{\nu}}}$ increases with the observing frequency, consistent with the findings discussed in this section. However, one exception is our steepest $\eta(\mathcal{M})$ model, which exhibits minimal change in $\sigma_{\delta_{S_{\nu}}}$ when the observing frequency is increased. Nonetheless, both high and low frequencies would render a radio relic notably patchy or with a high degree of patchiness.

In summary, our conclusions regarding the fresh injection model are twofold: 1) steeper $\eta(\mathcal{M})$ models consistently lead to a higher degree of patchiness, and 2) increasing $\eta(\mathcal{M})$ functions generally reproduce the previously discussed trend, wherein a radio relic appears patchier at higher frequencies. This is in alignment with the results presented in this section.


\subsection{Beam and resolution effects}
\label{sec:beam}

So far, we have shown that under the current accepted theories of particle acceleration, radio relics should appear to be patchier at very high frequencies. While this is a general consequence of the underlying physical processes along with the observing frequency, in this Section we study how much the surface brightness variations can change if one takes into account different beam resolutions. At the end of this Section, we also discuss the role of numerical resolution.

The linear size of radio relics is of the order of Mpc. Due to our periodic boundary conditions in the $y$ and $z$ dimensions (as described in Sec.~\ref{sec:PLUTO}) and the planar nature of the propagating shock, we can confidently stack our surface brightness maps to mock a larger radio shock without any loss of generality. We stacked our surface brightness maps in the $x$-direction to mock a 3 Mpc radio shock. That is, we stacked 15 surface brightness maps each with an extension of 200 kpc in the $y$-direction (see Fig.~\ref{fig:maps_all_otherstyle}). 
Taking as an example the CIZA J2242.8+5301 cluster at $z=0.1921$ (see Table~\ref{table:beams} for the references on different radio telescopes), we considered two beam sizes, 5"$\times$5" and 20"$\times$20". 
We stress that we do not aim to make a one-to-one comparison with any particular relic. Therefore, the FHWM values in Table~\ref{table:beams} are used only as representative values.

In Fig.~\ref{fig:new_stacked} we show the standard deviation of the $\delta S_{\nu}$ distribution normalised by that at 150 MHz at the shock-front for all our models at
all frequencies. We also show the effects of the 5"x5" (dot-dashed line) and 20"x20" (dashed line) beam sizes. As expected, the larger the beam
size, the less surface brightness fluctuations are seen along the shock front. We also observe that the trend of increased fluctuations or a patchier radio relic with higher frequencies diminishes considerably when we consider a beam size of 20"x20".
In particular, if the shock front is composed of a Mach number distribution that is not broad enough, e.g. a shock in its early propagation, the relic should be observed with the same degree of patchiness at low and high frequencies. This can be clearly observed in the second and fourth columns of Fig.~\ref{fig:new_stacked} where the dashed lines denote the results of our analysis with a beam size of 20"x20". Moreover, we note that even if the Mach number distribution is broad as expected from theory, a 20”×20” beam size would not be enough to capture the underlying surface brightness differences at different frequencies. 
The panorama is slightly better for a 5"x5" beam size, where the trend of having a patchier radio relic as frequency increases could be recovered. This suggests that the shock front should comprise a Mach distribution function with a spread $\sigma_{\mathcal{M}} \gtrsim 0.3--0.4$ (see Fig.~\ref{fig:mach_evol}) for us to be able to observe differences between low and high frequencies.

Next, we 
plot in Fig.~\ref{fig:sync_statistics_skew_new} additional statistics of the  
$\delta S_{\nu}$ distribution: skewness
(denoted as $g$)
and kurtosis
(denoted as $G$). 
We use the SciPy package \citep[][]{2020SciPy-NMeth} to compute both. The sample skewness and kurtosis are defined according the Fisher's definition as
\begin{equation}
    g = \frac{m_3}{m_2^{3/2}},
\end{equation}
and
\begin{equation}
    G = \frac{m_4}{m_2^{2}} - 3,
\end{equation}
respectively, where 
\begin{equation}
    m_i = \frac{1}{N} \, \sum_{n=1}^{N} \left( x[n] - \overline{x} \right)^i
\end{equation}
is the biased sample \textit{ith} central moment, and $\overline{x}$ is the sample mean.
These higher order statistics show that the higher the frequency the more the 
$\delta S_{\nu}$ distribution is asymmetric and heavy-tailed. In particular, a large kurtosis signals non-Gaussianity and super-Gaussian tails on the PDF. 
This trend is persistent both in the fresh-injection model and the re-acceleration model  
with power-law fossil electrons. Nevertheless, it is easy to see that the fresh-injection model leads to larger values of the kurtosis and skewness than the re-acceleration model. A larger difference between these two models can be better seen at lower $\mathcal{M}_i$ (see e.g. the $\mathcal{M}_i=2$ case in Fig.~\ref{fig:sync_statistics_skew_new}). As expected, a larger beam size smooths out the radio substructure leading to lower skewness and kurtosis values, i.e. the 
$\delta S_{\nu}$ distribution approaches a normal distribution. Finally, we note that in all cases, the distribution is positively skewed.

All these results imply the same findings as described above: 1) the higher the frequency, the larger the degree of patchiness, 2) the larger the initial Mach number of the shock, the less patchy the radio emission and 3) the larger the beam size, the smaller the degree of patchiness. Therefore, the measure of $\sigma_{\delta_{S_{\nu}}}$ at different frequencies would suffice to probe the degree of patchiness of radio relics. As such, we propose it as a potential tool for extracting merger shock properties as well as information about particle acceleration processes at shocks.


Finally, we investigate the influence of numerical resolution on our findings. We conducted simulations at three different resolutions: $128 \times 64 \times 64$ cells, $256 \times 128 \times 128$ cells, and $512 \times 256 \times 256$ cells, all for the fresh injection model under the conditions of $\mathcal{M}_i=3$ and $\mathcal{M}_{cr}=1$. It is important to note that we did not run simulations at higher resolutions across all our models due to the significant computational demands, particularly when dealing with the energy spectra of 25,165,824 Lagrangian particles in conjunction with the Eulerian code. However, as we will demonstrate, doubling the resolution from our current setting has a minimal impact on our primary outcomes. Interested readers can also refer to Appendix C of \citealt{dominguezfernandez2020morphology} for a detailed discussion of the differences between high and medium-resolution runs.

In the left panel of Fig.~\ref{fig:new_stacked}, we show that differences among the three different resolutions are only discernible at high frequencies.  At 18.6 GHz, the lowest resolution yields a percentage difference in $\sigma_{\delta S_{\nu}}/\sigma_{\delta S_{150}}$ compared to the highest resolution of approximately 16\%. 
However, doubling the resolution results in a modest enhancement of approximately 8\%. This finding affirms that our chosen resolution is well-suited for the objectives of this study. In the right panel of Fig.~\ref{fig:new_stacked}, we present the corresponding trends obtained from smoothing the surface brightness maps for completeness. We remind the reader that this analysis was analysis was done using also a stacked image, which incorporates the surface brightness map featuring a 3 Mpc radio shock.

\section{Summary and conclusions}
\label{sec:conclusions}

In this work, we aimed at 
studying the fluctuations of radio surface brightness or what we define as patchiness of radio relics
numerically, 
at high ($\gtrsim 10$ GHz) and low ($\lesssim$ 5--6 GHz) frequencies
by assuming the two current accepted theories of CR acceleration that we have at hand: 1) the fresh injection model, in which electrons from the thermal pool are injected into the DSA, and 2) the re-acceleration model, in which already mildly relativistic electrons are injected into the DSA. 
We have studied shock waves with $\mathcal{M}=2, 2.5, 3$ propagating through a magnetized medium with decaying solenoidal subsonic turbulence representing a small fraction of the ICM. 
We performed hybrid simulations where the MHD Eulerian grid represents the thermal fluid, while Lagrangian particles represent CR electrons. The CR electrons are injected at the shock discontinuity assuming DSA and evolved according to a simplified cosmic-ray equation including adiabatic, synchrotron and Inverse Compton losses.

The premise of our work is simple, we show that, in the presence of pre-shock fluctuations in the ICM, the shock front part of a radio relic is bound to
have substructure 
both at low and high frequencies. However, we find that the radio substructures differ
under the assumption of the fresh injection or re-acceleration models as a consequence of the underlying differences between the two physical processes.\\

Our findings can be summarized as follows:

\begin{itemize}
    \item[i)] We find an increasing degree of patchiness with frequency both in the fresh injection and re-acceleration models. We suggest that quantifying the degree of patchiness with the standard deviation of the relative surface brightness fluctuations, $\sigma_{\delta S_{\nu}}$, suffices to probe this trend across different frequencies. 
    We find that in the fresh injection model, the standard deviation at 18.6 GHz is $\sim 7$ times larger than that at 150 MHz. In contrast, in the re-acceleration models, the standard deviation at 18.6 GHz is $\sim 2$--2.5 times larger than at 150 MHz.
    Therefore, the re-acceleration model produces smoother radio structures compared to the fresh injection model.

    \item[ii)] The Mach number distribution of the shock front plays the main role in the observed surface brightness substructure. We find that the shock front should comprise a Mach number distribution function with a spread $\sigma_{\mathcal{M}}\gtrsim0.3$--0.4 to enable the observation of differences between low and high frequencies at a 5"$\times$5" resolution.
    
    \item[iii)] We find that the larger the mean Mach number of the shock (or $\mathcal{M}_i$), the less patchy the radio emission. Patchiness arises if the distribution of Mach number characterizing the shock front encompasses a large percentage of low Mach numbers ($\mathcal{M} \lesssim 2.5$). 
    
    \item[iv)] The projected Mach number weighted with the synchrotron emissivity shows the largest discrepancies with respect to other weights such as the density, temperature or magnetic field. This is particularly shown in the standard deviation of the projected Mach number. This projection effect adds to the problem where larger Mach number values are inferred from radio compared to those inferred from X-ray.

    \item[v)] A steeper $\eta(\mathcal{M})$ model leads to a higher degree of patchiness or more substructure in the radio emission.
    
\end{itemize}

Our study shows that comparing the degree of patchiness within radio relics at across a broad range of frequencies allows us to retrieve information about particle acceleration processes at shocks in the ICM. For instance, we have explored its potential to unveil details about the dispersion of the Mach distribution function and the steepness of the acceleration efficiency function. While a more extensive parameter study, incorporating various pre-shock turbulent conditions, is required for formulating a definitive formula for the former, we propose that analyzing the degree of patchiness represents a promising, novel, and robust addition to the traditional observational techniques.

We have shown that, in general, the fresh-injection model can lead to very patchy radio relics while the re-acceleration model leads to smoother ones. 
Nevertheless, it shall be noted that our modelling is based on a uniformly spaced distribution of fossil electrons. There are few observational examples in which CRe may spread-out up to very large scales in the ICM \citep[see e.g. ][for observations showing the late evolution of multiple generations of CRe associated to active-galactic-nuclei bubbles]{2021NatAs...5.1261B,2022A&A...661A..92B}. 
Yet, this evidence is showing us how complicated and unlikely it would be to get a spatially uniform fossil population such that it reproduces the whole extend of Mpc relics. 
While this would pose a limitation of our work, this essentially means that it would also be difficult to numerically reproduce very smooth radio relics at low frequencies with the re-acceleration model.

Additionally, we did not consider an acceleration efficiency depending on $\theta_B$ in our fresh-injection model. While this dependency has been considered in other works \citep[see e.g.][]{va16scienzo,10.1093/mnras/sty1410,2020MNRAS.496.3648B}, we restricted ourselves to study only the Mach number dependency 
because adding another degree of freedom to the acceleration efficiency will further narrow down the locations of emission. As a consequence, this will lead to a more pronounced patchiness in the radio emission along the shock front, ultimately resulting in a heightened degree of patchiness in radio relics. Hence, we anticipate that this variable will not only leave our main result unaffected but also reinforce it.

It is important to highlight that the main focus of this paper has been to study the physical processes accountable for generating patchiness in radio relics. 
On top of that, observational factors like sensitivity, resolution image noise, etc., can influence the perceived surface brightness morphology. Such morphological differences caused by instrumental effects differ conceptually from the physical mechanisms we have explored in this study and therefore, should be studied carefully in future work.

Recent works have started to show high quality data where more statistics beyond integrated quantities are being taken into account \citep[see e.g.][]{2018ApJ...852...65R,2020A&A...636A..30R,2021A&A...646A..56R,2022ApJ...927...80R}. 
While at very high frequencies this may pose more observational challenges due to limited resolution, conducting similar studies at both low and high frequencies could provide a more robust constraint on radio fluctuations along the shock front. This approach will help us determine whether a frequency-dependent patchiness trend, specifically an increasing $\sigma_{\delta S_{\nu}}$ with frequency, is consistently observable in various radio relics.
In future work, we will study instrumental effects and make a careful comparison with some observations.

\section*{Acknowledgments}
%
We gratefully acknowledge the
PLUTO development group with special recognition to our collaborators A. Mignone and D. Mukherjee
who provided the particle module of the PLUTO code
and extremely helpful comments. The simulations
presented in this work made use of computational resources on the
JUWELS cluster at the Juelich Supercomputing Centre (JSC), under
project no. 24944 CRAMMAG-OUT and CRAMMAG-OUT-2 with P. Dom\'inguez-Fern\'andez as
principal investigator.

P. Dom\'inguez-Fern\'andez acknowledges the Future Faculty Leaders Fellowship at the Center for Astrophysics, Harvard-Smithsonian, and the financial support from the European
Union’s Horizon 2020 programme under the ERC Starting Grant
‘MAGCOW’, no. 714196 with F. Vazza as principal investigator.
All authors acknowledge the support of the National Research
Foundation (NRF) of Korea through grant Nos. 2020R1A2C2102800 and 2023R1A2C1003131. 
We acknowledge useful scientific discussions with R. J. van Weeren, M. Johnston-Hollitt and K. Rajpurohit during the initial phase of this study. Finally, we would like to extend our appreciation to P. Nulsen and A. Stroe for 
their insightful contributions during the last phase of this study. \\

\textit{Software}: The source codes used for
the simulations of this study, FLASH \citep{2000ApJS..131..273F, 2002ApJS..143..201C} and PLUTO \citep{pluto1} are freely available
on \url{https://flash.rochester.edu/site/} and  \url{http://plutocode.ph.unito.it}. The main tools for our analysis are:
        Python \citep{van1995python}, SciPy \citep{2020SciPy-NMeth} and  
          IDL \citep{1993ASPC...52..246L}.


\section*{Data Availability Statement}

The data underlying this article will be shared on reasonable request to the corresponding author.

\bibliographystyle{aa}

\bibliography{main}{}

\begin{thebibliography}{73}
\expandafter\ifx\csname natexlab\endcsname\relax\def\natexlab#1{#1}\fi

\bibitem[{{Akamatsu} {et~al.}(2015){Akamatsu}, {van Weeren}, {Ogrean},
  {Kawahara}, {Stroe}, {Sobral}, {Hoeft}, {R{\"o}ttgering}, {Br{\"u}ggen}, \&
  {Kaastra}}]{2015A&A...582A..87A}
{Akamatsu}, H., {van Weeren}, R.~J., {Ogrean}, G.~A., {et~al.} 2015, \aap, 582,
  A87

\bibitem[{{Banfi} {et~al.}(2020){Banfi}, {Vazza}, \&
  {Wittor}}]{2020MNRAS.496.3648B}
{Banfi}, S., {Vazza}, F., \& {Wittor}, D. 2020, \mnras, 496, 3648

\bibitem[{{Blandford} \& {Eichler}(1987)}]{be87}
{Blandford}, R. \& {Eichler}, D. 1987, \physrep, 154, 1

\bibitem[{{B{\"o}ss} {et~al.}(2022){B{\"o}ss}, {Steinwandel}, {Dolag}, \&
  {Lesch}}]{2022arXiv220705087B}
{B{\"o}ss}, L.~M., {Steinwandel}, U.~P., {Dolag}, K., \& {Lesch}, H. 2022,
  arXiv e-prints, arXiv:2207.05087

\bibitem[{{Botteon} {et~al.}(2020){Botteon}, {Brunetti}, {Ryu}, \&
  {Roh}}]{2020A&A...634A..64B}
{Botteon}, A., {Brunetti}, G., {Ryu}, D., \& {Roh}, S. 2020, \aap, 634, A64

\bibitem[{{Brienza} {et~al.}(2022){Brienza}, {Lovisari}, {Rajpurohit},
  {Bonafede}, {Gastaldello}, {Murgia}, {Vazza}, {Bonnassieux}, {Botteon},
  {Brunetti}, {Drabent}, {Hardcastle}, {Pasini}, {Riseley}, {R{\"o}ttgering},
  {Shimwell}, {Simionescu}, \& {van Weeren}}]{2022A&A...661A..92B}
{Brienza}, M., {Lovisari}, L., {Rajpurohit}, K., {et~al.} 2022, \aap, 661, A92

\bibitem[{{Brienza} {et~al.}(2021){Brienza}, {Shimwell}, {de Gasperin},
  {Bikmaev}, {Bonafede}, {Botteon}, {Br{\"u}ggen}, {Brunetti}, {Burenin},
  {Capetti}, {Churazov}, {Hardcastle}, {Khabibullin}, {Lyskova},
  {R{\"o}ttgering}, {Sunyaev}, {van Weeren}, {Gastaldello}, {Mandal}, {Purser},
  {Simionescu}, \& {Tasse}}]{2021NatAs...5.1261B}
{Brienza}, M., {Shimwell}, T.~W., {de Gasperin}, F., {et~al.} 2021, Nature
  Astronomy, 5, 1261

\bibitem[{{Br{\"u}ggen} {et~al.}(2011){Br{\"u}ggen}, {Bykov}, {Ryu}, \&
  {R{\"o}ttgering}}]{br11}
{Br{\"u}ggen}, M., {Bykov}, A., {Ryu}, D., \& {R{\"o}ttgering}, H. 2011, \ssr,
  71

\bibitem[{{Bykov} {et~al.}(2019){Bykov}, {Kaastra}, {Br{\"u}ggen},
  {Markevitch}, {Falanga}, \& {Paerels}}]{2019SSRv..215...27B}
{Bykov}, A.~M., {Kaastra}, J.~S., {Br{\"u}ggen}, M., {et~al.} 2019, \ssr, 215,
  27

\bibitem[{{Calder} {et~al.}(2002){Calder}, {Fryxell}, {Plewa}, {Rosner},
  {Dursi}, {Weirs}, {Dupont}, {Robey}, {Kane}, {Remington}, {Drake}, {Dimonte},
  {Zingale}, {Timmes}, {Olson}, {Ricker}, {MacNeice}, \&
  {Tufo}}]{2002ApJS..143..201C}
{Calder}, A.~C., {Fryxell}, B., {Plewa}, T., {et~al.} 2002, \apjs, 143, 201

\bibitem[{Caprioli \& Spitkovsky(2014)}]{Caprioli_2014}
Caprioli, D. \& Spitkovsky, A. 2014, The Astrophysical Journal, 783, 91

\bibitem[{{Chandran}(2005)}]{2005PhRvL..95z5004C}
{Chandran}, B. D.~G. 2005, \prl, 95, 265004

\bibitem[{{Clarke} \& {Ensslin}(2006)}]{ce06}
{Clarke}, T.~E. \& {Ensslin}, T.~A. 2006, \aj, 131, 2900

\bibitem[{{de Gasperin} {et~al.}(2021){de Gasperin}, {Rudnick}, {Finoguenov},
  {Wittor}, {Akamatsu}, {Bruggen}, {Clarke}, {Cotton}, {Cuciti},
  {Dominguez-Fernandez}, {Knowles}, {O'Sullivan}, \&
  {Sebokolodi}}]{2021arXiv211106940D}
{de Gasperin}, F., {Rudnick}, L., {Finoguenov}, A., {et~al.} 2021, arXiv
  e-prints, arXiv:2111.06940

\bibitem[{{Dedner} {et~al.}(2002){Dedner}, {Kemm}, {Kr{\"o}ner}, {Munz},
  {Schnitzer}, \& {Wesenberg}}]{ded02}
{Dedner}, A., {Kemm}, F., {Kr{\"o}ner}, D., {et~al.} 2002, Journal of
  Computational Physics, 175, 645

\bibitem[{{Di Gennaro} {et~al.}(2018){Di Gennaro}, {van Weeren}, {Hoeft},
  {Kang}, {Ryu}, {Rudnick}, {Forman}, {R{\"o}ttgering}, {Br{\"u}ggen},
  {Dawson}, {Golovich}, {Hoang}, {Intema}, {Jones}, {Kraft}, {Shimwell}, \&
  {Stroe}}]{2018ApJ...865...24D}
{Di Gennaro}, G., {van Weeren}, R.~J., {Hoeft}, M., {et~al.} 2018, \apj, 865,
  24

\bibitem[{Domínguez-Fernández {et~al.}(2020)Domínguez-Fernández, Brüggen,
  Vazza, Banda-Barragán, Rajpurohit, Mignone, Mukherjee, \&
  Vaidya}]{dominguezfernandez2020morphology}
Domínguez-Fernández, P., Brüggen, M., Vazza, F., {et~al.} 2020, Monthly
  Notices of the Royal Astronomical Society, 500, 795

\bibitem[{Domínguez-Fernández {et~al.}(2021)Domínguez-Fernández, Brüggen,
  Vazza, Hoeft, Banda-Barragán, Rajpurohit, Wittor, Mignone, Mukherjee, \&
  Vaidya}]{dominguez2021}
Domínguez-Fernández, P., Brüggen, M., Vazza, F., {et~al.} 2021, Monthly
  Notices of the Royal Astronomical Society

\bibitem[{{Drury}(1983)}]{1983RPPh...46..973D}
{Drury}, L.~O. 1983, Reports on Progress in Physics, 46, 973

\bibitem[{{Eswaran} \& {Pope}(1988)}]{1988PhFl...31..506E}
{Eswaran}, V. \& {Pope}, S.~B. 1988, Physics of Fluids, 31, 506

\bibitem[{{Federrath} {et~al.}(2010){Federrath}, {Roman-Duval}, {Klessen},
  {Schmidt}, \& {Mac Low}}]{2010A&A...512A..81F}
{Federrath}, C., {Roman-Duval}, J., {Klessen}, R.~S., {Schmidt}, W., \& {Mac
  Low}, M.~M. 2010, \aap, 512, A81

\bibitem[{{Fryxell} {et~al.}(2000){Fryxell}, {Olson}, {Ricker}, {Timmes},
  {Zingale}, {Lamb}, {MacNeice}, {Rosner}, {Truran}, \&
  {Tufo}}]{2000ApJS..131..273F}
{Fryxell}, B., {Olson}, K., {Ricker}, P., {et~al.} 2000, \apjs, 131, 273

\bibitem[{{Ginzburg} \& {Syrovatskii}(1965)}]{1965ARA&A...3..297G}
{Ginzburg}, V.~L. \& {Syrovatskii}, S.~I. 1965, \araa, 3, 297

\bibitem[{{Ha} {et~al.}(2021{\natexlab{a}}){Ha}, {Kim}, {Ryu}, \&
  {Kang}}]{2021ApJ...915...18H}
{Ha}, J.-H., {Kim}, S., {Ryu}, D., \& {Kang}, H. 2021{\natexlab{a}}, \apj, 915,
  18

\bibitem[{{Ha} {et~al.}(2021{\natexlab{b}}){Ha}, {Ryu}, {Kang}, \&
  {Kim}}]{2021arXiv211014236H}
{Ha}, J.-H., {Ryu}, D., {Kang}, H., \& {Kim}, S. 2021{\natexlab{b}}, arXiv
  e-prints, arXiv:2110.14236

\bibitem[{{Hoang} {et~al.}(2017){Hoang}, {Shimwell}, {Stroe}, {Akamatsu},
  {Brunetti}, {Donnert}, {Intema}, {Mulcahy}, {R{\"o}ttgering}, {van Weeren},
  {Bonafede}, {Br{\"u}ggen}, {Cassano}, {Chy{\.z}y}, {En{\ss}lin}, {Ferrari},
  {de Gasperin}, {Gu}, {Hoeft}, {Miley}, {Orr{\'u}}, {Pizzo}, \&
  {White}}]{Hoang2017}
{Hoang}, D.~N., {Shimwell}, T.~W., {Stroe}, A., {et~al.} 2017, \mnras, 471,
  1107

\bibitem[{{Hong} {et~al.}(2015){Hong}, {Kang}, \& {Ryu}}]{2015ApJ...812...49H}
{Hong}, S.~E., {Kang}, H., \& {Ryu}, D. 2015, \apj, 812, 49

\bibitem[{{Kang}(2020)}]{2020JKAS...53...59K}
{Kang}, H. 2020, Journal of Korean Astronomical Society, 53, 59

\bibitem[{{Kang} \& {Ryu}(2011)}]{2011ApJ...734...18K}
{Kang}, H. \& {Ryu}, D. 2011, \apj, 734, 18

\bibitem[{{Kang} \& {Ryu}(2013)}]{kr13}
{Kang}, H. \& {Ryu}, D. 2013, \apj, 764, 95

\bibitem[{Kang {et~al.}(2007)Kang, Ryu, Cen, \& Ostriker}]{Kang_2007}
Kang, H., Ryu, D., Cen, R., \& Ostriker, J.~P. 2007, The Astrophysical Journal,
  669, 729

\bibitem[{{Kang} {et~al.}(2019){Kang}, {Ryu}, \& {Ha}}]{2019ApJ...876...79K}
{Kang}, H., {Ryu}, D., \& {Ha}, J.-H. 2019, \apj, 876, 79

\bibitem[{{Kang} {et~al.}(2012){Kang}, {Ryu}, \& {Jones}}]{ka12}
{Kang}, H., {Ryu}, D., \& {Jones}, T.~W. 2012, \apj, 756, 97

\bibitem[{Keshet \& Waxman(2005)}]{PhysRevLett.94.111102}
Keshet, U. \& Waxman, E. 2005, Phys. Rev. Lett., 94, 111102

\bibitem[{{Kierdorf} {et~al.}(2017){Kierdorf}, {Beck}, {Hoeft}, {Klein}, {van
  Weeren}, {Forman}, \& {Jones}}]{2017A&A...600A..18K}
{Kierdorf}, M., {Beck}, R., {Hoeft}, M., {et~al.} 2017, \aap, 600, A18

\bibitem[{Landau \& Lifshitz(1987)}]{Landau1987Fluid}
Landau, L.~D. \& Lifshitz, E.~M. 1987, Fluid Mechanics, Second Edition: Volume
  6 (Course of Theoretical Physics), 2nd edn., Course of theoretical physics /
  by L. D. Landau and E. M. Lifshitz, Vol. 6 (Butterworth-Heinemann)

\bibitem[{{Landsman}(1993)}]{1993ASPC...52..246L}
{Landsman}, W.~B. 1993, in Astronomical Society of the Pacific Conference
  Series, Vol.~52, Astronomical Data Analysis Software and Systems II, ed.
  R.~J. {Hanisch}, R.~J.~V. {Brissenden}, \& J.~{Barnes}, 246

\bibitem[{{Lee} {et~al.}(2009){Lee}, {Deane}, \&
  {Federrath}}]{2009ASPC..406..243L}
{Lee}, D., {Deane}, A.~E., \& {Federrath}, C. 2009, Astronomical Society of the
  Pacific Conference Series, Vol. 406, {A New Multidimensional Unsplit MHD
  Solver in FLASH3}, 243

\bibitem[{{Loi} {et~al.}(2017){Loi}, {Murgia}, {Govoni}, {Vacca}, {Feretti},
  {Giovannini}, {Carretti}, {Gastaldello}, {Girardi}, {Vazza}, {Concu},
  {Melis}, {Paladino}, {Poppi}, {Valente}, {Boschin}, {Clarke},
  {Colafrancesco}, {En{\ss}lin}, {Ferrari}, {de Gasperin}, {Gregorini},
  {Johnston-Hollitt}, {Junklewitz}, {Orr{\`u}}, {Parma}, {Perley}, \&
  {Taylor}}]{Loi_2017}
{Loi}, F., {Murgia}, M., {Govoni}, F., {et~al.} 2017, \mnras, 472, 3605

\bibitem[{{Loi} {et~al.}(2020){Loi}, {Murgia}, {Vacca}, {Govoni}, {Melis},
  {Wittor}, {Kierdorf}, {Bonafede}, {Boschin}, {Brienza}, {Carretti}, {Concu},
  {Feretti}, {Gastaldello}, {Paladino}, {Rajpurohit}, {Serra}, \&
  {Vazza}}]{Loi_2020}
{Loi}, F., {Murgia}, M., {Vacca}, V., {et~al.} 2020, \mnras, 498, 1628

\bibitem[{{Mignone} {et~al.}(2007){Mignone}, {Bodo}, {Massaglia}, {Matsakos},
  {Tesileanu}, {Zanni}, \& {Ferrari}}]{pluto1}
{Mignone}, A., {Bodo}, G., {Massaglia}, S., {et~al.} 2007, \apjs, 170, 228

\bibitem[{{Miyoshi} \& {Kusano}(2005)}]{2005JCoPh.208..315M}
{Miyoshi}, T. \& {Kusano}, K. 2005, Journal of Computational Physics, 208, 315

\bibitem[{{Nuza} {et~al.}(2017){Nuza}, {Gelszinnis}, {Hoeft}, \&
  {Yepes}}]{2017MNRAS.470..240N}
{Nuza}, S.~E., {Gelszinnis}, J., {Hoeft}, M., \& {Yepes}, G. 2017, \mnras, 470,
  240

\bibitem[{{Owen} {et~al.}(2014){Owen}, {Rudnick}, {Eilek}, {Rau}, {Bhatnagar},
  \& {Kogan}}]{2014ApJ...794...24O}
{Owen}, F.~N., {Rudnick}, L., {Eilek}, J., {et~al.} 2014, \apj, 794, 24

\bibitem[{Pais {et~al.}(2018)Pais, Pfrommer, Ehlert, \&
  Pakmor}]{10.1093/mnras/sty1410}
Pais, M., Pfrommer, C., Ehlert, K., \& Pakmor, R. 2018, Monthly Notices of the
  Royal Astronomical Society, 478, 5278

\bibitem[{{Pinzke} {et~al.}(2013){Pinzke}, {Oh}, \&
  {Pfrommer}}]{2013MNRAS.435.1061P}
{Pinzke}, A., {Oh}, S.~P., \& {Pfrommer}, C. 2013, \mnras, 435, 1061

\bibitem[{{Rajpurohit} {et~al.}(2018){Rajpurohit}, {Hoeft}, {van Weeren},
  {Rudnick}, {R{\"o}ttgering}, {Forman}, {Br{\"u}ggen}, {Croston},
  {Andrade-Santos}, {Dawson}, {Intema}, {Kraft}, {Jones}, \&
  {Jee}}]{2018ApJ...852...65R}
{Rajpurohit}, K., {Hoeft}, M., {van Weeren}, R.~J., {et~al.} 2018, \apj, 852,
  65

\bibitem[{{Rajpurohit} {et~al.}(2020){Rajpurohit}, {Hoeft}, {Vazza}, {Rudnick},
  {van Weeren}, {Wittor}, {Drabent}, {Brienza}, {Bonnassieux}, {Locatelli},
  {Kale}, \& {Dumba}}]{2020A&A...636A..30R}
{Rajpurohit}, K., {Hoeft}, M., {Vazza}, F., {et~al.} 2020, \aap, 636, A30

\bibitem[{{Rajpurohit} {et~al.}(2021{\natexlab{a}}){Rajpurohit}, {van Weeren},
  {Hoeft}, {Vazza}, {Brienza}, {Forman}, {Wittor},
  {Dom{\'\i}nguez-Fern{\'a}ndez}, {Rajpurohit}, {Riseley}, {Botteon}, {Osinga},
  {Brunetti}, {Bonnassieux}, {Bonafede}, {Rajpurohit}, {Stuardi}, {Drabent},
  {Br{\"u}ggen}, {Dallacasa}, {Shimwell}, {R{\"o}ttgering}, {de Gasperin},
  {Miley}, \& {Rossetti}}]{Kamlesh_Abell2256}
{Rajpurohit}, K., {van Weeren}, R.~J., {Hoeft}, M., {et~al.}
  2021{\natexlab{a}}, arXiv e-prints, arXiv:2111.04449

\bibitem[{{Rajpurohit} {et~al.}(2022){Rajpurohit}, {van Weeren}, {Hoeft},
  {Vazza}, {Brienza}, {Forman}, {Wittor}, {Dom{\'\i}nguez-Fern{\'a}ndez},
  {Rajpurohit}, {Riseley}, {Botteon}, {Osinga}, {Brunetti}, {Bonnassieux},
  {Bonafede}, {Rajpurohit}, {Stuardi}, {Drabent}, {Br{\"u}ggen}, {Dallacasa},
  {Shimwell}, {R{\"o}ttgering}, {Gasperin}, {Miley}, \&
  {Rossetti}}]{2022ApJ...927...80R}
{Rajpurohit}, K., {van Weeren}, R.~J., {Hoeft}, M., {et~al.} 2022, \apj, 927,
  80

\bibitem[{{Rajpurohit} {et~al.}(2021{\natexlab{b}}){Rajpurohit}, {Wittor}, {van
  Weeren}, {Vazza}, {Hoeft}, {Rudnick}, {Locatelli}, {Eilek}, {Forman},
  {Bonafede}, {Bonnassieux}, {Riseley}, {Brienza}, {Brunetti}, {Br{\"u}ggen},
  {Loi}, {Rajpurohit}, {R{\"o}ttgering}, {Botteon}, {Clarke}, {Drabent},
  {Dom{\'\i}nguez-Fern{\'a}ndez}, {Di Gennaro}, \&
  {Gastaldello}}]{2021A&A...646A..56R}
{Rajpurohit}, K., {Wittor}, D., {van Weeren}, R.~J., {et~al.}
  2021{\natexlab{b}}, \aap, 646, A56

\bibitem[{{Riquelme} \& {Spitkovsky}(2011)}]{2011ApJ...733...63R}
{Riquelme}, M.~A. \& {Spitkovsky}, A. 2011, \apj, 733, 63

\bibitem[{{Roh} {et~al.}(2019){Roh}, {Ryu}, {Kang}, {Ha}, \&
  {Jang}}]{2019ApJ...883..138R}
{Roh}, S., {Ryu}, D., {Kang}, H., {Ha}, S., \& {Jang}, H. 2019, \apj, 883, 138

\bibitem[{{Ryu} {et~al.}(2019){Ryu}, {Kang}, \& {Ha}}]{2019ApJ...883...60R}
{Ryu}, D., {Kang}, H., \& {Ha}, J.-H. 2019, \apj, 883, 60

\bibitem[{{Ryu} {et~al.}(2003){Ryu}, {Kang}, {Hallman}, \& {Jones}}]{ry03}
{Ryu}, D., {Kang}, H., {Hallman}, E., \& {Jones}, T.~W. 2003, \apj, 593, 599

\bibitem[{{Sarazin}(1999)}]{1999ApJ...520..529S}
{Sarazin}, C.~L. 1999, \apj, 520, 529

\bibitem[{{Schmidt} {et~al.}(2006){Schmidt}, {Niemeyer}, \&
  {Hillebrandt}}]{2006A&A...450..265S}
{Schmidt}, W., {Niemeyer}, J.~C., \& {Hillebrandt}, W. 2006, \aap, 450, 265

\bibitem[{{Skillman} {et~al.}(2013){Skillman}, {Xu}, {Hallman}, {O'Shea},
  {Burns}, {Li}, {Collins}, \& {Norman}}]{sk13}
{Skillman}, S.~W., {Xu}, H., {Hallman}, E.~J., {et~al.} 2013, \apj, 765, 21

\bibitem[{{Stroe} {et~al.}(2014){Stroe}, {Rumsey}, {Harwood}, {van Weeren},
  {Rottgering}, {Saunders}, {Sobral}, {Perrott}, \&
  {Schammel}}]{2014MNRAS.441L..41S}
{Stroe}, A., {Rumsey}, C., {Harwood}, J.~J., {et~al.} 2014, \mnras, 441, L41

\bibitem[{{Stroe} {et~al.}(2016){Stroe}, {Shimwell}, {Rumsey}, {van Weeren},
  {Kierdorf}, {Donnert}, {Jones}, {R{\"o}ttgering}, {Hoeft},
  {Rodr{\'\i}guez-Gonz{\'a}lvez}, {Harwood}, \&
  {Saunders}}]{2016MNRAS.455.2402S}
{Stroe}, A., {Shimwell}, T., {Rumsey}, C., {et~al.} 2016, \mnras, 455, 2402

\bibitem[{{Stroe} {et~al.}(2013){Stroe}, {van Weeren}, {Intema},
  {R{\"o}ttgering}, {Br{\"u}ggen}, \& {Hoeft}}]{Stroe2013}
{Stroe}, A., {van Weeren}, R.~J., {Intema}, H.~T., {et~al.} 2013, \aap, 555,
  A110

\bibitem[{{Vaidya} {et~al.}(2018){Vaidya}, {Mignone}, {Bodo}, {Rossi}, \&
  {Massaglia}}]{2018ApJ...865..144V}
{Vaidya}, B., {Mignone}, A., {Bodo}, G., {Rossi}, P., \& {Massaglia}, S. 2018,
  \apj, 865, 144

\bibitem[{Van~Rossum \& Drake~Jr(1995)}]{van1995python}
Van~Rossum, G. \& Drake~Jr, F.~L. 1995, Python reference manual (Centrum voor
  Wiskunde en Informatica Amsterdam)

\bibitem[{{van Weeren} {et~al.}(2011){van Weeren}, {Br{\"u}ggen},
  {R{\"o}ttgering}, \& {Hoeft}}]{2011JApA...32..505V}
{van Weeren}, R.~J., {Br{\"u}ggen}, M., {R{\"o}ttgering}, H.~J.~A., \& {Hoeft},
  M. 2011, Journal of Astrophysics and Astronomy, 32, 505

\bibitem[{{van Weeren} {et~al.}(2019){van Weeren}, {de Gasperin}, {Akamatsu},
  {Br{\"u}ggen}, {Feretti}, {Kang}, {Stroe}, \&
  {Zandanel}}]{2019SSRv..215...16V}
{van Weeren}, R.~J., {de Gasperin}, F., {Akamatsu}, H., {et~al.} 2019, \ssr,
  215, 16

\bibitem[{{van Weeren} {et~al.}(2017){van Weeren}, {Ogrean}, {Jones}, {Forman},
  {Andrade-Santos}, {Pearce}, {Bonafede}, {Br{\"u}ggen}, {Bulbul}, {Clarke},
  {Churazov}, {David}, {Dawson}, {Donahue}, {Goulding}, {Kraft}, {Mason},
  {Merten}, {Mroczkowski}, {Nulsen}, {Rosati}, {Roediger}, {Randall}, {Sayers},
  {Umetsu}, {Vikhlinin}, \& {Zitrin}}]{2017ApJ...835..197V}
{van Weeren}, R.~J., {Ogrean}, G.~A., {Jones}, C., {et~al.} 2017, \apj, 835,
  197

\bibitem[{{van Weeren} {et~al.}(2010){van Weeren}, {R{\"o}ttgering},
  {Br{\"u}ggen}, \& {Hoeft}}]{vw10}
{van Weeren}, R.~J., {R{\"o}ttgering}, H.~J.~A., {Br{\"u}ggen}, M., \& {Hoeft},
  M. 2010, Science, 330, 347

\bibitem[{{van Weeren} {et~al.}(2012){van Weeren}, {R{\"o}ttgering}, {Intema},
  {Rudnick}, {Br{\"u}ggen}, {Hoeft}, \& {Oonk}}]{2012A&A...546A.124V}
{van Weeren}, R.~J., {R{\"o}ttgering}, H.~J.~A., {Intema}, H.~T., {et~al.}
  2012, \aap, 546, A124

\bibitem[{{Vazza} {et~al.}(2016){Vazza}, {Br{\"u}ggen}, {Wittor}, {Gheller},
  {Eckert}, \& {Stubbe}}]{va16scienzo}
{Vazza}, F., {Br{\"u}ggen}, M., {Wittor}, D., {et~al.} 2016, \mnras, 459, 70

\bibitem[{Virtanen {et~al.}(2020)Virtanen, Gommers, Oliphant, Haberland, Reddy,
  Cournapeau, Burovski, Peterson, Weckesser, Bright, {van der Walt}, Brett,
  Wilson, Millman, Mayorov, Nelson, Jones, Kern, Larson, Carey, Polat, Feng,
  Moore, {VanderPlas}, Laxalde, Perktold, Cimrman, Henriksen, Quintero, Harris,
  Archibald, Ribeiro, Pedregosa, {van Mulbregt}, \& {SciPy 1.0
  Contributors}}]{2020SciPy-NMeth}
Virtanen, P., Gommers, R., Oliphant, T.~E., {et~al.} 2020, Nature Methods, 17,
  261

\bibitem[{Wittor {et~al.}(2021)Wittor, Ettori, Vazza, Rajpurohit, Hoeft, \&
  Domínguez-Fernández}]{Wittor_2021}
Wittor, D., Ettori, S., Vazza, F., {et~al.} 2021, Monthly Notices of the Royal
  Astronomical Society, 506, 396–414

\bibitem[{{Wittor} {et~al.}(2019){Wittor}, {Hoeft}, {Vazza}, {Br{\"u}ggen}, \&
  {Dom{\'\i}nguez-Fern{\'a}ndez}}]{wittor2019}
{Wittor}, D., {Hoeft}, M., {Vazza}, F., {Br{\"u}ggen}, M., \&
  {Dom{\'\i}nguez-Fern{\'a}ndez}, P. 2019, \mnras, 490, 3987

\bibitem[{{Xu} {et~al.}(2020){Xu}, {Spitkovsky}, \&
  {Caprioli}}]{2020ApJ...897L..41X}
{Xu}, R., {Spitkovsky}, A., \& {Caprioli}, D. 2020, \apjl, 897, L41

\end{thebibliography}

\appendix

\section{Spectral features}
\label{app:extra_features}

\begin{figure*}
    \includegraphics[width=0.47\textwidth,height=7.5cm]{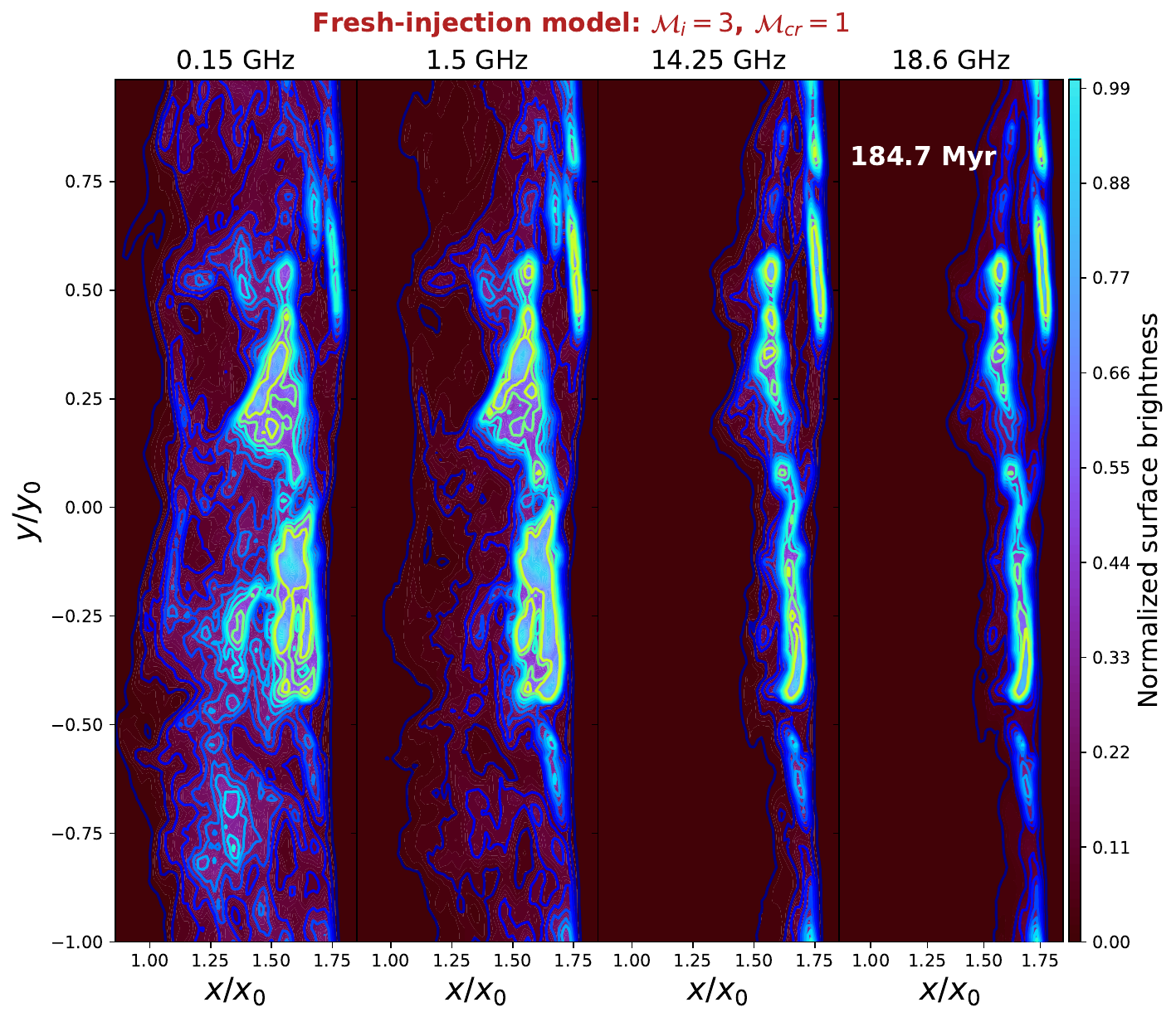}
    \includegraphics[width=0.47\textwidth,height=7.5cm]{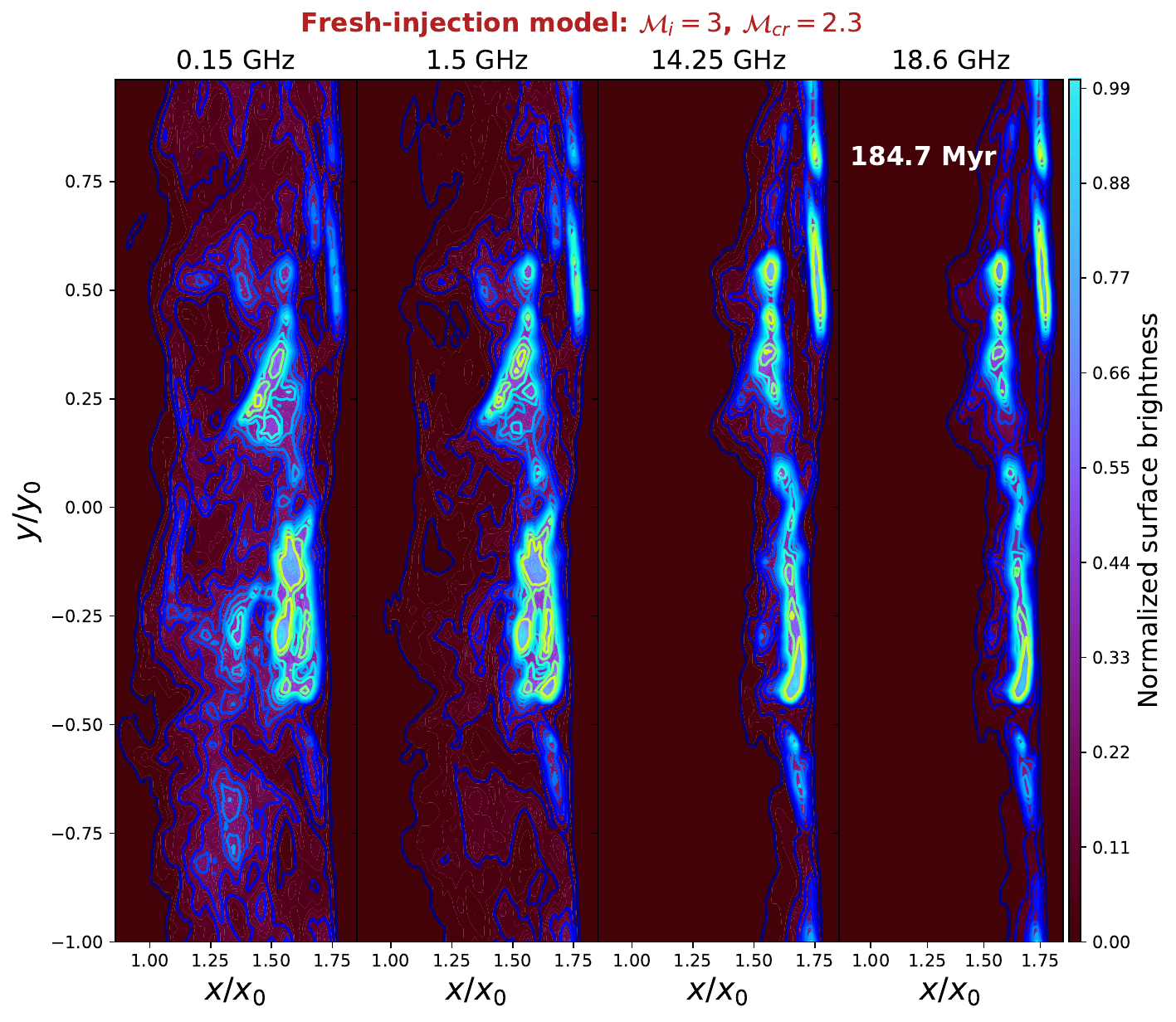}
  \includegraphics[width=0.47\textwidth,height=7.5cm]{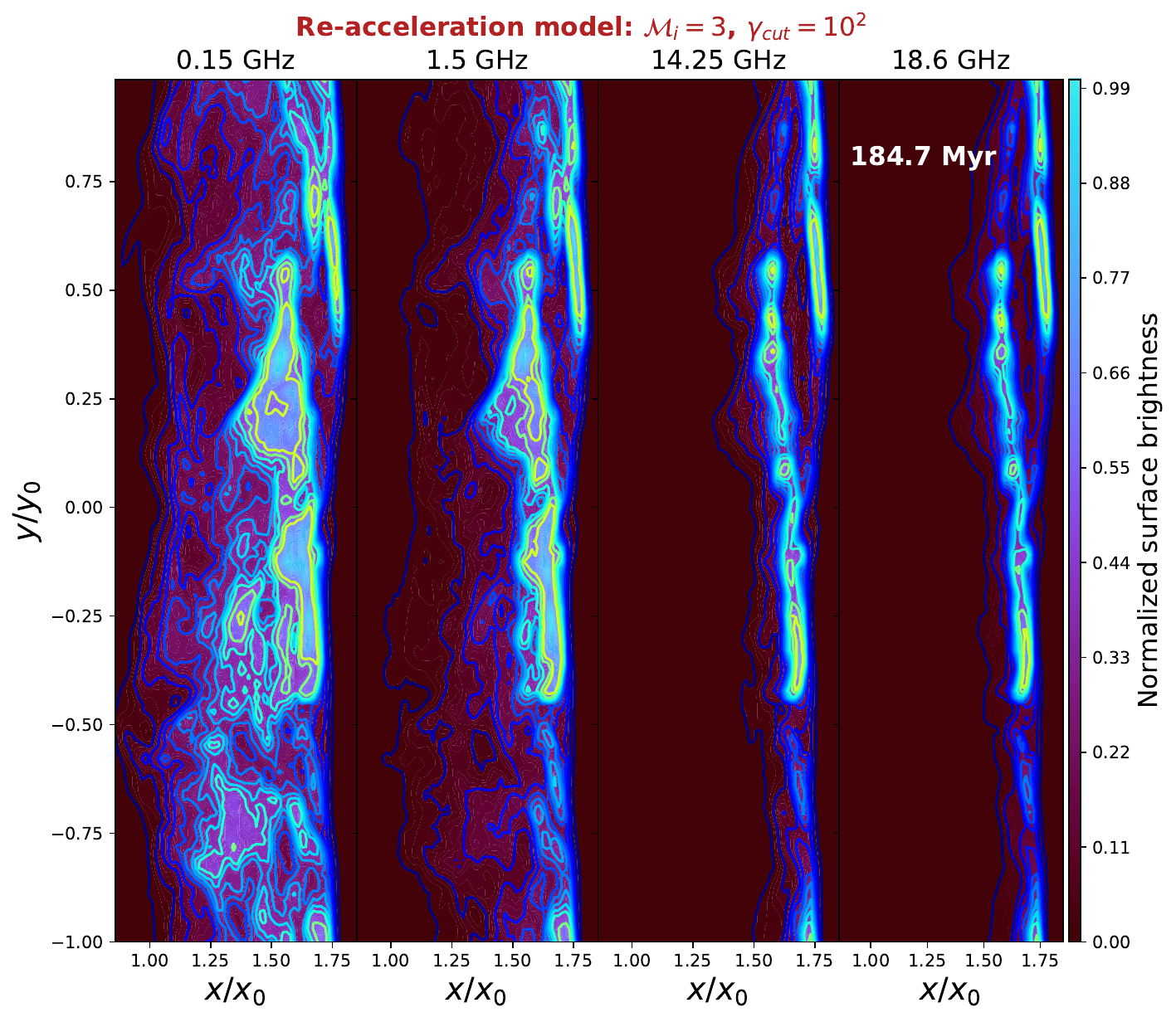}
    \includegraphics[width=0.47\textwidth,height=7.5cm]{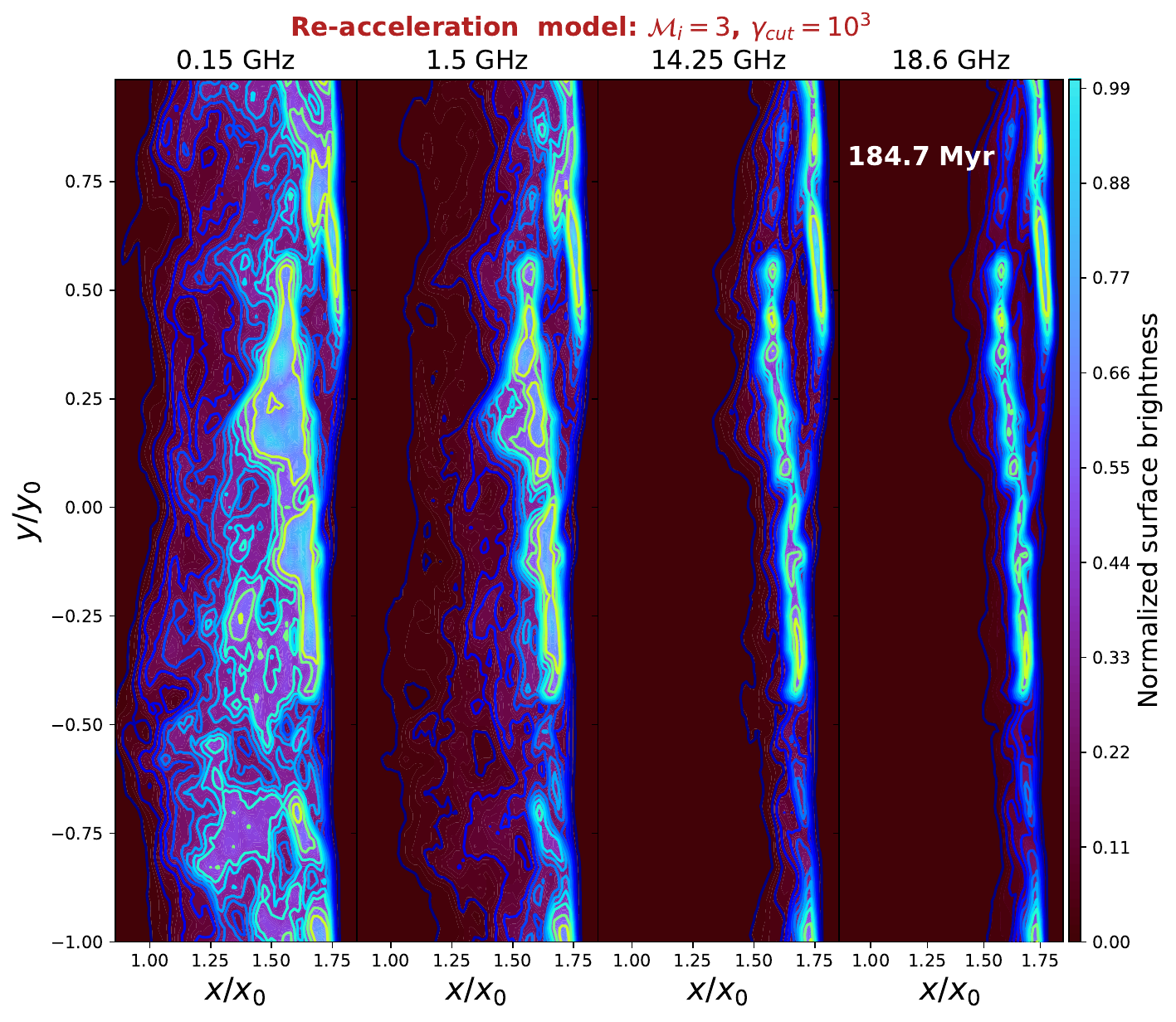}
    \includegraphics[width=0.47\textwidth,height=7.5cm]{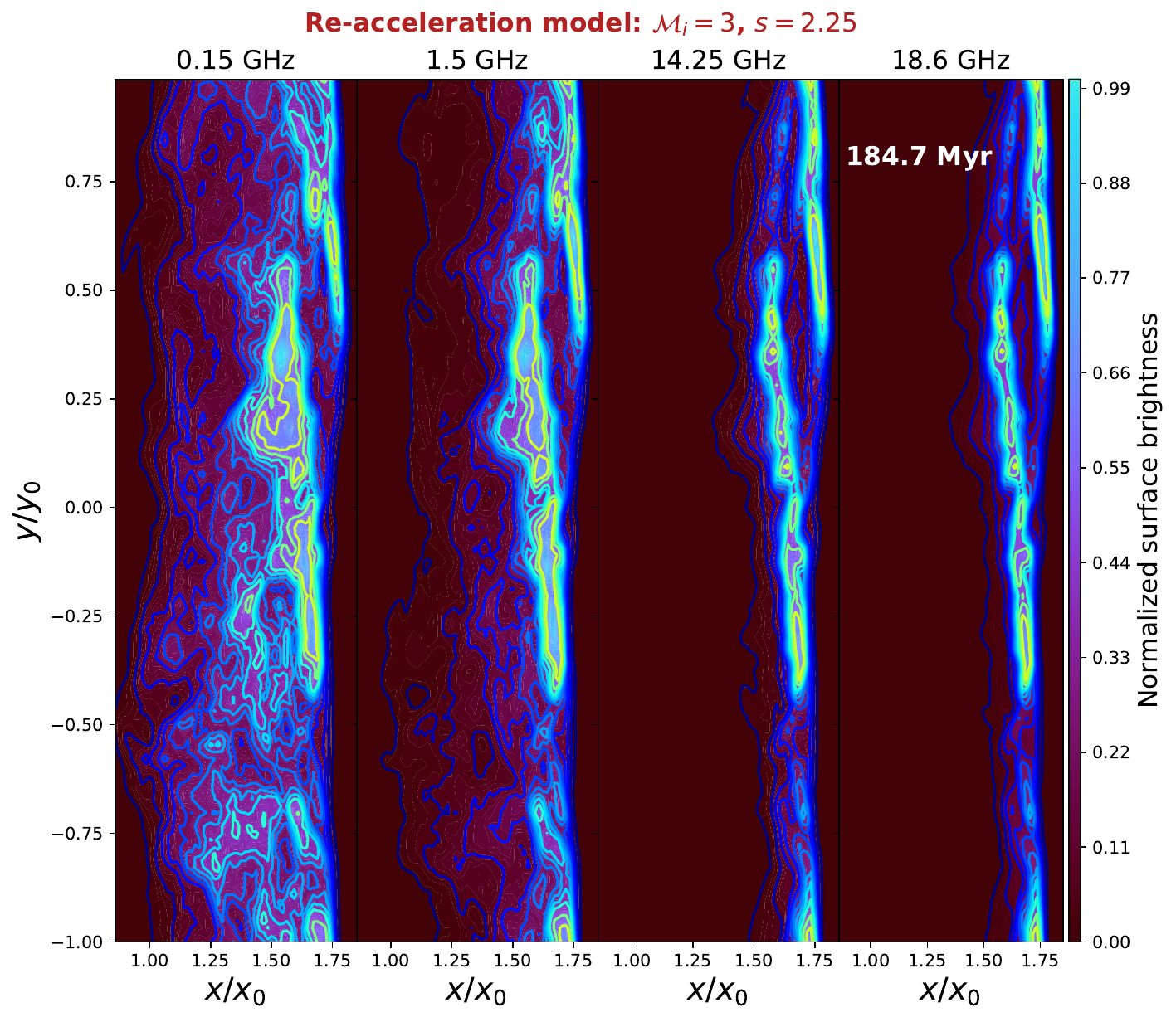} \hspace{5mm}
    \includegraphics[width=0.47\textwidth,height=7.5cm]{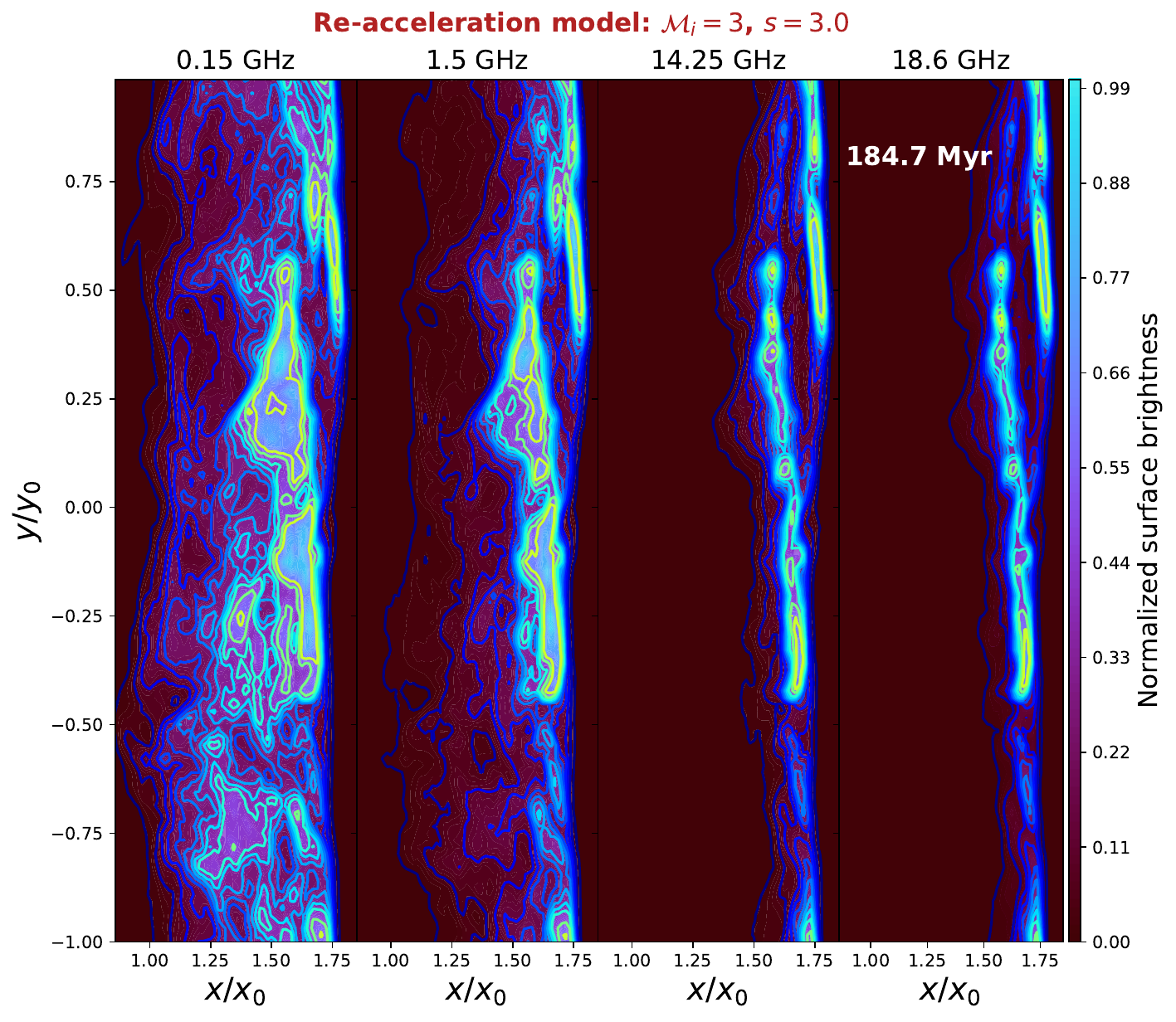}
\caption{Same as Fig.~\ref{fig:maps_all_otherstyle} but at the simulation time-step 187.4 Myr.
    }
    \label{fig:maps_donwstream}
\end{figure*}

\begin{figure}
    \centering
    \includegraphics[width=0.8\columnwidth]{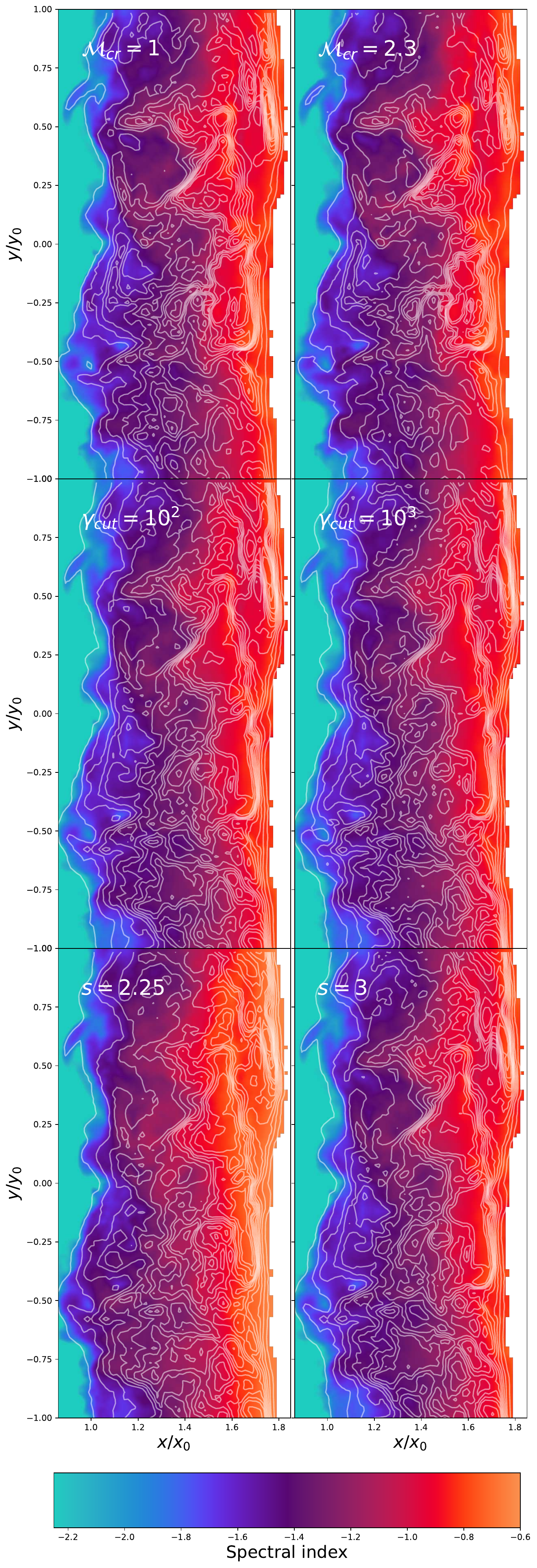}
    \caption{Spectral index maps obtained between 150 MHz and 650 MHz for the $\mathcal{M}_i=3$ at $t=184.7$ Myr. The first row corresponds to the fresh-injection model and the second and third rows to the re-acceleration model assuming Dirac Delta and power-law fossil electrons, respectively. The first (second) columns correspond to $\mathcal{M}_{cr} = 1$ ($\mathcal{M}_{cr} = 2.3$), $\gamma_{cut} = 10^2$ ($\gamma_{cut} = 10^3$) and $s = 2.25$ ($s=3$), respectively.
    }
    \label{fig:spec_idx}
\end{figure}

\begin{figure}
    \centering
    \includegraphics[width=\columnwidth]{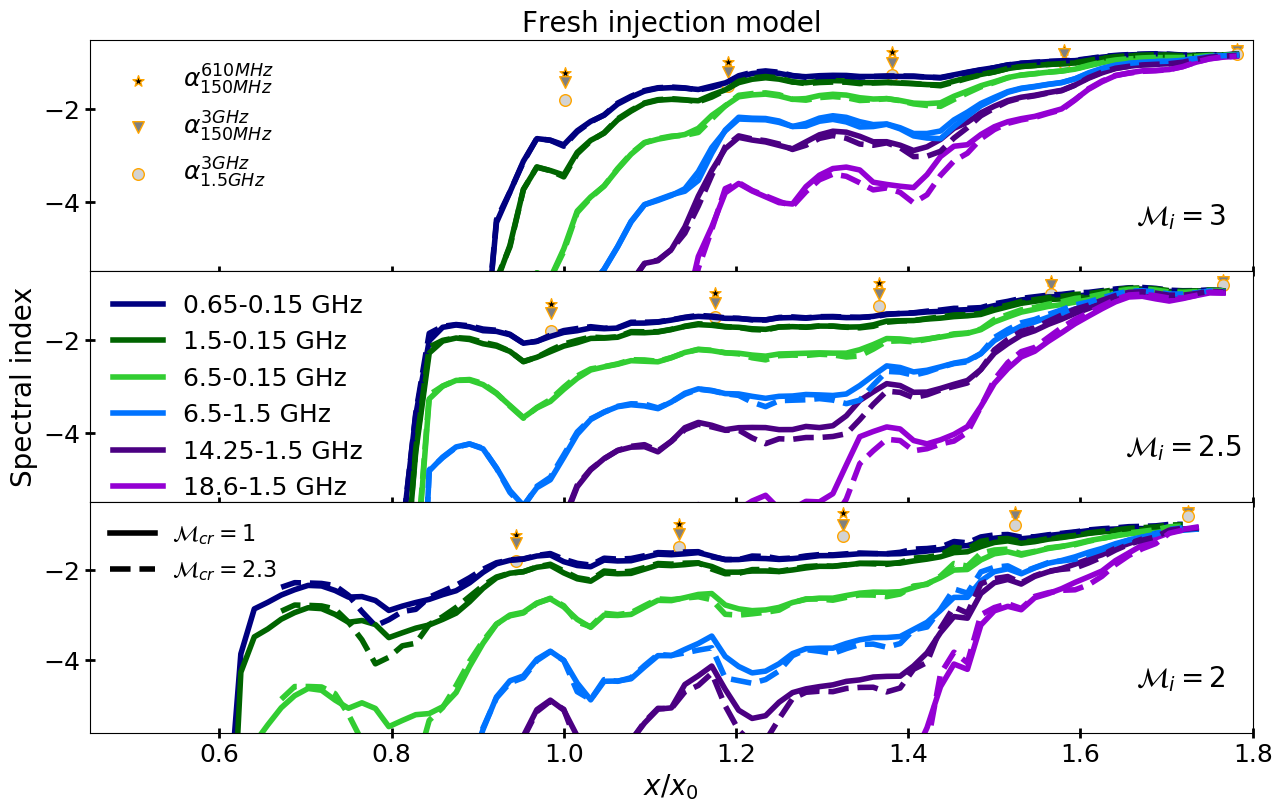}
    \includegraphics[width=\columnwidth]{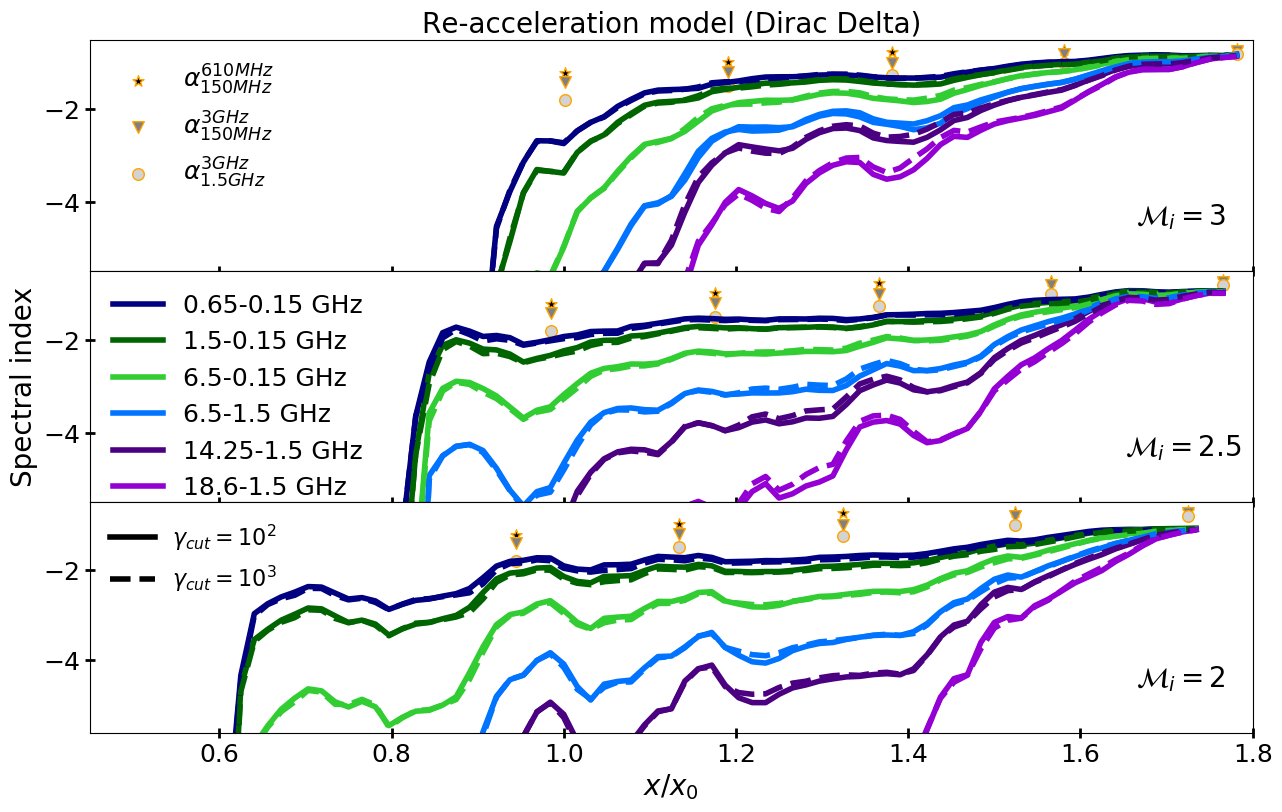}
     \includegraphics[width=\columnwidth]{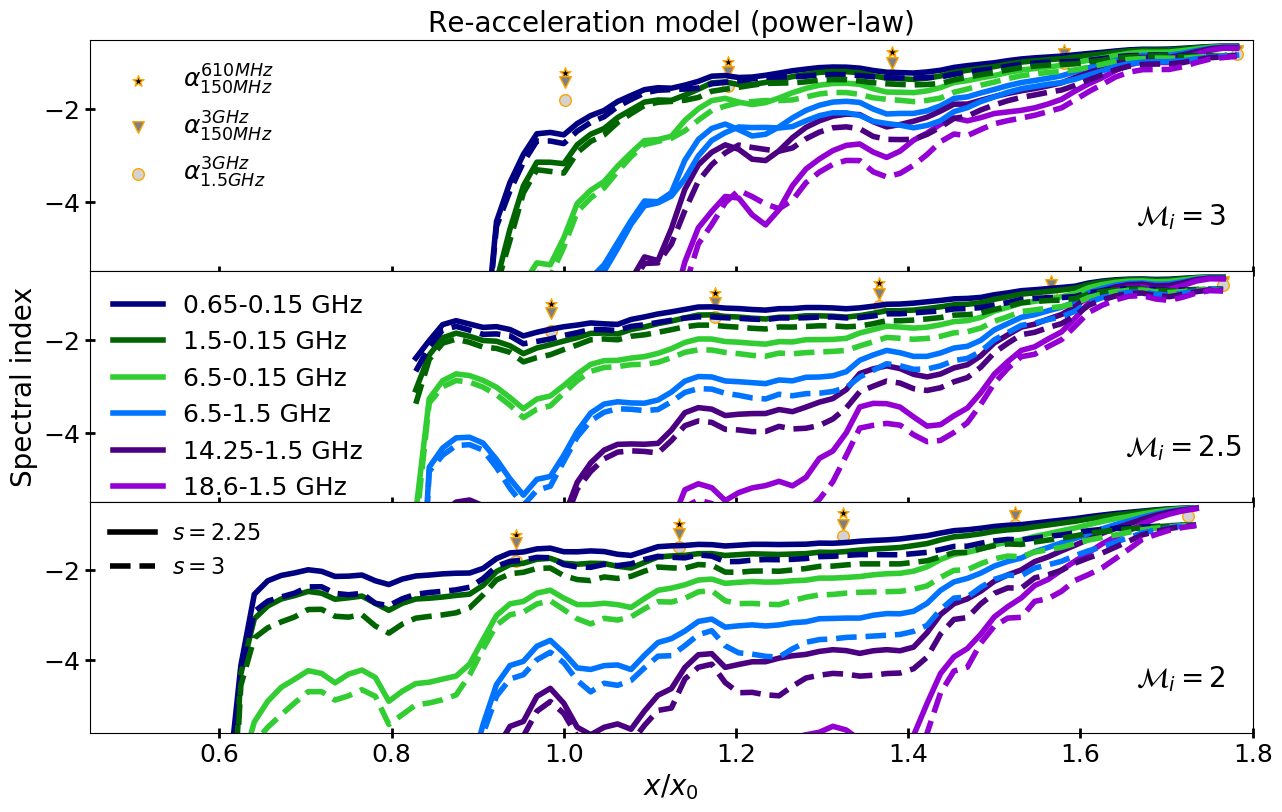}
    \caption{Spectral index profiles for all the models at $t=184.7$ Myr. The first row corresponds to the fresh-injection model and the second and third rows to the re-acceleration model assuming Dirac Delta and power-law fossil electrons, respectively. The data points are taken from Fig.~10 of \citealt{2018ApJ...865...24D}.
    }
    \label{fig:spec_idx_prof}
\end{figure}

In this Appendix, we show additional information about the modelled radio emission in our simulations with the purpose of providing a complete view of the simulated radio shock. In Fig.~\ref{fig:maps_donwstream}, we show the radio shock at $t=184.7$ Myr. In this case, the shock front is located almost at the right end of region III at $x\sim1.75$.

In Fig.~\ref{fig:spec_idx}, we show the spectral index maps of the $\mathcal{M}_i=3$ case. We computed these maps with the following equation:
\begin{equation}\label{eq:spectral_index}
    - \alpha(x,y) =
    \frac{\log \left[I_{\nu_{2}}(x,y)/I_{\nu_{1}}(x,y) \right]}{\log(\nu_{2} / \nu_{1})},
\end{equation}
where $I_{\nu}$ was defined in Eq.~\ref{eq:intensity}. In Fig.~\ref{fig:spec_idx}, we considered $\nu_1 = $150 MHz and $\nu_2=650$ MHz. We see the expected spectral index gradient starting from the shock
front (orange) to the end of the downstream region (blue). The spectral index values range between  $\alpha\sim -0.6$ and $\alpha \sim -2.1$.
In Fig.~\ref{fig:spec_idx_prof}, we show the spectral index profiles between various observing frequencies. We additionally plot the data points from Fig.~10 in \citealt{2018ApJ...865...24D}. While in this work we do not aim at reproducing any specific radio relic, we find it useful for the reader to see a comparison with observational data. We choose to show the data from this particular relic because it is believed to be observed almost edge-on \citep[see][]{2011JApA...32..505V}. The Mach number of the Sausage relic has been estimated to be $\mathcal{M}_{\mathrm{X-ray}}= 2.7$ from X-ray observations \citep{2015A&A...582A..87A}. On the radio band, the Mach number has been estimated to be $\mathcal{M}_{\mathrm{radio}}= 4.6$ \citep{vw10}, $\mathcal{M}_{\mathrm{radio}}= 2.9$ \citep{2014MNRAS.441L..41S}, $\mathcal{M}_{\mathrm{radio}}= 2.7$ \citep{Hoang2017} and $\mathcal{M}_{\mathrm{radio}}= 2.58$ \citep{2018ApJ...865...24D}. 
Some of these estimations would be comparable to the initial Mach number $\mathcal{M}_i$ used in some of our simulations (see upper or middle panels of Fig.~\ref{fig:spec_idx_prof}). Nevertheless, as mentioned in the main text, we stress that in this manuscript we do not aim at reproducing any particular observed radio relic. 


\section{Extra models}
\label{app:extra_models}

\begin{figure}
    \centering
    \includegraphics[width=\columnwidth]{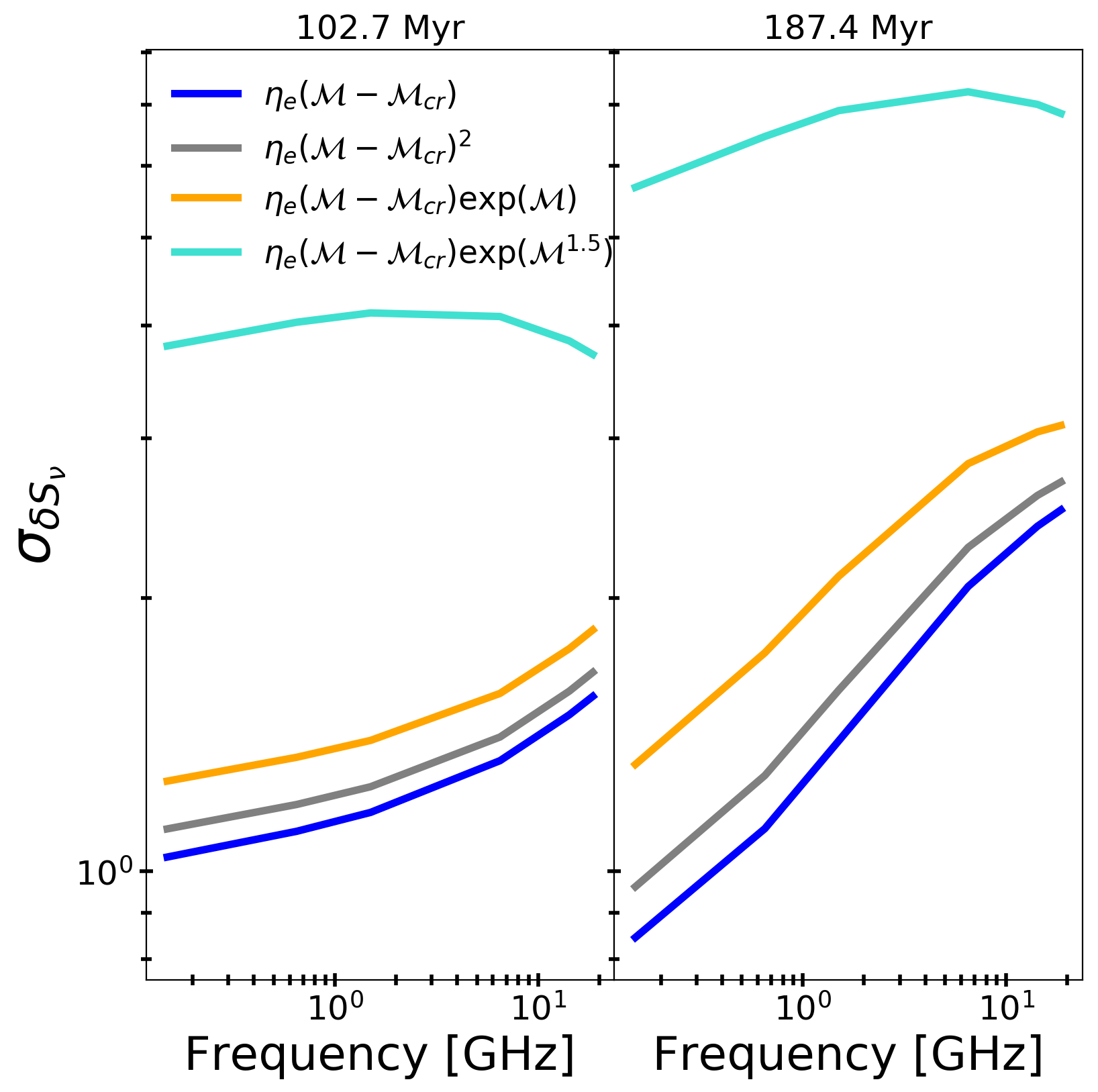}
    \caption{Standard deviation of the $\delta S_{\nu}$ distribution at the shock-front for the $\mathcal{M}_i=3$ case at all frequencies for the $\mathcal{M}_{cr}=1$ fresh injection model. In this case, we explore different $\eta({\mathcal{M}})$ functions (see legend).
    The left (right) panel corresponds to the simulation time-steps $t=102.7$ Myr and $t=187.4$ Myr.
    }
    \label{fig:std_extras}
\end{figure}

In this Appendix we show the results from extra runs in which we have explored different $\eta({\mathcal{M}})$ efficiency functions in the fresh-injection model. In principle, $\eta({\mathcal{M}})$ remains to be an unknown for weak shocks in the ICM. Nevertheless, various models depending on the Mach number have been proposed in the literature \citep[][]{Kang_2007,kr13,Caprioli_2014,2019ApJ...883...60R}. These models have one thing in common, 
$\eta(\mathcal{M})$ is an increasing function with the Mach number. In Fig.~\ref{fig:std_extras} we show the resulting standard deviation of the $\delta S_{\nu}$ for different runs with different $\eta(\mathcal{M})$ toy models. We considered only the $\mathcal{M}_i=3$ case. 
The steepest $\eta(\mathcal{M})$ model (shown in color turquoise in Fig.~\ref{fig:std_extras}) leads to the highest level of patchiness as seen by the highest values of $\sigma_{\delta_{S_{\nu}}}$. Yet, this very steep model reveals almost no change in 
the trend of the standard deviation with frequency.
Since steeper $\eta(\mathcal{M})$ models give even more weight to those regions at the shock front with the largest Mach numbers, the relic would look as patchy at high frequencies as it looks at low frequencies.
To summarize, we find that 1) a steeper $\eta(\mathcal{M})$ model leads to a higher the degree of patchiness or more substructure in the radio emission and 2) if the $\eta(\mathcal{M})$ model is too steep, the trend where $\sigma_{\delta_{S_{\nu}}}$ increases with frequency (see Sec.~\ref{sec:patchiness}) is erased.

\section{Dependence on the Mach number}
\label{app:Eradio_Mach}

\begin{figure}
    \centering
    \includegraphics[width=\columnwidth]{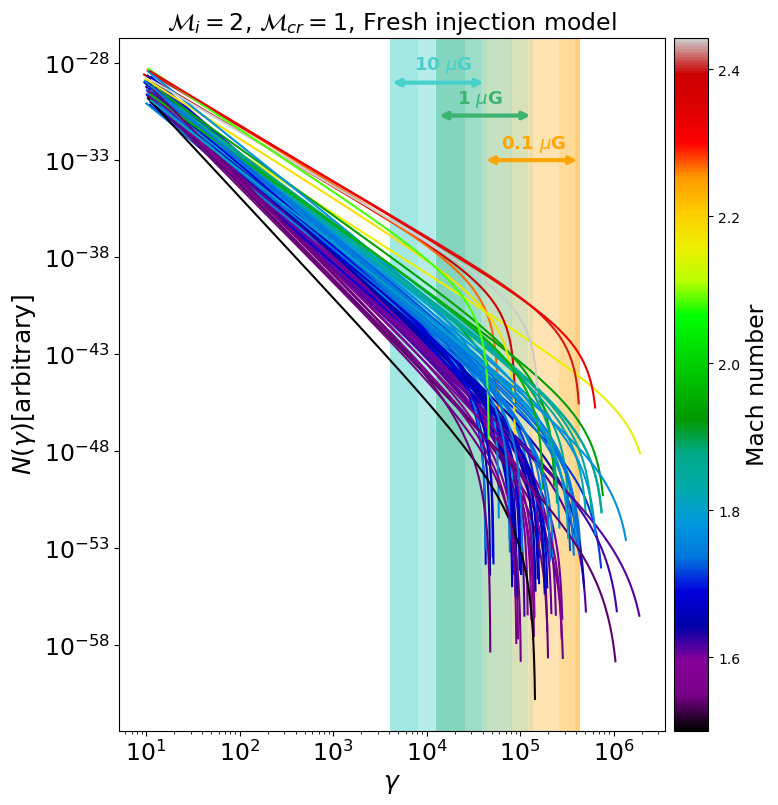}
    \caption{Energy spectra of 100 randomly selected particles near the shock front in the $\mathcal{M}_i=2$ case with $\mathcal{M}_{cr}=1$. The energy spectra are coloured by its corresponding Mach number. The shaded areas denote the energy range corresponding to the critical frequencies band 0.15--18.6 GHz (see Eq.~\ref{J_pol}) assuming 0.1 $\mu$G (\textit{orange}), 1 $\mu$G (\textit{green}) and 10 $\mu$ G (\textit{blue}).
    }
    \label{fig:spectra_100part}
\end{figure}

\begin{figure}
    \centering
    \includegraphics[width=0.65\columnwidth]{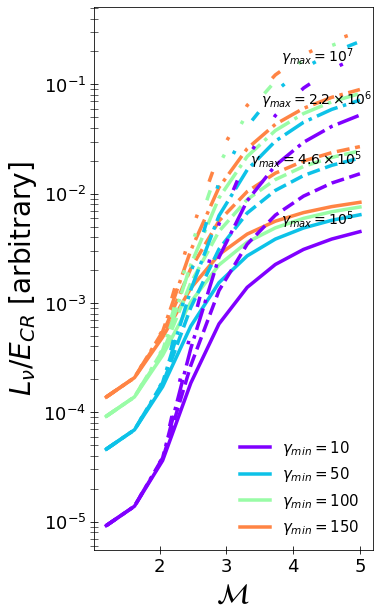}\\
    \includegraphics[width=0.65\columnwidth]{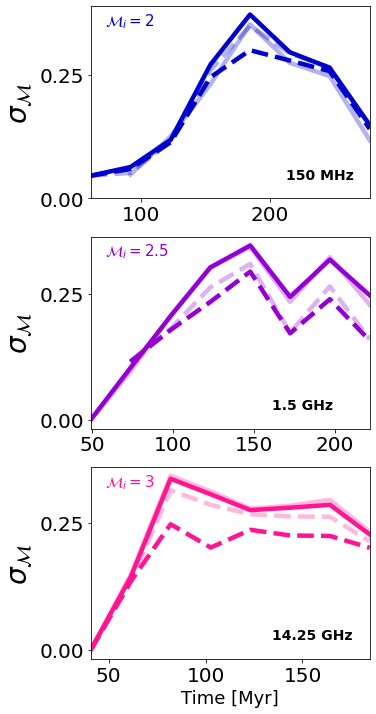}
    \caption{\textit{Left}: Ratio of synchrotron and CRe energies as function of the Mach number for different minimum and maximum energy limits. A magnetic field of 1 $\mu$G was assumed. \textit{Right}: Same as bottom row in Fig.~\ref{fig:mach_evol}, but using as weight both the fresh-injection model (solid lines; 
    $\mathcal{M}_{cr}=2.3$ with higher transparency) and the re-acceleration model (dashed lines; $s=3$ with higher transparency).}
    \label{fig:ratio_simplified}
\end{figure}

In this Appendix we show in a simple way how low Mach number shocks lead to the largest discrepancies in the energy spectra of the particles compared to high Mach number shocks.
As can be seen from Eq.~(\ref{eq:17}), the synchrotron emission mainly depends on the particle energy spectrum and on the Bessel function integral. Nevertheless, as a first order approximation, the radio luminosity produced by a given radio source can be estimated as
\begin{equation}\label{eq:Lnu}
    L_{\nu} \propto - \int \left( \frac{dE}{dt} \right)_{sync} N(E) dE,
\end{equation}
where the synchrotron power emitted per unit time is proportional to $E^2$.  
Thinking about the fresh injection model, the initial energy distribution of the particles is given by Eq.~\ref{eq:energy_dist} where the injection spectral index $p$ depends on the shock's Mach number. In Fig.~\ref{fig:spectra_100part} we show the energy spectra of various particles in one of our simulations. The lower the Mach number, the steeper the spectrum is. Apart from the fact that each Lagrangian particle representing an ensemble of CRe is subject to individual radiative losses or ageing, one can observe that the differences in energy between low and high Mach number spectra can be significant (compare black and red lines in Fig.~\ref{fig:spectra_100part}). Moreover, these differences are enlarged in the radio band which are shown as shaded areas. 
These differences along with the normalization factors are exactly what gives rise to a high degree of patchiness in the fresh injection model. 
The radio substructure will be brightest at those regions characterized by the highest Mach numbers of the shock's distribution, i.e. the high end tail of the Mach number distribution, and it will be patchier 
if the shock's Mach number distribution is encompassed by a large fraction of low Mach numbers. 

Following Eq.~\ref{eq:Lnu}, the ratio of radio emission to the CRe energy can then be estimated in a simplified way by
\begin{equation}
    \frac{L_{\nu}}{E_{CR}} \propto \frac{(2 - p)\left[ E_{max}^{3-p} - E_{min}^{3-p} \right]}{(3-p)\left[ E_{max}^{2-p} - E_{min}^{2-p} \right]}.
\end{equation}

In Fig.~\ref{fig:ratio_simplified} we show how this ratio changes considering different energy limits as a function of the shock Mach number. 
We can see that a steeper energy spectrum (corresponding to low Mach numbers) leads to lower radio luminosity in general. Focusing in a fixed maximum energy with Lorentz factor of $\gamma_{max}=10^7$ and a minimum of $\gamma_{min}=10$ (solid purple lines in the left panel of Fig.~\ref{fig:ratio_simplified}), the difference between $\mathcal{M}\sim2$ and $\mathcal{M}\sim3$  can be as high as $\sim4$ orders of magnitude. 

Finally, in the left panel of Fig.~\ref{fig:ratio_simplified} we show the standard deviation of the projected Mach number of the shock weighted with the radio emission coming from the fresh-injection and the re-acceleration models. This figure is complementary to the panels in the bottom row of Fig.~\ref{fig:mach_evol}.
As mentioned before, the fresh injection model produces more patchy radio relics than the acceleration model. 
As a consequence, the $\sigma_{\mathcal{M}}$ obtained weighting with the emission from the fresh injection model is larger than the  $\sigma_{\mathcal{M}}$ where the radio emission from the re-acceleration model is used.

\end{document}